\documentclass[aps,showpacs,onecolumn,floatfix,amsmath,amssymb,nofootinbib]{revtex4}
\usepackage{graphics}
\usepackage{epsfig}
\usepackage[english]{babel}
\usepackage{amsmath} 
\usepackage{verbatim} 
\usepackage{mathrsfs}
\usepackage{amsfonts}
\usepackage{amssymb}
\usepackage{soul} 
\usepackage[latin1]{inputenc}
\usepackage{hyperref}
\usepackage{float}
\usepackage{graphicx}
\usepackage{subfig}
\usepackage{color}
\usepackage{tabulary}
\newcolumntype{K}[1]{>{\centering\arraybackslash}p{#1}}

\DeclareMathOperator{\csch}{csch}
\newcommand{\beq}{\begin{equation}}
\newcommand{\eeq}{\end{equation}}
\newcommand{\barr}{\begin{eqnarray}}
\newcommand{\earr}{\end{eqnarray}}
\newcommand{\bea}{\begin{eqnarray*}}
\newcommand{\eea}{\end{eqnarray*}}

\begin{document}


\title{Equivalence Principle in Chameleon Models}

\author{Lucila Kraiselburd$^{1,2}$}
\email{lkrai@fcaglp.unlp.edu.ar}

\author{Susana J. Landau$^{2,3}$}
\email{slandau@df.uba.ar}

\author{Marcelo Salgado$^{4}$}
\email{marcelo@nucleares.unam.mx}

\author{Daniel Sudarsky$^{4,5}$}
\email{sudarsky@nucleares.unam.mx}

\author{H\'{e}ctor Vucetich$^{1}$}
\email{vucetich@fcaglp.unlp.edu.ar}

\affiliation{$^1$Grupo de Astrof\'{i}sica, Relatividad y Cosmolog\'{i}a, Facultad de Ciencias Astron\'omicas y Geof\'{\i}sicas, 
Universidad Nacional de La Plata, Paseo del Bosque S/N (1900) La Plata, Argentina \\ $^2$CONICET,Godoy Cruz 2290, 1425 Ciudad Aut\'onoma de Buenos Aires, Argentina \\
$^3$Departamento de F\'{\i}sica, Facultad de Ciencias Exactas y Naturales, Universidad de Buenos Aires and IFIBA, 
Ciudad Universitaria - Pab. I, Buenos Aires 1428, Argentina \\ $^4$Instituto de Ciencias Nucleares, Universidad Nacional Aut\'{o}noma de M\'{e}xico,
A.P. 70-543, M\'{e}xico D.F. 04510, M\'{e}xico\\
$^5$Department of Philosophy, New York University,  New York, NY 10003,   United States of América
}


\date{\today}


\begin{abstract}
Most theories that predict time and/or space variation of fundamental constants also predict violations of the Weak Equivalence Principle (WEP).  In  2004 Khoury and Weltman~\cite{KW04} proposed the  so called  {\it chameleon field} arguing that it could help avoiding experimental bounds on the WEP while having a non-trivial   cosmological impact. In this paper we revisit  the  extent  to  which   these  expectations  continue  to hold as  we  enter the regime of high precision tests. The basis  of the study is  the  development of  a {\it new} method  for computing the force  between  {\it two  massive bodies  induced  by the  chameleon field} which takes into account the   influence  on  the field  by  both,   the {\it large} and the {\it test} bodies. We confirm that in the {\it thin shell} regime the  force  does  depend  non-trivially  on the {\it test} body's composition,  even when the chameleon coupling constants $\beta^i=\beta$ are universal.  We also propose a simple criterion based on energy minimization, that  we  use  to determine which  of  the  approximations used in  computing  the    scalar  field  in a two  body  problem   is   better in   each  specific regime.  As  an application  of our analysis we  then compare the  resulting differential acceleration of two {\it test} bodies with the corresponding  bounds obtained from E\"{o}tv\"{o}s type experiments.  We consider two setups: 1) an Earth based experiment where the {\it test} bodies are 
made of {\rm Be} and {\rm Al}; 2) the Lunar Laser Ranging experiment. We find that for some choices of the free parameters of the chameleon model 
the predictions of the  E\"{o}tv\"{o}s parameter are larger  than some of the previous estimates. As a consequence, we put new constrains on 
 these free parameters. Our conclusions strongly suggest that the   properties  of immunity from  experimental tests of the WEP,  usually  attributed to the   chameleon and related  models,   should be   carefully reconsidered. An important result of our analysis is that our approach leads to new constraints on the  parameter space of the chameleon models.
\end{abstract}

\maketitle

\section{Introduction}
\label{sec:introduction}

The Weak Equivalence Principle (WEP) is  a cornerstone of the Einstein's Theory of General Relativity (hereafter GR), and  can  be  broadly seen as implying the  universal coupling between matter and gravity.  That is  usually taken as  translating  directly into the universality of free fall (UFF), and  thus   as a  feature easily  analyzed and  directly testable.   The first point we  want  to   make  is that the  situation is  substantially  more complex 
 than previously thought. This is  due  to the fact that  the  WEP  is, strictly speaking,
only meant to hold  for test point-like objects (i.e.  a  principle meant to apply locally) that do not affect the gravitational environment around them. 
However, in reality no such objects exist in nature, even if the test-point-like assumption is in occasions a good approximation. All objects have, in principle,  a non-vanishing energy-momentum tensor,  and  as  such, they all modify  the  geometry  of space-time. Those  modifications,  in turn, affect the motion of the object themselves. In fact,  taking into account the  so  called ``back reaction effects" is a highly nontrivial task (see for instance \cite{Wald1972}). Moreover, 
given that all  real objects have a finite  spatial extension, one requires,  for a rigorous description of  their  motion, the determination of something akin to 
the notion of the ``center of mass  world-line", a task that is highly nontrivial  in  general space-times \cite{Papapetrou1951}. The results of  the detailed analyses  indicate that those  world-lines do not, in general, correspond to geodesics. Quantum aspects further  complicate the    analysis  of the motion of even  the simplest  things,   which one  might  want to  take as    paradigmatic   point like objects, such as photons\cite{Drummond}.   Even though  such  UFF violating effects  are normally  very small, they are always  there in principle, and therefore these considerations  should   serve as  warnings  when  we go on to analyze more  complex  situations. 

Of  particular interest  for us  here   will be any theory in which the local coupling constants   are taken to be effectively  space-time dependent, while respecting the principles of locality and  general covariance,  as those  entail  some  kind of fundamental field  controlling the  spatial dependence,  with the field,  generically,   coupling in different manners to the various types of matter.  Such  field, usually taken to be a  scalar field, generically  mediates new  forces between  {\it macroscopic} objects, which  would look, at the empirical level,  as modifications of gravitation which  might,    in principle,  lead to effective violations of the UFF  for  {\it test} bodies in external gravitational fields.  For this  reason, most theories that predict variations of fundamental constants also predict effective  violations of the WEP~\cite{Bekenstein82,Barrow02,Olive02,DP94,Palma03}. For instance,  the  rest  energy  of a macroscopic  body is made of many contributions related to  the energies  associated   with various  kinds of  interactions  (strong, weak, electromagnetic)  and such components   would  be  affected   differently    by   a  light scalar field. From the experimental point of view,  one needs to   confront   the very strong  limits on possible violations of the WEP that come from E\"{o}tv\"{o}s-Roll-Krotkov-Dicke and Braginsky-Panov experiments and their modern  reincarnations~\cite{Adel,RKD64,Braginski72,KeiFall82,1994PhRvD..50.3614S,2008PhRvL.100d1101S} which explore the differential acceleration of {\it test} bodies in  gravitational contexts. In fact current  bounds  reach  sensitivities  of order    $\frac{\Delta a}{a} \simeq 10^{-13}$ (or more),  and thus  can, in principle, constrain the viability of many  models. 

Recently,  there  has  been a great   level of interest in models that  claim to be able to avoid the stringent  bounds  resulting from    experimental tests looking for   violations of the  WEP
based on schemes  where  the effects  of the fields are hidden by suitable non-linearities, as in  the  chameleon models and the Dilaton-Matter-gravity model with strong coupling~\cite{Damour02}. 
Chameleon models were introduced by Khoury and Weltman in 2004~\cite{KW04} and have been further developed 
by  several  authors~\cite{Brax04,MS07,Brax07,Hui2009,Brax2010,BB11,Brax2012,Upadhye12,K13}. Generically a {\it chameleon} consists of a scalar field 
that is coupled non-minimally to matter and minimally to the curvature (or the other way around via a conformal transformation),  
and where the  field's  effective mass depends on the  density and pressure of the matter that  constitutes  the  {\it environment}. This, in turn,   is  the result of the  nontrivial    coupling  of   the  scalar field with the trace of the energy-momentum tensor of  the matter sector of the theory~\footnote{In this context by {\it matter} we mean any field other that the scalar-field at hand, which in the situation of interest would be the ordinary matter making up the objects  present in the  experiment  including the  atmosphere. }. The  coupling  of the scalar field  with   matter  might  be {\it non-universal} (i.e.  it may depend  on the  particle's species), like in the original model introduced in~\cite{KW04}, or universal as  in  some  simpler models.

In their work Khoury and Weltman~\cite{KW04} analyzed the chameleon field associated with a single body using a ``linear'' approximation in the equation, 
and corroborated that the corresponding solution looks very similar to the numerical solution of the full non-linear equation

This approximation to the one body problem will be referred hereafter to as {\it the standard approach} \footnote{See  Appendix \ref{sec:OBP} for a short review of the standard approach for the one body problem and a discussion of the {\it thin shell} condition}. Their conclusion was that the bounds imposed by the experiments   testing the WEP can be satisfied (even if the coupling constants are of order 
unity) provided that the bodies involved in the relevant experiments generate  the so called {\it thin shell effect}. In this {\it thin shell} regime, the spatial variations of the  scalar field take place  just  on a small region near the body's surface thus preventing the scalar field from actually  exerting any force on most of the body. A further analysis by  Mota and Shaw~\cite{MS07} strengthened the conclusion  that the non linearities 
inherent to the chameleon models are, in fact, responsible for  suppressing the effective  violation of  WEP  in the actual experiments even when the coupling constants are very large, and not  just when they  are of order unity as it was  initially  thought~\cite{KW04}
\footnote{When analyzing the two-body problem, Mota \& Shaw~\cite{MS07} assumed that both bodies can be treated as two semi-infinite ``blocks" 
and then solved the non-linear equation for the chameleon. After  making several approximations 
they  concluded that the force between the bodies is composition-independent.  We  thus  worried that  in  such approximations  the mass of both bodies would become infinite, and  the  magnitude of  some  potentially problematic  terms was not  estimated.}.
Moreover, these authors argued  that   the predicted  effective  violations of the   WEP in low density environments (like in space-based laboratories) 
would be further  suppressed for some  adjusted  values of  the parameters of the scalar-field potential.

Based  on such  analyses,  a good   part  of the community  working  on the   area~\cite{KW04,MS07,TT08,Hui2009,Brax2010,Saidi11,Brax2012,K13,PO15} has come to believe  the chameleon fields should not lead to any relevant  forces in experiments on Earth designed  to  search for   a dependence on the composition of the  acceleration of a  falling {\it test} body, or in general, to any   observable interaction between  ordinary bodies  mediated by chameleon-type fields. 

However, it should be noted that there is no universal  consensus  in the community regarding  even  qualitative  aspects of  the theoretical  predictions,  with   some    arguments  indicating   the chameleon force on the {\it test} body is  composition  dependent~\cite{KW04,MS07,TT08,Brax2010,Hui2009,Saidi11,K13} and others indicating  that it is  not~\cite{PO15}. Defining a ``screened"  {\it test} body one in which  the {\it thin shell} condition is satisfied and an ``unscreened"  {\it test} body one in which it is not satisfied, most authors make different predictions on the composition dependence of the chameleon mediated force \cite{KW04,MS07}. It is also important to recall that there are two different expressions for the {\it thin shell} condition in the literature: the first one  offered  by  Khoury \& Weltmann \cite{KW04} using the linear approximation to the field equation, and  the second one, proposed by Mota \& Shaw \cite{MS07}, based on  considerations of   the complete nonlinear equation,  and that this latter  condition also depends on the value of the environment's density as well as  the value of the chameleon potential parameters and that  of $\beta$.

We do  not  find  the   arguments   based on  the  {\it thin shell}  phenomena   to be  sufficiently persuasive. The  basic observation is that  one  could, equally well, have claimed something not  entirely  different by considering two  macroscopic and charged  conductors. It is well known that the mobility of  charges  in  conducting materials  ensure that in static situations the charges are distributed on the conductors' surface in such a way that the electric field inside, vanishes exactly. Thus, except for a  {\it thin  shell} on the surface of each body,  one could have argued (following a similar logic as the one  used in the  context of the  chameleon models),  the   external electric  field  could not exert forces  on the  conductor's material. However, we  of course know that large macroscopic forces between such macroscopic  bodies are the rule. The  resulting   force   could therefore   be attributed  solely to   the {\it thin shell} effects. 

In view of this, we  proceeded to study this issue for  the case of the chameleon in more detail, in order to understand what, if  any, is the fundamental difference  between the  two situations (i.e. between the electromagnetic case  mentioned  above  and the one  concerning the theory at hand). That is, we  want to find out if the {\it thin  shell} arguments  are valid at the level of accuracy   that  would  actually  ensure the ``disappearance" of the expected forces. The basic  point is that  when the  contribution to the scalar field by the presence of the {\it test} body in  an actual  experiment is sufficiently large to create  a  {\it thin shell}  effect, then,
 the body cannot 
really  be considered as a simple {\it test} particle, and the  problem needs to be treated as a two body problem, i.e., one requires to analyze  the 
full effect of  the two bodies on the scalar field in order to evaluate the effective force  exerted by the field on the body of interest. 
The main contribution of this paper consists in analyzing the chameleon field generated by two spherical bodies of finite size (a large one and a smaller one) 
and the computation 
of the effective force on the {\it test} (smaller) body using first principles. To this end a {\it linear} approximation for the field equation 
was used in all our analysis. This linear approximation is different from the standard linear approximation used by Khoury and Weltman~\cite{KW04}
\footnote{  In a future work we plan to analyze other types of approximations which we expect can lead to more accurate solutions.
}. 
As far as we are aware, this approach is novel.
Now before moving forward, it is  worth stressing two important aspects related to chameleon models: 
the first one is a matter of principle,  and refers to the fact  that,  effectively  the WEP is violated by construction in  this kind of models, as opposed to  purely metric-based theories (like GR) where the WEP  is  incorporated {\it ab initio.} That is, the WEP (for {\it test point particles} moving  on geodesics) is satisfied with infinite precision in  purely metric  theories of gravity  involving no  other long range fields  coupling to matter. 
Nevertheless, even in  such  metric theories the absolute   validity of the WEP   refers  only to  {\it test point particles}  which, 
as stressed above, do not really exist in nature (as a result of Heisenberg's uncertainty principle). In any event,   when ignoring  those  quantum  complications,   it is  clear  that the WEP is respected  for  {\it point test particles} in metric theories, while, in 
chameleon models, the  WEP  is violated a priori (i.e. by construction in models  with non-universal coupling) {\it even} when referring to {\it point test particles}. As we shall see, this violation can   be exacerbated 
when considering actual {\it extended test  objects}. 
The second aspect concerns the experiments themselves. Even if chameleon models violate by construction the 
WEP, those violations might not be observable   in a laboratory experiment if the precision is not adequate. This, as is well known, means that some of the effects that  might reflect   these  violations of the WEP can be suppressed by 
the {\it thin shell} phenomenon associated with the chameleon field or by the Yukawa dependence of the  force associated with the effective 
mass of the field.  Thus,  as we shall illustrate, within a framework of a two body problem, these violations  of the WEP  might be   large, 
in contexts where the two bodies are  embedded  in a very   
a light medium (e.g. the vacuum or the air atmosphere) (see Fig.~\ref{resM} in Sec.~\ref{sec:results}) whereas they could  be  strongly   suppressed when part of the setting (e.g. the {\it test} body) 
is encased in  some  shell of a dense material like a metal vacuum chamber. 

In order to explore  these issues in detail, we  evaluate the  static  chameleon  field   associated  with two  extended spherical bodies, one large and one small,  of 
different chemical composition. 
For simplicity, only the larger one will be taken to be  {\it the source} of the gravitational field. We shall see that under these and other 
simplifying assumptions concerning the equation of motion for the chameleon, even when considering a universal coupling and the suppression due to the metal encasing of the experimental setup,   violations of the WEP  do arise for some values of the chameleon field parameters 
(see Figs.~\ref{competaconshell2}, \ref{resetaconshell} and~\ref{resLLR3} in Sec.~\ref{sec:results}). This result, although {\it previously  known},  may seem rather counter-intuitive as in this scenario all the various metrics $ g_{\mu\nu}^{(i)} $ coalesce into a single 
(Jordan) metric $g_{\mu\nu}^{\rm J}$ (see Sec.~\ref{sec:modelo} for the details). These violations are then put in perspective with the results found by various 
authors using different approaches and approximations (see Sec.~\ref{sec:results}). 

 The paper is organized as follows. After briefly reviewing the original chameleon model in Section~\ref{sec:modelo}, we present  the method for computing the two body problem proposed in this work. Then, in  Section \ref{sec:energy} we develop a simple criterion based on  the  use  of an  energy  functional   for  which  the minimization   corresponds  to the  field  configuration  of the  physical solution,     to determine  when   the method proposed in this paper is better than the one employed  in  the  standard approach  and  when it is  not. That method  is used to identify the   range of   the model's  parameters  where our  results  should be  considered  as more trustworthy that the   results obtained   in previous studies. 
In Section~\ref{sec:force} we compute the force on the {\it test} body  and show that the force is not negligible and, to the extent that  there is no 
protective symmetry  to prevent it,  it  will lead to  an  acceleration  that  depends  on the {\it test} body's composition.  In Section~\ref{sec:results} we   apply   our  results  to  two   concrete experiments,  the E\"{o}t-Wash   torsion balance and the Lunar Laser Ranging (LLR). 
For the first one,  our discussion  incorporates  the  characterization of the outside medium and the  effects   of  the vacuum chamber's encasing shell. In both cases we  provide numerical estimates for the E\"{o}tv\"{o}s parameter. In the  former  case  we  find  that the  actual   experimental  setting 
generates    complications  in  the numerical part of our analysis.  Nevertheless,  we provide   our estimates for the  bounds including   rough  estimates  of the corrections   resulting from the  specific  experimental  setup,    which  we   consider    can serve as motivation for  further   analytical and    experimental work. For instance, we  explore  the   estimates for the  experimental bounds  that arise  from   considering two different  modelings for the outside medium. We also explore  the modifications  that  arise  from    including the effect of the metal shell,  showing  that in  some regimes  the violations of the WEP can be  significatively suppressed for 
$\beta > 10^{-2}$ (we are  indebted to P.  Brax  for pointing this  specific  aspect to us).  
Finally, in 
Section~\ref{sec:conclusions} we present our conclusions and a discussion about their impact on other alternative theories of gravity. Several 
appendices complete the ideas of the main sections. 

\section{The Chameleon Model}
\label{sec:modelo}
In this section we briefly review the main aspects of the chameleon model. The model  involves  a scalar-field $\varphi$ that couples minimally to gravity via a fiducial metric $g_{\mu\nu}$, according to the following action
\begin{equation}
\label{action}
  S[g_{\mu\nu},\Psi_m^{(i)},\varphi] = \int d^4x \sqrt{-g} \left[ \frac{M_{pl}^2}{2} R 
- \frac{1}{2}g^{\mu\nu}(\nabla_\mu\varphi)(\nabla_\nu\varphi)- V(\varphi)\right] - \int d^4x L_m \left(\Psi_m^{(i)}, g_{\mu \nu}^{(i)}\right), 
\end{equation}
where $M_{pl}=1/{\sqrt{8\pi}}$ is the reduced Planck mass, $R$ is the Ricci scalar associated with $g_{\mu\nu}$, and $\Psi_m^{(i)}$ represents schematically the 
different matter fields (i.e. the fields other than the chameleon $\varphi$; for instance, all the fundamental fields of the standard model of 
particle physics). The potential $V(\varphi)$ is specified below. The particular feature of this action is that 
each specie $i$ of matter couples {\it minimally} to its corresponding metric $g_{\mu\nu}^{(i)}$, while the scalar field $\varphi$ 
couples {\it non-minimally}, and in general, {\it non-universally} to the matter through a conformal factor that relates each metric 
$g_{\mu\nu}^{(i)}$ with the so called Einstein metric $g_{\mu\nu}$:
\beq
\label{conftrans}
g_{\mu\nu}^{(i)} = \exp{\left[\frac{2 \beta_i \varphi}{M_{pl}}\right]} g_{\mu\nu}.
\eeq
Here $g_{\mu\nu}^{(i)}$ is the metric which is usually associated with the geodesics of each specie $i$ of matter and 
 $\beta_i$ is the corresponding coupling constant between each specie and the chameleon field. 
For instance, when the coupling constants are universal, $\beta_i=\beta$, then the (universal) metric $g_{\mu\nu}^{J}=g_{\mu\nu}^{(i)}$ is called the 
Jordan metric, and then  point particles  would  follow the geodesics of this metric. 
We  will see, however, that  a more detailed  analysis   indicates that   even in  the universal coupling case,  within the relevant  experimental situations,  that   simple conclusion  need  not  apply. Given the fact that, in general, the chameleon field couples differently to each 
specie of particle, this model leads  to potential violations of the WEP that could  in principle  be explored  experimentally. 

Following \cite{KW04,MS07,K13}, we consider the  potential for the chameleon field to be, 
\beq
\label{Vbare}
V(\varphi)= \lambda M^{4+n} \varphi^{-n},
\eeq
where $M$ is a constant, and $n$  is a free parameter that can be taken to be either a positive integer or a negative even integer 
[cf. Eq.~(\ref{Phiinout})] 
($\lambda=1$ for all values of $n$ except when $n=-4$ where $\lambda=\frac{1}{4!}$.).

The energy-momentum tensor (EMT), $T^{m\vspace{5mm}(i)}_{\mu\nu}$, for the {\it i}\,the  matter  component  can be written in terms of the EMT, $T^{m}_{\mu\nu}$ associated with the Einstein metric as follows:
\beq
T^{m\vspace{5mm}(i)}_{\mu\nu}=+
\frac{2}{\sqrt{-g^{(i)}}}\frac{\delta L_m}{\delta g^{\mu\nu}_{(i)}}=\exp\left[\frac{-2\beta_i\varphi}{M_{pl}}\right]T^{m}_{\mu\nu},
\eeq
where $T^{m}_{\mu\nu}$ is defined similarly from $L_m$ but using the metric $g_{\mu\nu}$ instead of $g_{\mu\nu}^{(i)}$.
\bigskip

Consequently, the traces of both EMT's are related by
\beq 
T^{m\vspace{5mm}(i)}=g^{\mu\nu}_{(i)}T^{m\vspace{5mm}(i)}_{\mu\nu}=\exp\left[\frac{-4\beta_i\varphi}{M_{pl}}\right]g^{\mu\nu}T^{m}_{\mu\nu}=\exp\left[\frac{-4\beta_i\varphi}{M_{pl}}\right]T^{m}.
\eeq
Eventually, we will assume a perfect-fluid  description for $T^{m}_{\mu\nu}$. Specifically, this perfect fluid will  be taken to characterize  each of two extended bodies, together   with  the   matter constituting the  environment, all of which  we consider as the  ``source''  of the chameleon field. Furthermore, in the analysis of the  resulting force between the two bodies, we will simply consider a universal coupling and set 
$\beta_i= \beta$, in order to simplify the calculations. Hereafter, and unless otherwise indicated, all the differential operators 
and tensorial quantities are associated with the Einstein metric.

The equation of motion for the the chameleon $\varphi$ which arises from the action (\ref{action}) is
\begin{equation}
\label{mov}
\Box \varphi = \frac{\partial V_{\rm eff}}{\partial \varphi},
\end{equation}
where $V_{\rm eff}$ represents the effective potential defined by:
\beq
V_{\rm eff} = V(\varphi) - T^{m} \frac{\beta \varphi}{M_{pl}}, 
\label{Veff}
\eeq
which depends on the energy-density and pressure of the matter fields via $T^{m}$. So for the perfect fluid model, 
$T^m=-\rho+3P$ which does not depend explicitly on $\varphi$.

\subsection{Modelling the experimental setup}
\label{Setup}

As we mentioned earlier, we are interested in the general static solution of Eq.~(\ref{mov}) in the presence of two extended bodies that for simplicity we  consider  as spherical. We take one of them  to be  a  very {\it large} body of mass ${\cal M}$ (e.g. Earth, Sun, a mountain), and 
the other one is a smaller body of mass ${\cal M}_2$ (${\cal M}_2\ll {\cal M}$; hereafter {\it test} body). 
Both bodies are the source of the chameleon, and each one is taken to have a different but uniform density, which 
formally can be represented by suitable Heaviside (step) functions. Moreover, we shall consider  the  {\it linear} approximation in Eq.~(\ref{mov}) obtained 
by linearizing Eq.~(\ref{Veff}) around $\varphi_{\rm min}$ (associated with the minimum of $V_{\rm eff}$) in each of the three mediums (i.e. the two bodies and the {\it environment}) and then match the solutions at the border of each of the two bodies. This is similar, 
but not identical to the  standard  method  of  analysis of these   models (see Appendix \ref{sec:OBP}).

Now prior to tackling the two body problem for the chameleon field, it is crucial to review some of the main relevant aspects of the 
one body problem, i.e., the situation where 
the {\it test} body does not back react on the scalar-field configuration, because some of those approximations will be used in the treatment of the two body problem. However, in our approach, it is essential that the back reaction of the field to the presence of the two bodies, be fully taken into account.

Let us consider a spherically-symmetric and {\it homogeneous} body of radius $R$ and density $\rho_{\rm in}$ immersed in an external medium of density $\rho_{\rm out}$. We will call this body, the {\it larger body}, as opposed to the {\it smaller body} that will be introduced later. 
The corresponding EMT's, $T^{m}_{\rm in (out)}$ are assumed to be of a perfect fluid:  $T^{m}_{\rm in (out)}=-\rho_{\rm in (\rm out)}+3P_{\rm in (out)}$, where the scripts ${\rm in (out) }$ 
will refer to the interior (exterior) of the body, respectively. In this case we will consider two regions (interior and exterior), and since $P\ll \rho$ for non-relativistic matter, we neglect the pressure. Consequently,
\begin{equation}
\label{densOneB}
\rho=
\begin{cases}
\rho_{\rm in} \hspace{1cm} r \leq R \\
\\
\rho_{\rm out} \hspace{1cm} r > R
\end{cases}
\,\,\,,
\end{equation}
where $R$ is the radius of the body.

\bigskip

In order to solve the chameleon equation Eq.(\ref{mov}) with the effective potential (\ref{Veff}) we can adopt several possible 
strategies. Clearly, one consists in solving numerically the full non-linear chameleon equation without any approximation. 
In the case of a one body problem in spherical symmetry (where the test body does not backreact on the chameleon field), the chameleon equation becomes an ordinary differential equation and it can be solved using a Runge-Kutta scheme. 
Indeed, this was done by Khoury \& Weltman~\cite{KW04} in their pioneering paper, in order   to check  the analytical solution for the one body problem. Besides, Elder et al \cite{Elder16} and Schlogel et al \cite{Schlogel16}, also solved numerically the chameleon equation to study atomic interferometry experiments. For this experimental situation, the test body is of the size of an atom, and therefore it is appropriate to calculate the force using  the one body problem solution.

However, in the two-body problem where the test body is considered not as a point particle but as an extended body 
which backreacts on the chameleon field, solving the full non-linear chameleon equation entails to solving a very complicated 
partial differential elliptic equation. In order to advance in this direction, we postpone this analysis for the future, and consider 
to solve a linear elliptic equation by approximating the effective potential. To this end, there are two possible approximations. 
These depend actually on the specific setup, like the size of the bodies, their density, the environment's density, and the actual value of the 
coupling $\beta$. The values of these parameters determine, for example, if the bodies have or not a {\it thin shell} or even an intermediate 
situation where the {\it shell} is not {\it thin} nor {\it thick}. The thin (thick) shell dimensionless parameter  $\frac{\Delta R}{R}$ (cf. Section~\ref{sec:OBP}) allows us to determine in which regime the large or the test body is. Namely, if $\Delta R/R\ll 1$ the body has a thin shell, whereas 
$1\lesssim \Delta R/R$ corresponds to a thick shell regime.   Thus, depending on each of the two regimes one can approximate differently the effective potential.

If the {\it thin shell} condition is  satisfied,  the  expansion  of   the effective potential  $V_{\rm eff}(\varphi)$ about its minimum up to the quadratic terms, for  each  one of the 
regions, as  all  higher order terms are ignored  provides a  good  approximation  for  the  determination of the scalar field profile (see for example, \cite{KW04,Brax07,Hui2009,Brax2012}). Such approximation  leads to a linear differential equation for the chameleon field.  In a forthcoming paper we  plan to  study  in detail  the corrections to  this approximation arising from higher order terms~\cite{Oursnonlin}. We will see that the corrections become relevant for $n>4$~\cite{Oursnonlin}\footnote{We thank A. Upadhye, B. Elder and J. Khoury for raising this issue after the first submission of this manuscript.}. Next, we will describe the {\it quadratic approximation}  to the effective potential and at the end of this subsection, we will discuss if it is appropriate  for  the experimental situations at hand . 

We now proceed to  deal with the simple,  one body  problem,  taken to be immersed  within  a  single medium  acting as    the environment. In  section IIB, we generalize this expansion for the  ``two body problem''.

The  expansion  of  the effective potential about the corresponding minimum in each region up to the quadratic term gives:
\begin{equation}
\label{expan}
V_{\rm eff}^{\rm in,out}(\varphi) \simeq V_{\rm eff}^{\rm in,out}(\varphi_{\rm min}^{\rm in,out}) +\frac{1}{2} 
\partial_{\varphi \varphi} V_{\rm eff}^{\rm in,out}(\varphi_{\rm min}^{\rm in,out})[\varphi - \varphi_{\rm min}^{\rm in,out}]^2, 
\end{equation}
The effective mass of the chameleon is then defined in the usual way:
\begin{equation}
m_{\rm eff}^{ 2 \rm in,out}(\varphi_{\rm min}^{\rm in, out}, \beta_i, T^{m}_{\rm in,out})=
  \partial_{\varphi \varphi} V_{\rm eff}^{\rm in,out}(\varphi_{\rm min}^{\rm in,out}). 
\end{equation}
In particular,  setting   $\beta_i=\beta$, the expression for the effective mass $m_{\rm eff}^{\rm in,out}$ turns out to be:
\begin{equation}
\label{meff}
m_{\rm eff}^{ 2 \rm in,out}(\varphi_{\rm min}^{\rm in, out}, \beta, T^{m}_{\rm in,out})=\lambda n(n+1)M^2\Big(\frac{M}{\varphi^{\rm in, out}_{\rm min}}\Big)^{n+2},
\end{equation} 
with
\begin{equation}
\label{Phiinout}
\varphi^{\rm in, out}_{\rm min}=M\Big(\frac{n\lambda M_{pl}M^3}{-\beta T^{m}_{\rm in, out}}\Big)^{\frac{1}{n+1}}.
\end{equation}

Expressions (\ref{meff}) and (\ref{Phiinout}) are valid when $n$ is a positive integer or a negative even integer, 
as otherwise a minimum does not exist. Furthermore, for $n=-2$ the effective mass does not depend on the composition of the  body or the environment.

\begin{figure}
\begin{center}
\includegraphics[width=8cm,height=8.2cm,angle=-90]{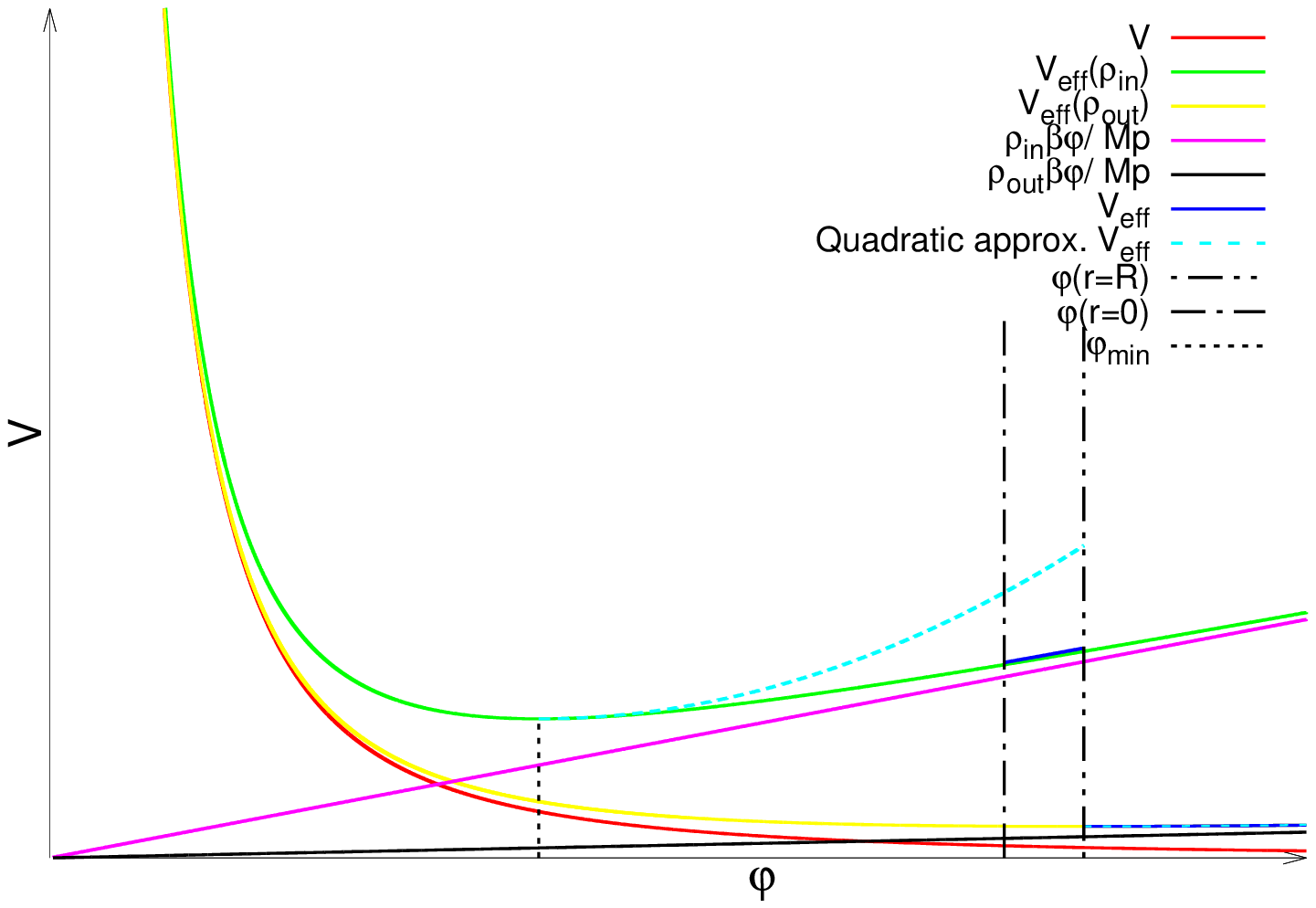}
\hspace{0.8cm}
\includegraphics[width=8cm,height=8.2cm,angle=-90]{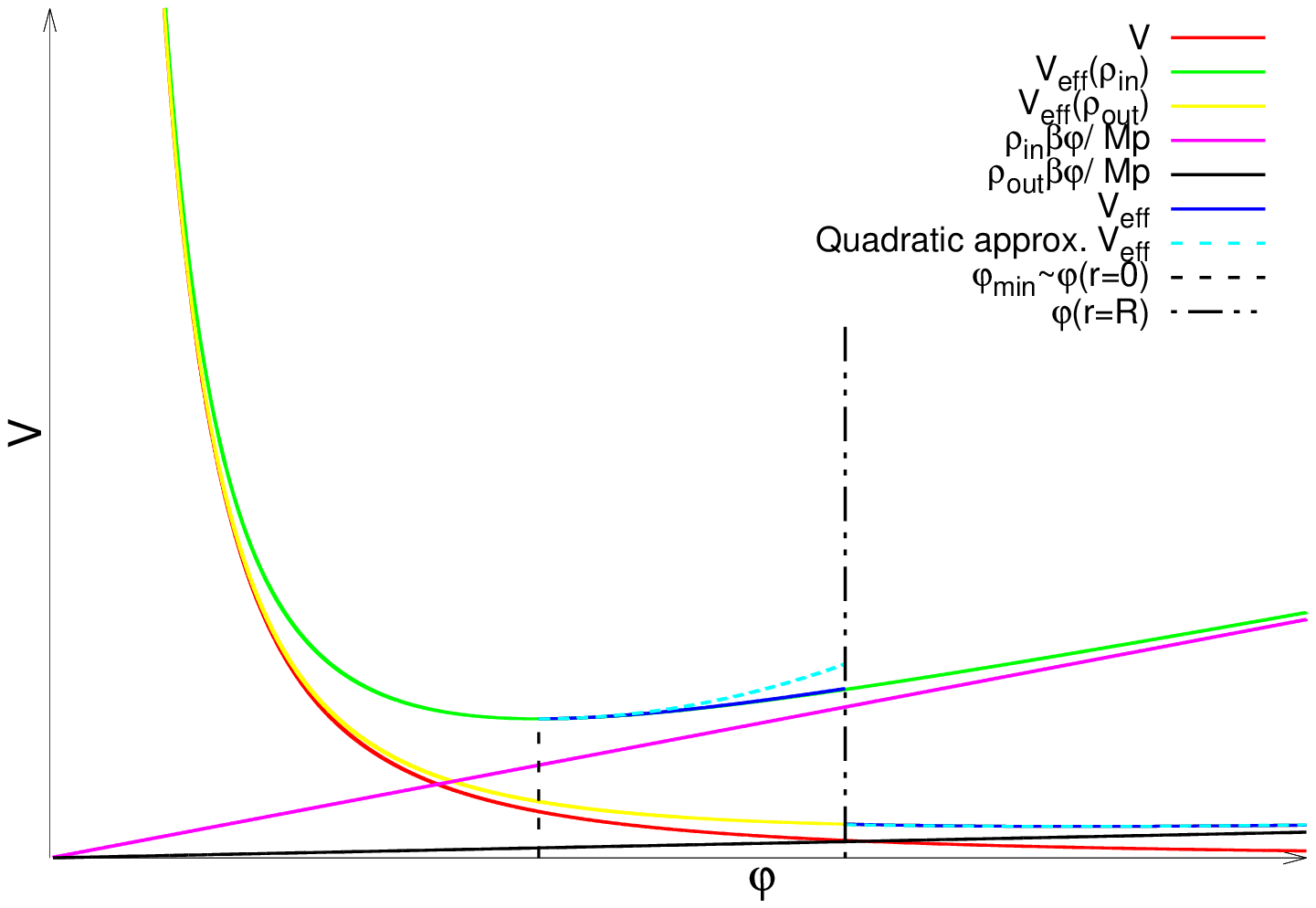}
\caption{Effective potential inside and outside the large body. The blue line shows the  potential $V_{\rm eff}(\varphi)$ for the one body problem, the red line shows $V_{\rm eff}(\varphi)$ for $\rho=\rho_{\rm out}$  in Eq.\ref{Veff}, the green line shows $V_{\rm eff}(\varphi)$ for $\rho=\rho_{\rm in}$ in Eq.\ref{Veff} and the light blue line shows the quadratic approximation Left: The {\it thick shell} case; Right: The {\it thin shell} case.} 
 
\label{Veff1}
\end{center}
\end{figure}

\begin{figure}
\begin{center}
\includegraphics[width=5.5cm,height=9.cm]{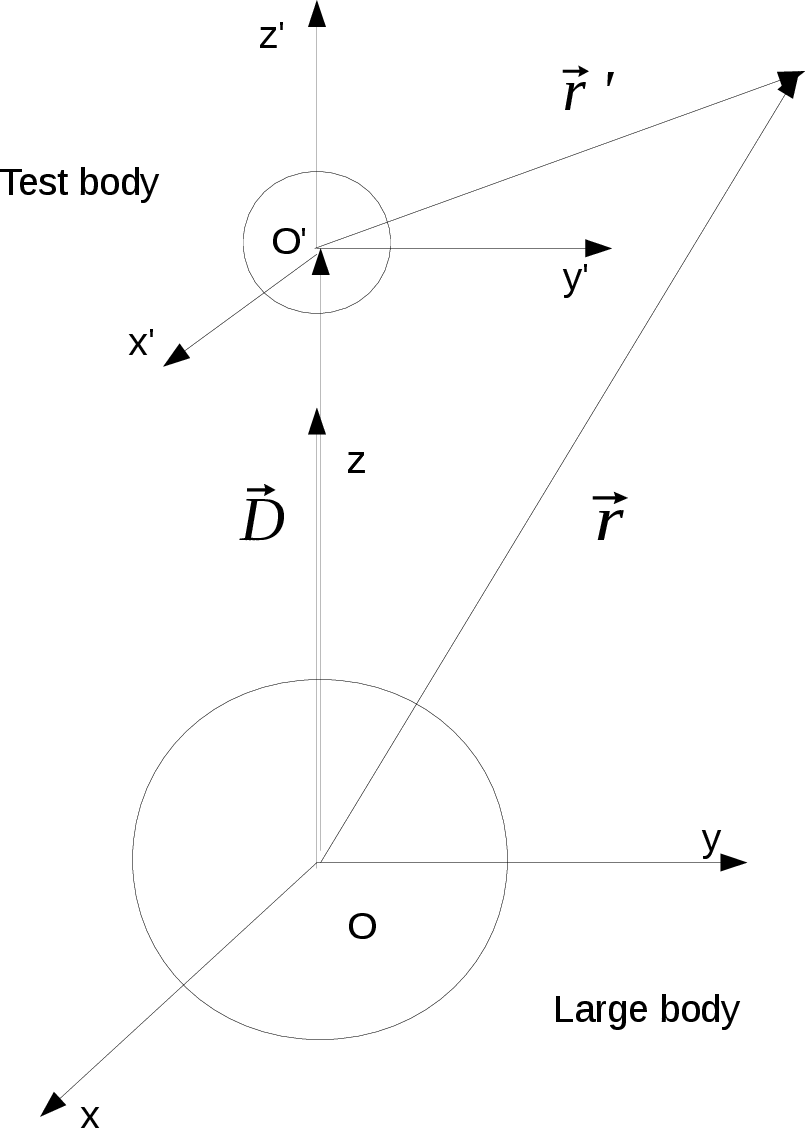}
\caption{Two body problem.}
\label{figura1}
\end{center}
\end{figure}

Now,  let us  consider the  interesting situation  where  we  have  two bodies, the large one and the test body.  The effective potential   must be   expended in  various regions: 
   the interiors of the two bodies  and their exterior which is associated with the  environment.
Typically the density of the large body is much larger than 
the density of the environment ($\rho_{\rm out} \ll \rho_{\rm in}$), thus  taking $T\sim -\rho$ leads to $\varphi_{\rm in}^{\rm min} \ll \varphi_{\rm out}^{\rm min}$.  The effective potential inside the large body develops a minimum 
$V_{\rm eff\,,\,in}^{\rm min}$ at $\varphi_{\rm in}^{\rm min}$ which is much larger than the 
minimum of the effective potential $V_{\rm eff\,,\,out}^{\rm min}$ at $\varphi_{\rm out}^{\rm min}$     associated with the environment (see Fig. \ref{Veff1}). 
The difference between the $\varphi_{\rm in}^{\rm min} $  and $  \varphi_{\rm out}^{\rm min}$,  which can be very large, depends  on the  density of the materials 
involved and the  value of    model's parameters.  This, in turn, can lead to  different types of approximations  used  to describe $V_{\rm eff}$. For instance, since we  take  the environment  to extend  to infinity, the chameleon field must  be such  that  it reaches  
$\varphi_{\rm out}^{\rm min}$ at spatial infinity. Therefore   the field must   interpolate between $\varphi_{\rm out}^{\rm min}$ at spatial infinity and 
$\varphi_{\rm in}^{c}$ at the centers of the each of the two bodies, both modelled as spherical (see Figure~\ref{figura1}).  We   emphasize that   this central values need  not coincide  with    $\varphi_{\rm in}^{\rm min} $ the actual  minima  of the function   $V_{\rm eff}^{\rm in} ( \varphi)$  in the bodies'  interior. 
Now,   roughly   speaking the chameleon field rolls up the potential $V_{\rm eff}^{\rm out}$ from its minimum 
at $\varphi_{\rm out}^{\rm min}$ to 
the value $\varphi_R$ where the field reaches the surface of the 
large body (respectively, the test body), where $R$ stands schematically for the location of the border of the body. 
However,  and this  is  an essential aspect of the present analysis,  the  field  configuration  is not  really    constant   and  it is  precisely  the  resulting  directional dependence  that  is  associated  with a  non-vanishing force  (in the direction of the large body)  on the  test body.  
 
 Now,  when  $\varphi_R\sim \varphi_{\rm out}^{\rm min}$ we can  rely on the quadratic  expansion about its minimum  as a good approximation  to  $V_{\rm eff}^{\rm out}$  since the 
field outside the bodies takes values that are  always  close to the minimum $V_{\rm eff\,,\,out}^{\rm min}$. 
We report results only for those situations where the energy criterion favors our approximations over that of the standard approach. 
 Now, at the border of the bodies, 
the effective potential experiences a discontinuous jump (due to the jump between the body's and the environment's densities) 
and at $\varphi_R$ we have to use an   approximate expression  for   $V_{\rm eff}^{\rm in}$ instead of  one for $V_{\rm eff}^{\rm out}$ 
in order to solve the chameleon equation inside the  body. However,  as is  standard  in these cases, the field itself $\varphi$ ( together with its  first derivative)  must  be continuous everywhere,  
and in particular  at the surface of the bodies. As  one  moves towards the interior of the bodies, the field $\varphi$ starts to 
``roll down" the potential $V_{\rm eff}^{\rm in}$ towards its minimum, but   without  necessarily reaching it. 
More specifically,  the  field  configuration interpolates   from its  surface value and  the value  $\varphi_{\rm in}^c$ at the center of the body. 
If $\varphi_R\sim \varphi_{\rm in}^{c}\sim  \varphi_{\rm in}^{\rm min}$, it means that the field is already near the minimum of the potential inside the body  and the  quadratic  expansion about the minima  is  a  very  good one.  
This situation is emblematic of a {\it thin shell} regime.
On the other hand, when $\varphi_{\rm in}^{\rm min}\sim \varphi_{\rm in}^{c} \ll \varphi_R$, we  have a situation where  the field at the body's  surface  
is very far from its minima, and thus, the {\it quadratic} approximation 
for $V_{\rm eff}^{\rm in}$ is  not necessarily  a very  good one.  
In that case, however, $V_{\rm eff}^{\rm in}$ is  usually   dominated by the 
matter  dependent part, and then  one may  use the  approximation  $V_{\rm eff}^{\rm in}\sim  \rho_{\rm in} \frac{\beta \varphi}{M_{pl}}$ 
~\cite{KW04}. This happens in some situations in which  the large body falls  within  the   {\it thick shell regime};  as  for the example the  Sun with $n=1$, $M=10$ {\rm eV}  and $\beta \ll 1$. 

In the present paper we shall focus mainly in scenarios where the bodies are within  {\it thin shell regime}, and  thus, where 
the quadratic approximation to the effective potential is adequate. In Section \ref{sec:energy} we shall develop a criteria to assess 
the extent to which this approximation is in fact a good one using a minimization of a suitable energy functional.

\subsection{The two body problem}
\label{FModel} 

We proceed with the calculation of the chameleon in the 
presence of two bodies without, in principle,  neglecting any contribution, but still within the 
 {\it quadratic} approximation for the effective potential discussed before.  The geometry of the problem is depicted in Fig. \ref{figura1}.

\begin{figure}
\begin{center}

\includegraphics[width=6.5cm,height=10.cm,angle=-90]{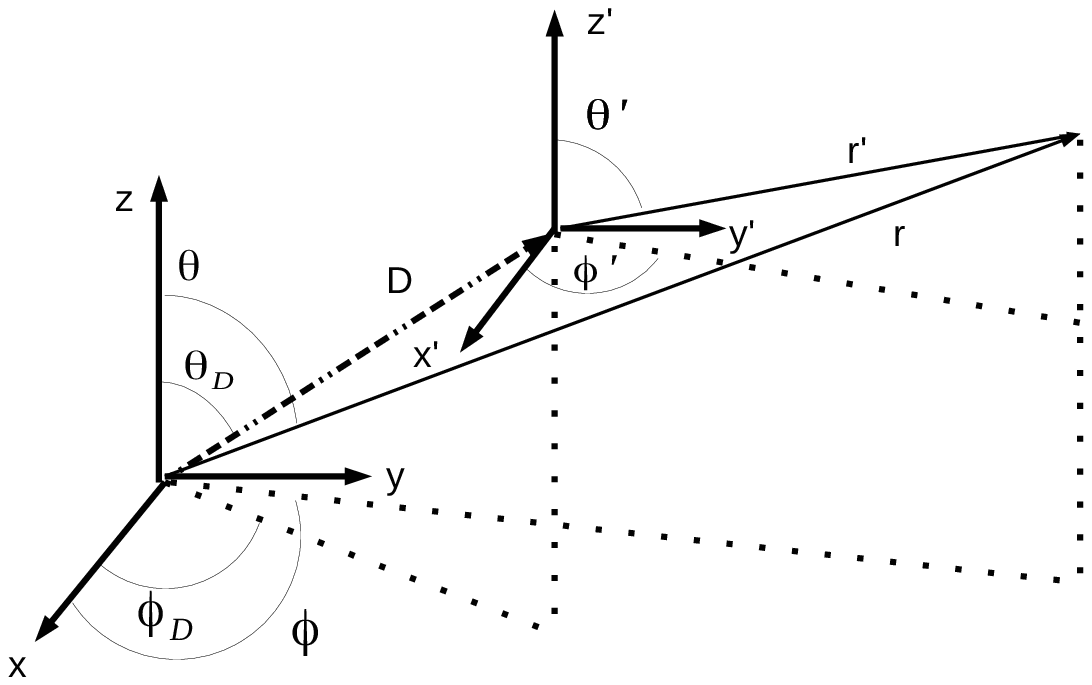}
\caption{Coordinates transformation.}
\label{figura1bis}
\end{center}
\end{figure}

In order to do so, we expand the most general solution in complete sets of solutions  of the   differential  equation  in  the  inside and outside regions  determined  by  the two bodies. Thus  we  write,

\begin{equation}
\label{fullsol}
\quad \varphi=
\begin{cases}
 \varphi_{\rm in1}= \sum\limits_{lm} C_{lm}^{\rm in1} i_l(\mu_1 r) Y_{lm}(\theta,\phi)+\phi_{1\rm min}^{\rm in} \hspace{1.3cm} 
 (r \le R_1) \\
 \varphi_{\rm out}=\sum\limits_{lm} C_{lm}^{\rm out1} k_l(\mu_{\rm out} r) Y_{lm}(\theta,\phi)+ 
 C_{lm}^{\rm out2} k_l(\mu_{\rm out} r') Y_{lm}(\theta',\phi')+\varphi_{\infty} 
 \hspace{0.7cm} (\rm exterior\,\,solution) \\
 \varphi_{\rm in2}=\sum\limits_{lm} C_{lm}^{\rm in2} i_l(\mu_2 r') Y_{lm}(\theta',\phi')+\varphi_{2\rm min}^{\rm in} \hspace{0.7cm} (r' \le R_2)
\end{cases}
\end{equation}
where  $\mu_1=m_{\rm eff}^{\rm large\,\ body}$, $\mu_{\rm out}=m_{\rm eff}^{\rm out}$, $\mu_2=m_{\rm eff}^{\rm test \,\ body}$ and 
$R_1,R_2$ are the radii of the {\it large} and {\it test} bodies, respectively, and $i_l$ and $k_l$ are Modified Spherical Bessel Functions (MSBF). In the above equation, the 
Cartesian coordinate system $x$,$y$,$z$ is centered in the {\it large} body, while the coordinate system $x'$,$y'$,$z'$ is centered in the {\it test} body (see Fig. \ref{figura1}). 
 Notice that in writing Eq.~(\ref{fullsol}) we have taken into account the regularity conditions of the scalar field at the center of both 
bodies. That is, the MSBF used in the expansions for the solutions inside both bodies are well behaved within their corresponding compact supports. 
Conversely, for the exterior solution (i.e. the solution outside both bodies) we employ the set of MSBF functions associated with each body 
which are well behaved at infinity. 
Furthermore, the coefficients $C_{lm}$ of Eq.~(\ref{fullsol}) are calculated using the following continuity conditions for the field 
and its derivative at the boundaries of the two bodies: 
\begin{equation}
\varphi_{\rm in1}(r=R_{1}) = \varphi_{\rm out}(r=R_{1}),\qquad
\partial_r\varphi_{\rm in1}(r=R_{1})=\partial_r\varphi_{\rm out}(r=R_{1}),
\nonumber
\end{equation}
\begin{equation}
\varphi_{\rm in2}(r'=R_{2}) = \varphi_{\rm out}(r'=R_{2}),\qquad
\partial_{r'}\varphi_{\rm in2}(r'=R_{2}) = \partial_{r'}\varphi_{\rm out}(r'=R_{2});
\nonumber
\end{equation} 
In order to describe properly the two body problem in a single coordinate  system  we can use the following relationship ( see Figure \ref{figura1bis}) that links the special functions in the two coordinate systems~\cite{Scatt,Scatt2}:
\beq
\label{k1}
k_l(\mu_{\rm out}r)Y_{lm}(\theta,\phi)=\sum_{vw}\alpha^{*lm}_{vw}i_v(\mu_{\rm out}r')Y_{vw}(\theta',\phi'),
\eeq
where the coefficients $\alpha^{*lm}_{vw}$ can be expressed as follows:
\beq
  \begin{split}
    \alpha^{*lm}_{vw} =& (-1)^{m+v} (2v+1)\sum_{p=|l-v|}^{|l+v|}(-1)^{-p}(2p+1)\Big[\frac{(l+m)!(v+w)!(p-m-w)!}{(l-m)!(v-w)!(p+m+w)!}\Big]^{1/2} \\
    &\times \begin{bmatrix} l & v & p\\ 0 & 0 & 0 \end{bmatrix} \begin{bmatrix} l & v & p\\ m & w & -m-w \end{bmatrix} k_p(\mu_{\rm out}|D|)Y_{p(m-w)}(\theta_D,\phi_D),
   \end{split}
\label{sum1}
\eeq
being $\theta_D$ and $\phi_D$ the angular coordinates associated with the vector $\vec D$ (see Fig. \ref{figura1bis}) and we remind the reader that $D$ is the distance between the center of the two bodies. 

Moreover, we can use a similar relationship between the special functions 
given in terms of the coordinates centered in the {\it test} body with the ones centered in the {\it large} body:
\beq
\label{k2}
 k_l(\mu_{\rm out}r')Y_{lm}(\theta',\phi')=\sum_{vw}\alpha^{lm}_{vw}i_v(\mu_{\rm out}r)Y_{vw}(\theta,\phi),
\eeq
where
\beq
   \begin{split}
    \alpha^{lm}_{vw}=& (-1)^{m+v} (2v+1)\sum_{p=|l-v|}^{|l+v|}(2p+1)\Big[\frac{(l+m)!(v+w)!(p-m-w)!}{(l-m)!(v-w)!(p+m+w)!}\Big]^{1/2} \\
    &\times \begin{bmatrix} l & v & p\\ 0 & 0 & 0 \end{bmatrix}\begin{bmatrix} l & v & p\\ m & w & -m-w \end{bmatrix} k_p(\mu_{\rm out}|D|)Y_{p(m-w)}(\theta_D,\phi_D),
   \end{split}
\label{sum2}
\eeq
valid for $|r|\le |D|$ and $|r'|\le |D|$. We make use of  the axial symmetry of the problem and thus   set  the $z-$axis as containing the centers of the two bodies (see Fig.\ref{figura1}). Thus, the coordinate transformation is, in this case, a translation along this axis. Therefore, $\theta_D=0$ and $\varphi_D$ becomes irrelevant do to the axial symmetry.
An approximate expression for the chameleon  field  is found by truncating the infinite series Eqs.~(\ref{k1}) and (\ref{k2}). 
We further note  that  as   shown in  Refs.\cite{Gume1,Gume2},  the series  of this type  can be estimated by truncating  the sum after   the first $N$  terms, with $N$   given  by the integer part of $N_0=\frac{e\mu_{\rm out}|D|}{2}$ where  $e$  is  Euler's number.
 Using this last result we can  solve the following two equations for $C^{\rm out1}_{l}$ and $C^{\rm out2}_{l}$; the first one reads:
\beq
 b_1\delta_{l0}=C^{\rm out1}_{l}a_l+\sum_{w=0}^{N}C^{\rm out2}_{w}z_{wl},
\label{eq1}
\eeq
where
\begin{eqnarray}
 a_l &=& \frac{k_l^\prime(\mu_{\rm out}R_1)i_l(\mu_1R_1)}{i_l^\prime(\mu_1R_1)}-k_l(\mu_{\rm out}R_1),\\
 z_{wl} &=& \alpha^{w0}_{l0}\Big[i_l(\mu_{\rm out}R_1)-\frac{i_l^\prime(\mu_{\rm out}R_1)i_l(\mu_1R_1)}{i_l^\prime(\mu_1R_1)}\Big],\\
 b_1 &=&\sqrt{4\pi}(\varphi_{\infty}-\varphi_{1\rm min}^{\rm in }).
 \end{eqnarray}
where a {\it prime} `$\,^\prime\,$' indicates differentiation of the MSBF with respect to its argument.

The second equation reads:
\beq
  b_2\delta_{l0}=C^{\rm out2}_{l}x_l+\sum_{w=0}^{N}C^{\rm out1}_{w}y_{wl},
\label{eq2}
\eeq
where
\begin{eqnarray}
  x_l &=& \frac{k_l^\prime(\mu_{\rm out}R_2)i_l(\mu_2R_2)}{i_l^\prime(\mu_2R_2)}-k_l(\mu_{\rm out}R_2),\\
  y_{wl} &=& \alpha^{*w0}_{l0}\Big[i_l(\mu_{\rm out}R_2)-\frac{i_l^\prime(\mu_{\rm out}R_2)i_l(\mu_2R_2)}{i_l^\prime(\mu_2R_2)}\Big],\\
  b_2 &=& \sqrt{4\pi}(\varphi_{\infty}-\varphi_{2\rm min}^{\rm in }).
\end{eqnarray}
We can now write the system of equations for the coefficients $C^{\rm in1}_{l}$ and $C^{\rm in2}_{l}$ associated with the interior solutions in terms of the coefficients of the exterior solution:
\begin{eqnarray}
\label{Cinfull1}
C^{\rm in1}_{l}i_l^\prime(\mu_1R_1) &=& C^{\rm out1}_{l}k_l^\prime(\mu_{\rm out}R_1)+\sum_{w=0}^{N}C^{\rm out2}_{w}\alpha^{w0}_{l0}i_l^\prime(\mu_{\rm out}R_1),\\
\label{Cinfull}
C^{\rm in2}_{l}i_l^\prime(\mu_2R_2) &=& C^{\rm out2}_{l}k_l^\prime(\mu_{\rm out}R_2)+\sum_{w=0}^{N}C^{\rm out1}_{w}\alpha^{*w0}_{l0}i_l^\prime(\mu_{\rm out}R_2).
\end{eqnarray}

In this way, by solving the system of equations (\ref{eq1}) and (\ref{eq2}) we can obtain a solution of the two body problem for the chameleon where we perform an approximation (within the quadratic approach of the effective potential) consisting in the truncation of the series used for the transformation of coordinates. The dependence of the field on the composition of the {\it test} body appears through the constants $C^{\rm out 1}_{l}$ and $C^{\rm out 2}_{l}$.

\begin{figure}
\begin{center}
\subfloat[Chameleon  field within a neighbourhood of the {\it large} body. ]
{\includegraphics[width=8.5cm,height=8.5cm,angle=-90]{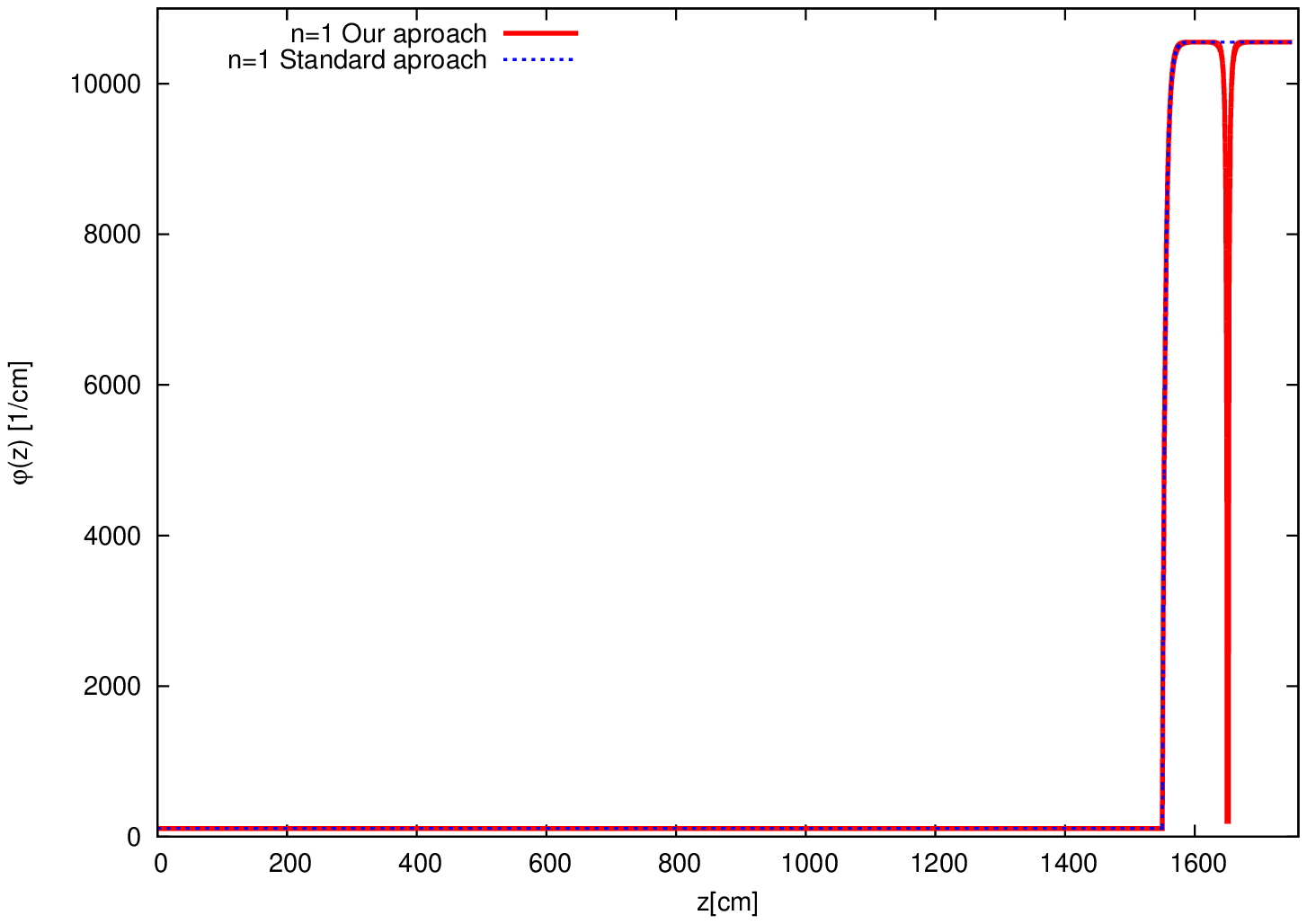}\label{f:grafico3a}}
\subfloat[Chameleon  field within a neighbourhood of the {\it test} body.]
{\includegraphics[width=8.5cm,height=8.5cm,angle=-90]{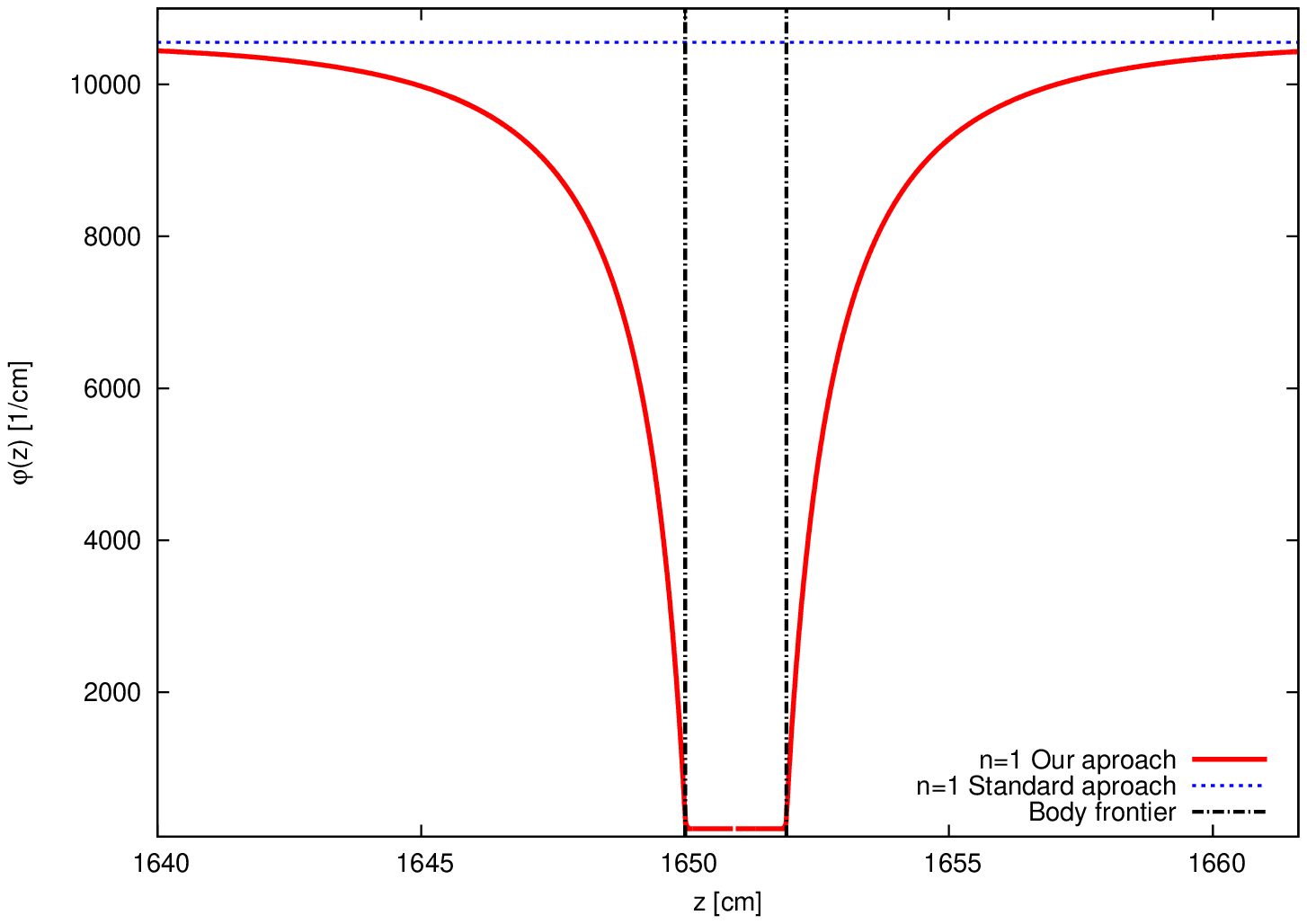}\label{f:grafico4a}}
\caption{The chameleon  field $\varphi$ as a function of coordinate $z$ 
inside the bodies and in their outskirts ($z$ is the distance to the center of the {\it large} body in  ${\rm cm}$, then $\varphi$ is measured in ${\rm cm^{-1}}$). The {\it large} body consists of a sphere of radius $R_1=1550$ $\rm cm$ and $\mu_1=197.2244$  $\rm cm^{-1}$. This body mimics roughly 
the actual hill where the  experiment of the E\"ot-Wash group takes place and which produces the 
desired combined effects (gravitational and ``fifth" force if any). The {\it test} body consists of a $10$ gr sphere of aluminium and has 
a radius $R_2=0.96$ $\rm cm$ and $\mu_2=78.1940$ $\rm cm^{-1}$ . In both figures $n=1=\beta=1$ and $M=2.4$ ${\rm meV}$ and $\rho_{out}=\rho_{\rm atm} = 10^{-3} \hspace{0.1cm}{\rm g \hspace{0.1cm} cm^{-3}}$ Red line: Our approach, Blue Line: Standard Approach~\cite{KW04}.}
\label{f:graficos3y4a}
\end{center}
\end{figure}

Figure~\ref{f:graficos3y4a} depicts the field around the centers of the two bodies and outside them along the $z-$axis which   results  from  our   method  by   cutting off  the  series   expansions  after   taking  the first three terms in each one\footnote{The same result is obtained if only the first term of the series is used as this turns out to be many orders of magnitude larger than the following terms.}. We also show, for comparison, the field obtained by Khoury \& Weltman  \cite{KW04} using the analytical solution  to the linearized time-independent  chameleon field equation for one body (we will refer to this solution as that of the standard approach)\footnote{We have verified that this solution agrees to a very good approximation with  the numerical solution to the non-linear chameleon field equation for one body.}. Note that under the conditions 
$1 \ll \mu_{1,2} R_{1,2}$, which are valid within each body, the {\it thin shell} effect appears in both bodies. Namely, 
the field is almost constant with values close to $\varphi_{\rm min\,1,2}$ in each body, respectively, and grows exponentially 
near the surfaces.  As $\mu R$ becomes smaller, 
the  {\it thin shell}  disappears.  Moreover, Figure \ref{f:graficos3y4a} (right panel)  compares the chameleon field 
obtained within the framework of the standard approach with the corresponding field obtained with our two-body method in the region where the {\it test} body is located. In order to check our method,  we compute the solution for the one body problem using our approach, and verify that we recover the same results as in Ref. \cite{KW04}. 

\bigskip

\section{Minimum Energy criterion}
\label{sec:energy}

 It  is well known that  the  classical solution of  a field   equation   corresponds to an extreme of the action  (with fixed boundary conditions). When we are interested  in   static solutions,  such  extreme  coincides  with the field  configuration    that minimizes   the energy  functional. 
 That  well known  result  is  used  in this section to   construct  an  appropriate  energy functional $U[\varphi,\rho,\beta]$ associated with the class of  test  field configurations  which  can be   used   to compare the  quality of  various  approximations  to the  true solution of the differential   equation   satisfied by the scalar  field. 
   This  will empower us to  determine based   on an objective quantitative criteria  when  our analysis  should   be trusted  over other treatments and  when   the opposite  is true. In the next section, we   obtain the chameleon  mediated  force  from the variation with distance of this functional.

A first naive attempt to find $U[\varphi,\rho,\beta]$  consists in considering the time-time component of the energy-momentum tensor 
associated with the system, which in turn, is the source of the Einstein's 
field equations of the model. Thus,  we  focus   on the gravitational  field equations that are obtained from varying the action (\ref{action}) 
with respect to $g^{\mu\nu}$:~\footnote{From the Bianchi identities followed by the use of Eq.~(\ref{mov}) one can see that the EMT of matter is not conserved in the Einstein frame: $\nabla^\mu T_{\mu\nu}^m= T^m(\partial_\varphi {\rm ln}A) \nabla_\nu \varphi$,where $A^2=e^{2\beta\varphi/M_{pl}} $ is the conformal factor between the Einstein-frame metric and the geodesic metrics [cf. Eq.~(\ref{conftrans})]. We then obtain $\nabla^\mu T_{\mu\nu}^m= T^m \beta/M_{pl} \nabla_\nu \varphi$. The right-hand side of this equation is precisely related with the chameleon force [cf. Eq.~(\ref{forcepointp})].}

\begin{eqnarray}
G_{\mu\nu} &=&  \frac{1}{M_{pl}^2} \left( T_{\mu\nu}^\varphi +  T_{\mu\nu}^m\right) \,\,\,,\\
T_{\mu\nu}^\varphi &=& (\nabla_\mu\varphi) (\nabla_\nu\varphi) - g_{\mu\nu} \left[\frac{1}{2}g^{\alpha\beta}
(\nabla_\alpha\varphi) (\nabla_\beta\varphi) + V(\varphi)\right] \,\,\,,
\end{eqnarray}
where, as before, we assume $T_{\mu\nu}^m = u_\mu u_\nu (\rho + P) + g_{\mu\nu} P$ for the EMT of matter. 
Now the energy associated with the total EMT under the assumptions of staticity and flat space-time is 
\begin{eqnarray}
U &=& \int_V \left(T^{\varphi}_{00}+T^m_{00} \right) dV =
\int_V \left[\frac{1}{2}(\nabla^i\varphi) (\nabla_i\varphi) + V(\varphi) + \rho \right] dV  \,\,\,,
\end{eqnarray}

where the time components of both EMT's are taken with respect to an observer that is static relative to the two-body 
configuration~\footnote{That is, we take a reference frame defined by the unit time-like vector (4-velocity) 
$n^\mu= \left(\frac{\partial}{\partial t}\right)^\mu$ as to coincide with $u^\mu$, such that $T_{00}^{\rm tot}
= n^\mu n^\mu T_{\mu\nu}^{\rm tot}$, where $T_{\mu\nu}^{\rm tot}= T_{\mu\nu}^{\varphi } + T_{\mu\nu}^{m}$. 
The staticity assumption translates into $n^\mu \partial_\mu \varphi =0$.}. 

While this energy functional certainly contains the total energy of the system from the Einstein-frame point of view, the extremization of 
this functional with respect to $\varphi$ with fixed densities (i.e., as considered independent of $\varphi$) does not lead to the 
static chameleon equation that is obtained from Eq.(\ref{mov}). That is, we require an energy functional extremized by the actual 
field configuration. As a consequence we turn our attention to the more appealing functional given by 

\begin{eqnarray}
\label{Ueff}
U_{\rm eff} &=& \int_V \left[\frac{1}{2}(\nabla^i\varphi) (\nabla_i\varphi) + V_{\rm eff}(\varphi) \right] dV  \,\,\,,
\end{eqnarray}
 where the integral is computed over the whole Euclidean three-dimensional space (i.e. ${\mathbb{R}}^3$)  and  where 
the effective potential $V_{\rm eff}(\varphi)$ at each point of space is the one  corresponding to the  
density of matter at that point (associated  with  each of the bodies and the media respectively) and which is given by (\ref{Veff}). 

We can rewrite the functional (\ref{Ueff}) integrating by parts , which leads to
\begin{equation}
\label{Eint}
U_{\rm eff} = \int_V \left[-\frac{1}{2}\varphi\nabla^2\varphi +V_{\rm eff}(\varphi) \right] dV \,\,\,,
\end{equation}
where we discarded the  irrelevant surface term which  must  vanish  at infinity.

One  issue that  is   extremely important to have in mind  is the fact that   the validity of our analysis  does  depend 
on the   accuracy  of  approximation used  in  the expansion of the   effective potential  for the scalar field 
in each region  of  space. In the present work, as we have explained earlier,  we   have used quadratic expansions  around the minimum  of the
effective potential  in each one of the   regions involved. This  should clearly  be a   very good  approximation  when all the  
bodies involved   (including the  media  between the  solid   bodies)    are in the  so called  deep {\it thin  shell} regime. 
However,  that regime   is only identified heuristically,  and sharp   boundaries   delimiting the exact  regions  in 
parameter   space   simply do not exist, and   in fact should not  exist  as the transition  between   {\it thin} and {\it thick}  shell 
regimes   must undoubtedly  be a smooth one.    This   faces us   with a  problem  when trying to  establish  
when   our results   are    more trustworthy  that the existing ones.   The problem would   of course be resolved  
if   one  had    exact  solutions   for   the complete    specific  problem  at hand,  involving  all the bodies present
in the actual  experimental situation, or  one that included  at least the  most relevant ones, namely the  source  body, 
the test body,  and the media  in between  them.  We note that  the  relatively simple treatments that  consider exact solutions 
for   just  one  body  surrounded  by a media  and  then  use  the gradient of the 
resulting   scalar  field  at the location of the second body  while ignoring the effect of such body on the  field  itself, and then attempt to 
correct for the  so  called   {\it thin shell} effect  by introducing a simple multiplicative factor extracted  from  consideration of another  one body 
problem,    cannot be      considered  {\it a priori} more reliable   than  our method. However, we must recognize,  of course, that  it is possible that  under some  circumstances,     those   analyses  might  provide  better  estimates  than ours. That 
would  of course correspond   to situations where our  second order expansion   fails to provide  an accurate enough  characterization of the  
potential in the regime   explored by the  actual    scalar field  configuration. 
That can lead to situations  where it is not clear  which  results  should one trust.   
   
Fortunately there is  a simple  method  to discriminate   between   two approximated solutions  to  the static  field  configuration 
corresponding to a given distribution of  sources and  media.   We refer here to the fact that  the  field 
configuration  corresponding  to   the solution  is  an  extreme    of the action  functional  which, for static  situations, correspond to the  
minima of the energy functional.   The point is then that  when faced  with two approximations to a given problem  one 
can   determine which one  is a better approximation by comparing the  value of the energy functional of the two
configurations.   Of course  in making  this comparison   it   is  essential that one  uses  the same energy functional and 
fixes the  relevant bodies and interpolating media  to be   exactly the  same  when  making the energetic
comparison. The   configuration with the lowest  value of the   energy functional   provides
a better approximation and thus  should  be  better  trusted. Of course ideally
one would  prefer  an exact solution  but lacking that,  we  must rely on  the  
better   one of the approximations.
    
  We have  carried  out precisely such analysis  in order to  compare  the field 
configurations  emerging from our analysis   with  the  field  configurations 
obtained   by  Khoury \& Weltman~\cite{KW04}.  As mentioned above, the relevant energy functional is given by Eq.(\ref{Eint}).  
We note  by looking  at   this equation  that  the second term of the integrand  
 will contain a constant corresponding to the minimum of the effective potential which is determined by the environment. This constant does not affect the calculation of the chameleon force between the bodies because  the force  is    derived  from  the variation of the  potential $U_{\rm eff}$ with respect to the    bodies'  separation  and the   contribution  from  such  a constant,  as  long as the bodies are  not deformed, is independent of  the separation. On the other, the integral of such constant in the whole  space,   extending to  infinity, would   lead to an infinite contribution for the energy. In any event it is clear   that such   term is  irrelevant  in the   determination of the  force  between  two  bodies
\footnote{ In the realistic context of the full fledge chameleon theory  that   term   would have to be regarded as   a contribution to  the ``cosmological constant''.}. In view of this, and in order to work with only finite energy functionals, we proceed to ``renormalize" our expressions   by subtracting the divergent term, and  considering  just,  

\begin{equation}
U_{\rm eff}^{*}=\int_V \left[-\frac{1}{2}\varphi\nabla^2\varphi +V_{\rm eff}(\varphi)-V_{\rm eff}(\varphi_{\rm min}^{\rm out})   \right] dV
\,\,\,.
\label{Energydef}
\end{equation}\\ 

In this paper, we  analyse two experimental situations: i) the E\"{o}t-Wash torsion balance:  
an Earth based experiment  where the large body is a mountain and the test bodies are centimeter-size metal sphere and ii) the Lunar Laser Ranging experiment where the 
large body is represented by the Sun and the test bodies by the Earth and the Moon. In each scenario the bodies are surrounded by an environment, 
which is described in more detail in Sections \ref{sec:EotWash} and \ref{sec:LLR}.

 For each of the experimental situations  considered in  detail in this   work we have computed  the value of the above functional Eq.(\ref{Energydef})  corresponding to  the two body problem using  the field   configuration   obtained with our  method  (described in Section \ref{FModel}) 
 and  that corresponding  to   the  field  configuration  obtained in  the standard 
approach.  We remind the  reader that in  such approach (as  exemplified by  the 
work of Khoury and  Weltman)  the  effect of the small body is  ignored when
determining the scalar  field  configuration,  and  the force on the latter is 
estimated  by  simply considering the gradient of the field corresponding to the large body and 
the environment at the location of the small body (see Appendix~\ref{sec:Plimit}) with some additional factors which are related with the thin shell parameter 
$\Delta R/R$. Moreover it  often  relies  
on  a approximated  expression   for the  effective potential that differs  from  
that the one we  employ (see Appendix \ref{sec:OBP}).\footnote{As regards the expressions obtained by Mota \& Shaw \cite{MS07} for the chameleon field, we could not apply the energy criterion since the explicit expression for the chameleon field is not reported in their work 
but they just present an expression of the derived force.}
On the other hand, a  solution considering the contribution of the test body  was  proposed by Hui et al \cite{Hui2009} and used in Refs. \cite{Brax2010,Burrage15} to calculate the E\"otv\"os parameter. In that work, a superposition of the one body problem solution  of both   the {\it large } and the  {\it test} bodies is considered. Furthermore, the authors  claim that this solution is valid outside both bodies. However, no expression for the field inside the bodies is provided. Therefore, it is not possible to apply the energy criterion proposed in this section to the above mentioned solution \footnote{ Nevertheless, we applied the energy criterion to the region where the field is defined and found that our solution has lower value of the energy functional}. The values of $\eta$ shown in Section \ref{sec:results} are computed from Eqs. \ref{fkhoury1} and \ref{fkhoury2}. However the reader should take into account that using such expressions for $\eta$ within the Khoury and Weltmann approach yields strictly speaking $\eta=0$ for universal $\beta$ given the fact that in such analysis the test body is treated as a point particle.

\begin{figure}
\begin{center}
\subfloat[$M=2.4$ {\rm meV } and $n=1$]
{\includegraphics[width=7.3cm,height=8.5cm,angle=-90]{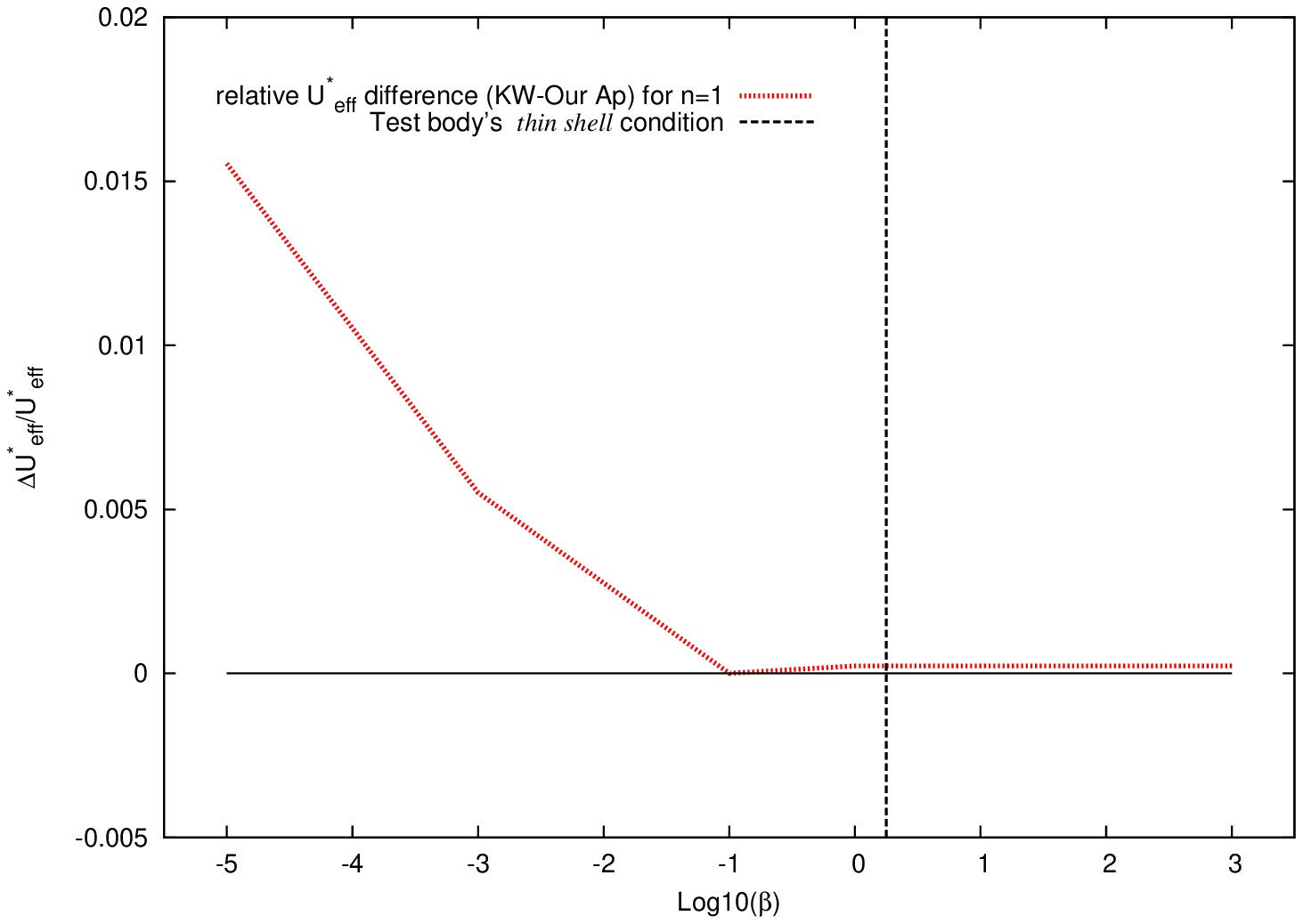}}
\subfloat[$M=10$ {\rm eV } and $n=1$.]
{\includegraphics[width=7.3cm,height=8.5cm,angle=-90]{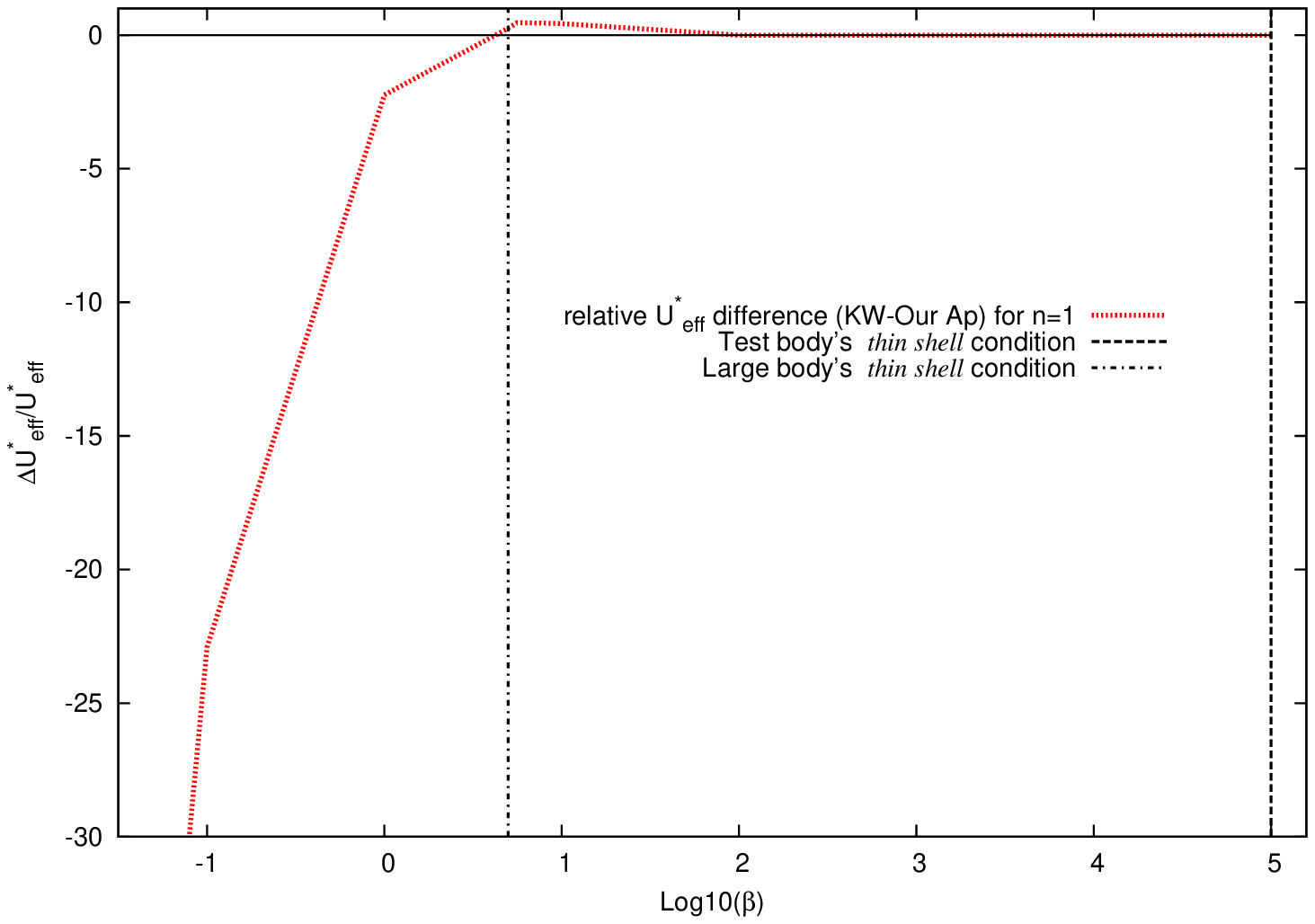}}
\caption{ Relative difference of energy ($\frac{U_{\rm standard }-U_{\rm our \,\ approach}}{U_{\rm standard}}$) computed for the E\"otv\"os experiment. The density of the environment surrounding both the test body (Al) and the source (the mountain) is assumed to be the same. Left: $M=2.4$ {\rm meV} and $\rho_{out} = 10^{-7} {\rm g \,\ cm^{-3}}$ for the density of the vacuum chamber. The {\it thin shell} condition for the mountain is always satisfied, while for $\beta$ 
lower than the value indicated by the vertical dotted line the {\it test} body does not have a thin shell. 
Right: $M=10$ {\rm eV} and  $\rho_{out} = 10^{-3} {\rm g \,\ cm^{-3}}$ for the density of the atmosphere. The vertical dotted lines show the thin shell condition  limits for the {\it test} and the large bodies (i.e. for values of $\beta$ 
lower than those indicated by the vertical dotted lines the {\it thin shell} condition does not hold) .}
\label{cenergy1}
\end{center}
\end{figure}

\begin{figure}[H]
\begin{center}
\subfloat[$n=1$]
{\includegraphics[width=7.2cm,height=7.5cm,angle=-90]{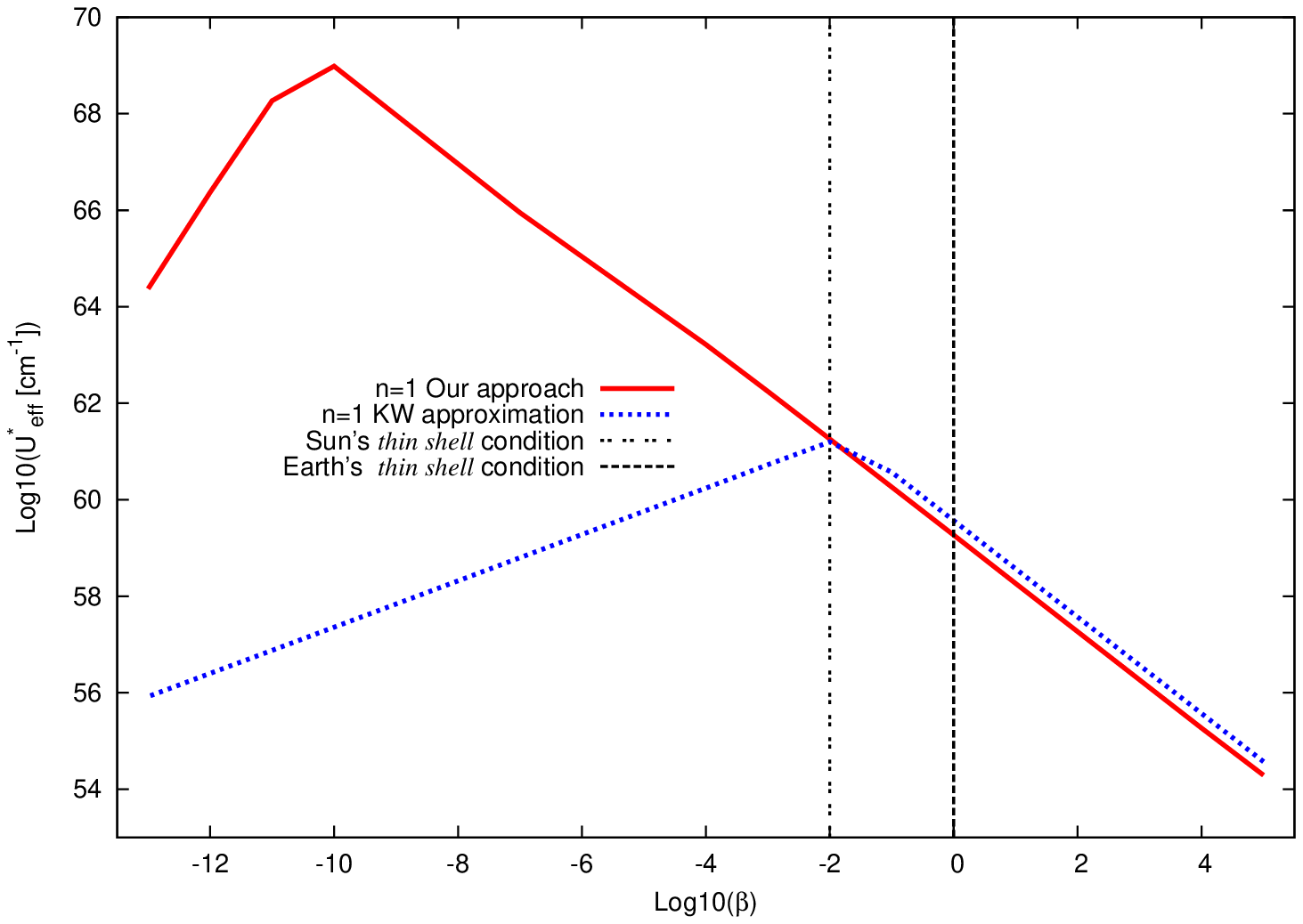}}
\subfloat[$n=2$.]
{\includegraphics[width=7.2cm,height=7.5cm,angle=-90]{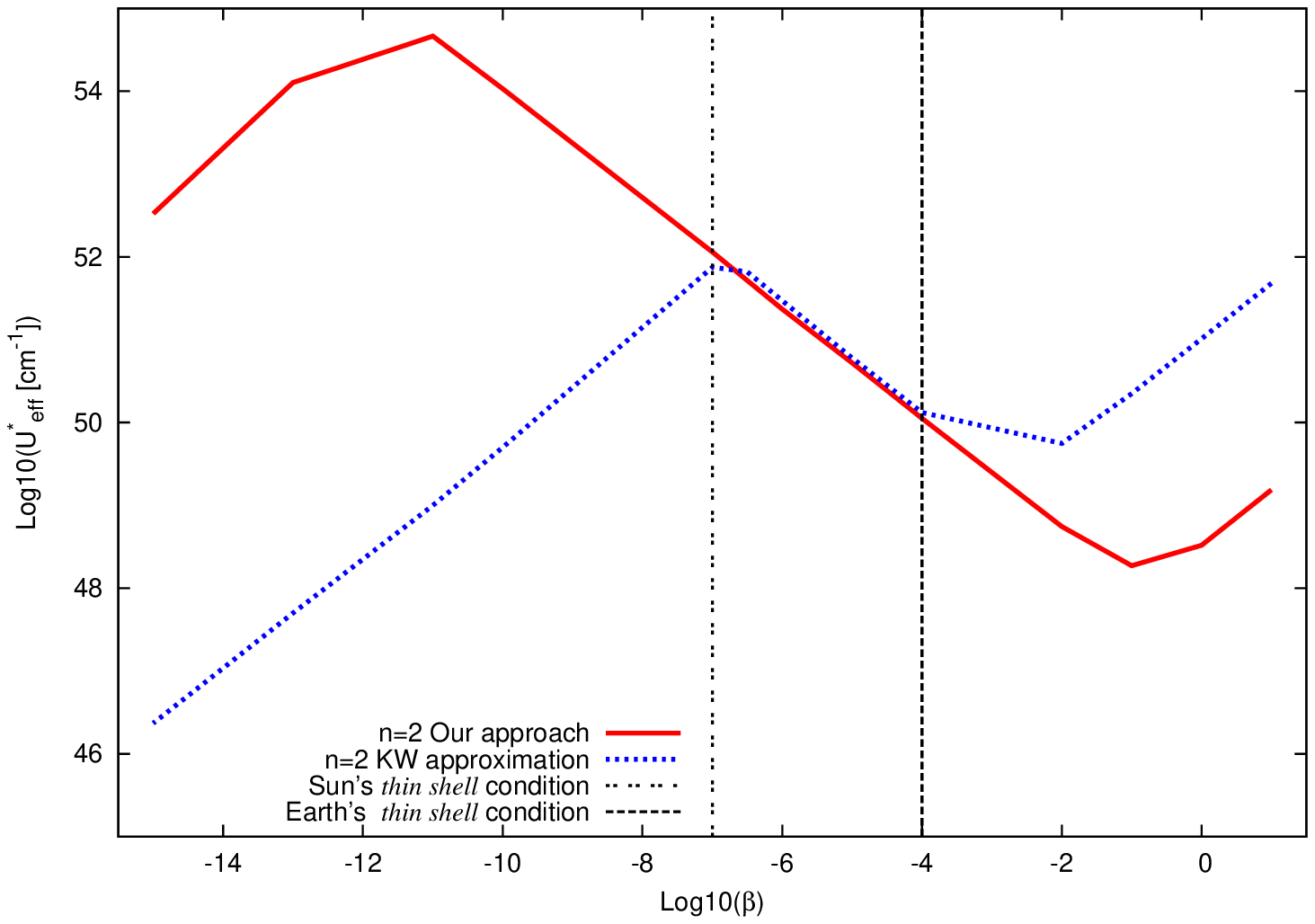}}
\caption{ Comparison of the energy computed for the LLR experiment using our approach (red) and the standard one (blue) taking $M=10$ {\rm eV}. The test body is assumed to be the Earth and the environment corresponds to the interstellar medium with density $\rho_{out}=10^{-24} {\rm g \,\ cm^{-3}}$ . The vertical dotted lines show the thin shell condition for Earth and Sun (for values of $\beta$ lower than those indicated by the vertical dotted lines the  bodies have no thin shell).}
\label{cenergy2}
\end{center}
\end{figure}

 Figure~\ref{cenergy1} shows the results  for the E\"{o}t-Wash torsion balance. The test body (Al) and the source (the mountain)  are  taken  as immersed 
 in the same   environment when  computing  the  energy of the field configurations.  We considered the cases $M= 2.4 $ {\rm meV} (the cosmological chameleon) with $\rho_{out} $ as  the density of the  vacuum-chamber while for  the  case  $M=10$ {\rm eV}, $\rho_{out} $ was taken as  the atmosphere's density. For the case $M=2.4$ {\rm meV} (left panel) the relative  difference of the energy functional is small but the energy of the field  configuration  obtained  with our method is always smaller. We have also checked that similar  results are obtained for various  values of $n$. 
On the other hand, for  the case   $M=10$ {\rm eV}, the situation is different.  Figure~\ref{cenergy1} (right panel) shows that for $n=1$ and $\beta<10$ the energy criterion indicates that  our approach yields an energy that is much larger than the corresponding energy obtained 
in the standard approach which indicates that the latter is much trustworthy. On the other hand, for $\beta>10$ the two body problem with the {\it quadratic} approximation for $V_{\rm eff}$ is a better solution to the exact problem than the one obtained from the standard approach\footnote{For $\beta>10^2$ the relative difference is of order $10^{-3}$ but always positive. This cannot be appreciated from the left panel of Figure~\ref{cenergy1} 
due to the scale of the plot. For  each of the values of  $n$  considered here such transition takes place  at  a different value of $\beta$ (see Table \ref{TS3} in Section \ref{sec:results})}.  It should be noted, that we could not apply  our  energy criterion to the  most realistic case where one  includes   the metal shell of the vacuum chamber  in the modeling for the E\"{o}tv\"os torsion balance   experiment.  This is because we cannot compare with the  standard approach since, as far as we are aware, it does not usually include the explicit determination of the chameleon field's profile  corresponding  to  this aspect  of the experimental device. \footnote{Ref.~\cite{Upadhye12} obtains a numerical solution for the torsion pendulum which has a different geometry than the torsion balance analyzed in this paper and estimates the effect of including a shell in the value of the force. Most analyses  focused in the E\"{o}tv\"os torsion balance do not include the metal encasing in the calculation of the field and/or the force \cite{KW04,Brax2010,MS07,Hui2009,Brax2012}.} \footnote{In  section \ref{sec:EotWash}, we estimate the value of $\eta$ for the standard approach including  the  effect of the metal shell   through   multiplication of a  correction  term  as suggested by \cite{Upadhye12}.}.  Figure \ref{cenergy2} 
 compares the values of the energy functional computed for the LLR experiment obtained from our method and from the standard approach 
taking $M=10$ {\rm eV}. The density of the environment is assumed to be the one of the interstellar medium $\rho_{out}=10^{-24} {\rm g \,\ cm^{-3}}$ and the vertical dotted lines show the  onset  of the  thin shell condition for Earth and Sun  respectively  (for  values of $\beta$ lower than the  corresponding  to vertical lines,  the thin shell condition does not hold). For the cosmological chameleon, the energy criterion indicates that our solution is a better approximation than the  one obtained  in standard approach for $n=1,2,3,4$ and for the values of $\beta$ considered here \footnote{ For $M=2.4$  {\rm meV}  (the cosmological chameleon) and the $\beta$ values that we test, the {\it thin shell} holds always for the Sun, Earth and Moon. }.  Moreover, for $M=10$ {\rm eV}  we  find that for each of the tested  values of $n$,   there is a region (which corresponds to the case where  the large body satisfies  the {\it thin} shell condition, while  the {\it test} body may not)  where the energy criterion indicates that the field  profile obtained  by our method   is a better solution than the one obtained in the standard approach.  On the other hand,  for the $\beta$ values excluded from that region (this is the case where  the large body fails to satisfy  the {\it thin shell} condition), the KW approach gives a better result (see Fig.\ref{cenergy2}) .

 We will limit the results of this paper to the case where the {\it quadratic} approximation to the effective potential is a good one and leave for a future work \cite{Krai17} the case where another approximation has to be considered. It should be stressed, that for both experimental situations  there  are ranges of the model's  parameters  for which  despite the fact  that  the {\it test} body  does not satisfy  {\it thin shell}  condition, the energy criterion indicates that the   scalar  field  profile  obtained  by  considering the  full two body problem within  the {\it quadratic} approximation for the effective potential yields  a better approximation to the exact solution than the one used within the standard approach. On the other hand, when the {\it thin shell} condition does not hold for the  {\it large} body, the energy criterion always indicates that the standard approach yields  a  better  approximation than the one  obtained  by the method proposed in this work, i.e., indicating  that  it is essential  to employ something  like  the {\it thick shell} approximation for the effective potential.

\section{Chameleon   mediated force  between two  spherical objects}
\label{sec:force}

In this section we calculate from first principles the effective chameleon force between the {\it large} and the {\it test} bodies. This force, together with the gravitational force on the {\it test} body, will be  considered as characterizing  a ``free falling" {\it test} body in laboratory conditions under the influence of both interactions. 
The expression for this force together with the chameleon field computed in  sections \ref{sec:EotWash} and \ref{sec:LLR}  will allow us to compute numerically the 
E\"otv\"os parameter associated with the acceleration of two {\it test} bodies of different composition and to estimate the magnitude of the predicted 
violation of WEP.

 As we mentioned before, the variation of the functional of Eq. (\ref{Eint}) with respect to $\varphi$ with the densities of the bodies fixed, 
leads to the actual equation for the chameleon for a static problem:
\begin{eqnarray}
\label{chamEqstat}
\nabla^2 \varphi= \frac{\partial V_{\rm eff}}{\partial \varphi},
\end{eqnarray}
where $\nabla^2$ stands for the Laplacian operator in three dimensional Euclidean space.

We can rewrite the functional (\ref{Ueff}) integrating by parts and using Eq.(\ref{chamEqstat}), which leads to
\begin{equation}
\label{Eint2}
U_{\rm eff} = \int_V \left[-\frac{1}{2}\varphi\nabla^2\varphi +V_{\rm eff}(\varphi) \right] dV = 
\int_V \left[-\frac{1}{2}\varphi\partial_\varphi V_{\rm eff}(\varphi) + V_{\rm eff}(\varphi) \right] dV \,\,\,,
\end{equation}
where  as usual, we discarded the surface term.

Using Eq,(\ref{Veff}) we obtain after some simplifications

\begin{equation}
\label{Eint3}
U_{\rm eff} = \int_V \left[ \frac{1}{2} V_{\rm eff} (\varphi)(2+n)+ \frac{(1+n) \beta \varphi T^m}{2M_{pl}} \right] dV
\,\,\,.
\end{equation}
At this point the energy functional is exact. However, we  now use in this functional the quadratic approximation 
for $V_{\rm eff}$ about its minimum in each region of space (i.e. inside and outside the bodies).

Our goal is to work with the difference ${\hat \varphi}= \varphi-\varphi_{\rm min}$ rather with 
$\varphi$ alone given that ${\hat \varphi}$   vanishes at infinity. According to this, Eq. \ref{Eint3} becomes,
\begin{equation}
\label{Ueffmin3}
U_{\rm eff} = \int_V \left[ \frac{(2+n) m_{\rm eff}^2}{4} {\hat \varphi}^2 + \frac{(1+n) \beta T^m \hat \varphi}{2M_{pl}} +\frac{(1+n) \beta T^m \varphi_{\rm min}}{2M_{pl}}+\frac{1}{2} V_{\rm eff}^{\rm min} (2+n)\right] dV
\,\,\,.
\end{equation}

The only approximation we have made so far for the energy of the whole system is to replace the effective potential of the chameleon with the corresponding expansion around its minimum in each of the three regions (i.e. the two bodies and the environment).


We can  now compute  the chameleon mediated force using  $F_{\rm z \varphi}=-\frac{\partial U_{\rm eff}^{\rm trunc}}{\partial D}$, where $D$ is the distance between the center of the two bodies. To this aim, we first notice that the last two terms within the integral of Eq.~(\ref{Ueffmin3}) 
are independent of the separation $D$ of the bodies, and thus, do not contribute to the force, i.e., 
$\frac{\partial}{\partial D} \lbrace\int_V [\frac{(n+2)}{2}V_{\rm eff}(\varphi_{\rm min}) + \frac{(1+n) \beta \varphi_{\rm min} T^m}{2M_{pl}} ] dV\rbrace=0$
~\footnote{As usual, the integral $\int_V  V_{\rm eff}(\varphi_{\rm min}) dV$ (in all the space) is an infinite constant.}.

In order to calculate the total energy, we have to consider  the contributions due to the chameleon in the regions inside the two bodies and in the region exterior to both bodies. To do so, let us define $V_1$ as the region corresponding to the {\it large} body, $V_2$ as the region corresponding to the {\it test} body, 
and  $V_3$ is the region exterior to the {\it large} body in the coordinate system centered in the {\it large} body:
\begin{equation}
\quad V_3=
\begin{cases}
 R_1\leq r\leq \infty \\
 0\leq\theta\leq\pi\\
 0\leq\varphi\leq 2\pi
\end{cases}
\end{equation}
 
Thus, the relevant energy functional can be written as: 
\begin{eqnarray}
 U_{\rm eff} & = & \int_{\rm large \,\ body} t_{00}[\hat{\varphi}(V_1)] dV_1 + \int_{\rm test \,\ body} t_{00}[\hat{\varphi}(V_2)] dV_2 + \int_{\rm outside\,\ bodies} 
t_{00}[\hat{\varphi}({\rm out})] dV_{out} \nonumber \\ 
&=&\int_{V_1} t_{00}[\hat{\varphi}(V_1)] dV_1 + \int_{V_2} t_{00}[\hat{\varphi}(V_2)] dV_2 + \int_{V_3} t_{00}[\hat{\varphi}({\rm out})] dV_3-  
\int_{V_2} t_{00}[\hat{\varphi}({\rm out})] dV_2
\label{volumenes}
\end{eqnarray}
where $t_{00}= \frac{(2+n)}{4}m^2_{\rm eff}\hat{\varphi}^2  + \frac{({\color{blue}1}+n) \beta \hat{\varphi} T^m}{2M_{pl}}$ corresponds to the first two terms 
of Eq.~(\ref{Ueffmin3}) which are the responsible for the force between the two bodies. Here $\hat{\varphi}({\rm out})$ refers to the field outside the 
two bodies,  and because of this definition and the arrangement of the integrals, we have to subtract the last term. In consequence, the total force can be expressed as:

\begin{eqnarray}
\label{fuerzaC}
F_{\rm z\varphi} &=& \frac{\partial}{\partial D}\int_{V_2}\Big\lbrace\hat{\varphi}_{\rm in2}(\vec{r}^{\,\prime})\Big[\frac{(n+1)}{2}\frac{\rho_2\beta}{M_{pl}}\Big]-\frac{(2+n)\mu_2^2\hat{\varphi}^2_{\rm in2}(\vec{r}^{\,\prime})}{4}\Big\rbrace 
d^3{\vec r}^{\,\prime} \nonumber\\
&+& \frac{\partial}{\partial D} \int_{V_1}\Big\lbrace\hat{\varphi}_{\rm in1}(\vec{r})\Big[\frac{(n+1)}{2}\frac{\rho_1\beta}{M_{pl}}\Big]-\frac{(2+n)\mu_1^2\hat{\varphi}^2_{\rm in1}(\vec{r})}{4}\Big\rbrace 
d^3{\vec r} \nonumber\\
&+&\frac{\partial}{\partial D}\int_{V_3}\Big\lbrace\hat{\varphi}_{\rm out}(\vec{r})\Big[\frac{(n+1)}{2}\frac{\rho_{\rm out}\beta}{M_{pl}}\Big]-\frac{(2+n)\mu_{\rm out}^2\hat{\varphi}^2_{\rm out}(\vec{r})}{4}\Big\rbrace 
d^3{\vec r} \nonumber\\
&-& \frac{\partial}{\partial D}\int_{V_2}\Big\lbrace\hat{\varphi}_{\rm out}(\vec{r}^{\,\prime})\Big[\frac{(n+1)}{2}\frac{\rho_{\rm out}\beta}{M_{pl}}\Big]-\frac{(2+n)\mu_{\rm out}^2\hat{\varphi}^2_{\rm out}(\vec{r}^{\,\prime})}{4}\Big\rbrace d^3{\vec r}^{\,\prime};
\end{eqnarray}
where we used $T^m \approx -\rho$, and $\rho_{1}$ and $\rho_{2}$ are the densities of the {\it large} and the {\it test} bodies, respectively, and 
$\rho_{\rm out}$ is the density of the {\it environment} (i.e. the density outside both bodies).  We stress that the terms quadratic in the field appearing in Eq.~(\ref{fuerzaC}) are {\it not} negligible 
in comparison to the corresponding linear terms, as  our detailed numerical calculations show. In fact both terms turn out to be 
of the same order of magnitude.


 In the last two integrals of Eq.~(\ref{fuerzaC}), we have to rewrite the chameleon in terms of the corresponding coordinates 
adapted to each body~\cite{Gume1,Gume2,Scatt,Scatt2}. Moreover, we neglect the contribution of the second part of 
$\hat{\varphi}_{\rm out}$ given by $\sum_{lm}C_{lm}^{\rm out2}k_l(\mu_{\rm out}r')Y_{lm}(\theta',\varphi')$, to the last integral [i.e. we take $\hat{\varphi}_{\rm out}(\vec{r'})\approx \sum_{lm} C_{lm}^{\rm out1}\sum_{w=0}^N \alpha^{*l0}_{w0}i_w(\mu_{\rm out} r') Y_{w0}(\theta',\varphi')$] 
because that contribution turns out to be negligible in  comparison with the other one. 

Each integral of the above equation has a linear term in $\varphi$ and a squared term ($\varphi^2$). Therefore, we separate $F_{\rm z\varphi}$ in two terms, a linear one $F_{1\rm z\varphi}$ and a squared one $F_{2\rm z\varphi}$ such that $F_{\rm z\varphi}=F_{1\rm z\varphi}+F_{2\rm z\varphi}$. The final expressions for such terms that provide the 
chameleon mediated effective 
force on the {\it test} body are not very enlightening, and the steps leading to them are
rather cumbersome and technical (see Appendix~\ref{sec:detailsforce}). However, the most important point to be stressed, and which 
constitute the  basic  result of this paper, is that the expressions for $F_{1\rm z\varphi}$ and $F_{2\rm z\varphi}$ 
show an explicit  dependence with the composition and size of the {\it test} body. Therefore, and  as already noted, the WEP is violated in principle
in these kind of models even in the case  of an universal coupling. 

In the next section we use these results in order 
to evaluate numerically the extent to which this effective force is suppressed by the {\it thin shell} effects in 
the two bodies, and also by other effects induced by the presence of additional objects in the setup. 
Then we estimate the E\"otv\"os parameter and confront the outcome with the bounds imposed by the current data 
associated with the Earth based experiments and the LLR.

The general  formulae that  we  have provided  can be used  for the evaluation of the force  to 
any desired   degree of precision.  This  can be   done by  following the   steps we have presented in previous sections  and  by   adjusting  the level of  approximations  we have used.  The     approximations   involved  are of course  the   cut-off  in the series expansion,  which  can  straightforwardly be  continued to any desired order.

\section{Applications}
\label{sec:results}

 It is important to point out that  the geometry of the actual Earth based experiments that have been  used  to test the WEP, such as  
those using torsion balances  is much more  complex  than what our simple model   depicts. In particular,  the inclusion of the other material  bodies  that  are present in the   laboratory  near the 
torsion balance could  drastically modify  the effective chameleon force acting on the {\it test} bodies, and their incorporation  might led  to important changes  in  the theoretical predictions 
of the E\"otv\"os parameter. For instance, the environment that separates the 
torsion balance (that contains the {\it test} bodies) from the hill (which in the  simple one body approach is the only source of the chameleon field) 
does not consist of just  a vacuum and the Earth's atmosphere, but  includes  also   a  metal case and  other  objects located in the proximity of the experimental equipment.
Those   complications can often be ignored when one   is  considering  linear fields which   couple   in a  non-universal  way to matter, and  whose presence can   led to  effective violations  of the  WEP    just as in  the  original Fifth  Force proposals~\cite{Will14}.  However,   when  dealing with highly non-linear models  as the one  studied here,  the situation might be  more complicated. 
Therefore  one  might need, in principle,   a  very  detailed model for the actual  $\rho$  in the laboratory in order to take into account the true  effect of the environment  around the two bodies (the {\it large} body and the {\it test} body) used in the  test of  the WEP.
This  would  require  the modeling of  the  matter  distribution in   great detail, and  then  facing  the  very  difficult task of 
solving a complicated non-linear partial differential equation with a rather intricate  boundary conditions. In order to advance in this direction and to make the calculations feasible, we are forced to ignore  some of these complications 
while incorporating in the modeling  of the situation some  of the most important features of the experiment. For instance, we  have analyzed the extent to which different environments 
affect the resulting chameleon force between the two bodies. 

In the following  we  will  proceed in two stages. In the first one we consider the simplest model for $\rho$ as given 
by Eq.~(\ref{densOneB}). First we take two type of environments represented by 
$\rho_{\rm out}$: one is given by the chamber's vacuum density,  and the other is given by the Earth's atmosphere. 
 In the second stage we reanalyzed the results of the first stage,  after  taking into account,  in a self-consistent manner, the  encasing's  material (which we consider to be of spherical shape) around the {\it test} body,  as a manner to  more accurately  characterize  the vacuum chamber used
in the E\"ot-Wash experiment that shields the test bodies (see Sec.~\ref{mencasing})~\footnote{We thank P. Brax for calling our attention to this aspect of the experiment.}. Regarding this  point, we  need to  emphasize that,  in previous works \cite{Upadhye12}, the 
contribution of this metal encasing has been  modelled together with the two bodies, the low  density  environments  and the metal encasing 
as  planar slabs (in contrast   to  our   more  realistic modeling   based on 
spherical like objects). That is, the setting  used  in~\cite{Upadhye12} corresponds to a ``one dimensional planar''  model\footnote{It should be noted that  the  analysis  of Ref. \cite{Upadhye12}  is devoted to the torsion pendulum experiment rather than the torsion balance here considered. Needless is to say that the two are not exactly the same.}. 

Now,   in  the scenario where   one  does not consider such 
metal casing, the  model could    be  applied  to at least  two kind of experiments: i) a laboratory  experiment 
similar to the E\"ot-Wash experiment but without the metal shell. This scenario, as  far as we  know,  has not been implemented in  
an actual   high precision Earth based experiment, and  therefore our estimates cannot be  used directly,  at this time,   to set  relevant bounds. However,   combined  with rough 
 estimates   of the suppression  effects, these already show that the appearance of the {\it thin shell} effects are by themselves 
 insufficient to suppress  the  observable violations of the WEP for some  values of the parameters; ii) an actual space-based experiment like the LLR 
where the Sun plays the role of the {\it large} body and the Earth or Moon represent the {\it test} bodies. For this second situation  the   use of our  analysis   would mean  one 
 is  neglecting  the effects of an actual {\it three body problem} where  the Sun, the Moon and the Earth  collectively   determine   the scalar-field. A much    more  precise   study would  require  taking the effects of these three  bodies into account simultaneously and 
self-consistently. We have  not done  so  and instead have relied   on an analysis  based on two  body  system  as  an approximation.  We consider this to  be a reasonable   approximation  due to the fact that  despite the intrinsic non-linearity of the  model,  the  change due to the presence of the  Earth   on the value of the scalar field at the Moon's  surface is   sub-leading to that of the Sun.  Thus,  in our estimates,  when computing the acceleration of the Earth towards the Sun we neglect the presence of the Moon and viceversa. This  by the  way,   is a standard  approximation used  widely  in the   community  studying these questions. 

Now we illustrate the usefulness of the analytic expressions for the force in situations of  experimental relevance.

\subsection{The  E\"ot-Wash Torsion Balance Experiments}
\label{sec:EotWash}

\subsubsection{An idealized experiment without a vacuum chamber}

One  the  most  precise  experiments  on  this line  consists of a continuously rotating torsion balance 
 which is used to measure the acceleration difference  toward  a large source (like the Earth, a lake,  or  a mountain) of {\it test} bodies with the same mass but  different composition. In addition,   as  we  have already mentioned, the E\"ot-Wash experiment includes a vacuum chamber 
that encases the test bodies,  and which is shielded by a metal encasing. We shall first consider a model of a simplified ``E\"ot-Wash'' scenario where the metal encasing is absent (or ignored). The set up  considered  in previous sections correspond  to this idealized experiment.
  Despite the fact that   such  scenario does  not  represent the  realistic  experimental set up   at present, we have decided to consider it  in order to motivate the realization of an 
experiment that avoids the inclusion of such  vacuum chamber  encasing,  providing the theoretical bounds one might be  able to achieve   by a relatively  simple  modification  of the current   experiment.

Thus  we will compute the E\"otv\"os parameter associated with the differential acceleration of two bodies of different composition. This parameter is given by $\eta\sim2\frac{|\vec{a_1}-\vec{a_2}|}{|\vec{a_1}+\vec{a_2}|}$, where $\vec{a_i}=\vec{a}_{i\varphi}+\vec{g}$ ($i=1,2$) is the acceleration of the $i-${\it test} body due to the chameleon force $\vec{a}_{i\varphi}$, 
and the force of gravity $\vec{g}$, which is basically due to the gravitational field produced by the {\it large} body. 
The acceleration $\vec{a}_{i\varphi}$ on the small body is computed using Eq.~(\ref{fuerzaC}).

 As we emphasized before, in all the cases that we have analyzed  we find that, the linear and  the quadratic terms of the chameleon field in the expression of the total energy (and hence the force) are comparable, so neither of them can  be ignored.  In most cases, the  ``optimal " cutoff $N$ in the series expansion,  estimated  as the integer part of $N_0$ turns out to be zero (cf. Section~\ref{FModel}). Thus, we keep only the first term of each sum so as to avoid numerical instabilities. The value  for $N$ required to improve the accuracy in the solution increases with increasing $n$ and $\beta$, the effect   being more  pronounced   for   dependence on $n$. Furthermore, the value of $N$ also increases with the  density of the outside medium $\rho_{out}$. In fact, in some cases, the   value  needed for $N$ is  so large that it effectively  impedes  the calculation.

Moreover, regarding  the summations over the infinite ranges for  $l$ and $w$ in  the series of Eq. (\ref{fuerzaCf31f}) 
we  have checked the  rate  of convergence  and  found  that  one  
 obtains basically the same result when cutting off the sums at  $w=4$ and $l=4$ than when taking only the  dominant  terms  $l=0$ and $w=0$ 
(the relative difference between both calculations in the predicted  value of the E\"otv\"os parameter, namely on the value of 
  $\frac{\Delta\eta}{\eta}$ is of order $5 \times 10^{-3}$) . 
The reason for this is that  when $l$ increases, $k_l$ also increases  but $i_l$ decreases very rapidly. The main  point is that 
the  quantity $C_l$ decreases faster   with $ l$   than the  corresponding  rate of  growth of $k_l$.  In  Appendix~\ref{NL} we show, that the results obtained for $N=4$  and taking the summations up to $l=4$, are the same as those presented in this 
section. We have been able to perform calculations up to N = 20 and we have observed that the first terms, those corresponding to $ l = w = 0$ are the ones representing the main  contributions to the results.

On the other hand, in the Appendix~\ref{sec:Plimit} we study the {\it test-particle} limit by taking $R_2\to 0$ as $\rho_2$ is kept constant, and 
analyze how the violations of the WEP in the E\"otv\"os parameter are suppressed by the presence of a thin shell. 
In particular, this occurs even when the coupling is {\it not} universal.

One of the  most stringent bounds regarding the   possible violations of  the WEP  is found when  comparing the differential acceleration of two {\it test} bodies (e.g. two test-balls of Beryllium and Aluminium) using the Earth  or  its local  inhomogeneities, as the source of the acceleration; the experimental value is  $\Delta a_{\rm Be-Al}=(-2.5\pm2.5)\times10^{-15}{\rm m/s^2}$ \cite{Adel}. As discussed in  Refs.~\cite{Adel} and \cite{Adel2}, there are two sources for the relevant signals in short-range effects: a hillside of 28 m located close to the laboratory, and a layer of cement blocks added to the wall of the laboratory of\\
 1.5 m (we estimate 1 m distance between the wall and the device).  We model the short range sources by spherical masses at a given distance and consider only the contribution of the hillside, such as in Ref.~\cite{LTBS12}. For such a configuration, the differential acceleration on two {\it test} bodies due to the hill produces an E\"otv\"os parameter $\eta_{\rm Be-Al}^{\rm hill}=(-3.61\pm3.47)\times10^{-11}$. In order   to characterize  such a configuration in our expression for the force, we make the following assumption for the {\it test} bodies: for the masses we 
take $m_{\rm Be}=m_{\rm Al}\approx 10\hspace{0.1cm}{\rm g}$
and for the densities  $\rho_{\rm Be} \simeq 1.85  {\rm g}\hspace{0.1cm}{\rm cm^{-3}}$, $\rho_{\rm Al} \simeq 2.70 {\rm g}\hspace{0.1cm}{\rm cm^{-3}}$,  and $\rho_{\rm hill} \simeq 9.27 {\rm g}\hspace{0.1cm}{\rm cm^{-3}}$. We consider the bodies as surrounded by an environment of constant density $\rho_{\rm out}$ 
and take two values for this density (so  that  people may compare actual and possible experiments): i) the density of the vacuum-chamber $\rho_{\rm out }= 10^{-7}  {\rm g}\hspace{0.1cm}{\rm cm^{-3}}$; ii) the density of the Earth's atmosphere $\rho_{\rm out }= 10^{-3}  {\rm g}\hspace{0.1cm}{\rm cm^{-3}}$. 

 In addition, for the  mass  $M$ that appears in the chameleon potential [cf. Eq.~(\ref{Vbare})] we take also the value $M= 2.4 \,\ {\rm meV}$ which, 
 as mentioned before, corresponds to the cosmological chameleon.

In Figure~\ref{resM} we show the predictions for the E\"otv\"os  parameter $\eta$ calculated with the method developed in this paper together with the predictions of the standard approach .   For this latter, and following \cite{KW04,Hui2009,Burrage15} (among others), 
we use the following expression for the total force (gravitational plus chameleon field) between two bodies $A$ and $B$,

\begin{subequations}
\begin{equation}
F_{AB}=\left(1+2Q_{A}Q_{B}\right) F_N,
\label{fkhoury1}
\end{equation}
\begin{equation}
Q_C=min\left(\beta,\frac{|\varphi_{\infty}-\varphi^{\rm in}_{C\rm min}|}{2M_{pl}\Phi_N}\right),
\label{fkhoury2}
\end{equation}
\end{subequations}
where $F_N$ refers to the gravitational force between $A$ and $B$, $\Phi_N=\frac{G\cal{M}_C}{R_C}$ refers to the Newtonian potential of the body $C$.
We  considered the two different values for the density of the environment mentioned above\footnote{In the latter case  and  for $\beta >1$  we  were unable  to  estimate the effective  violation of the WEP because,  in this case the value of $N$ (the cutoff in the  series expansion used in  relating the two   coordinate systems)   becomes  very large . The same problem arises for the case $n<0$.}. The difference between our prediction and those calculated by other authors are due to two main reasons: i) other authors do not compute the   field  configuration for  the two body problem as  done   in this paper, instead  they estimate the  chameleon  mediated  force  of the {\it large} body   on the {\it test} body by considering the effects of the field  only due to the {\it large} body, and then they compute this force by using the gradient of the field at the location of the point-like test body and the characterization of that  effect of the field on the small   object  through an  estimation of  an   effective coupling \footnote{For instance, Khoury \& Weltmann \cite{KW04} use the {\it test particle} limit to estimate the force,  Brax et al .\cite{Brax2010} use the superposition principle to calculate the force, Puetzfeld \& Obukov \cite{PO15} consider the {\it test particle} limit, Mota \& Shaw \cite{MS07} consider two infinite parallel planes, Tamaki \&  Tsujikawa \cite{TT08} consider the {\it test} bodies as {\it test particles} with the acceleration of the {\it test particle} proportional to  an effective coupling of the field to the {\it test} body; and therefore miss the contribution of the {\it test} body on the  chameleon field.} ii) other authors  split the prediction into two cases: screened or unscreened {\it test} body considering only the effect of the {\it test body} when computing the force in an approximate way as described above.
In  the present work    we  amply the   energy  functional  criteria  described in Sec III    in order to establish  which of the  approximate   configurations provides  a  more  accurate  characterization of the situation in  each   case. In  particular our own  treatments are  based on the   quadratic approximation to
the  effective potential in  each region,  but as  indicated by the energy 
functional criteria, the   results    so   obtained  are not always  the most
accurate  among the  existing ones,  and   for some  situations  and   values of the
  model's parameters   the    standard  results are more trustworthy than ours. However, we have  also found that  in  various  situations  our   analysis   yields 
approximations  to the    chameleon field  profile  that    are better  that  those
obtained in  the standard  approaches.  In those cases  our   results  are more
trustworthy  than the previously existing ones.
Tables \ref{TS1} and \ref{TS2} show,  for various  values of the exterior medium's  density, the values of $n$ and $\beta$ for which  the  {\it test} body (made of  {\rm Al} and {\rm Be})  transitions from  satisfying  the  {\it thin shell} condition  to  failing  to  do  so.   Furthermore, we use the energy criterion developed in Sec.\ref{sec:energy} to determine when  the solution   obtained  by the methods of  this paper is better than the one obtained by the standard approach to the problem at hand.  We note  that  these approximations can be improved by turning to  something like a {\it thick shell} approximation for the {\it test} body potential,   while  still   working   with  the   scalar field's   equation for the full   two body problem \footnote{ For the  situations  we have  studied   in this work, the {\it large} body is always screened for $M=2.4$ {\rm meV} and  the   considered  values of $n$ and  $\beta$. }.  We  will  present   the results  obtained  by such refined  calculations in a forthcoming paper \cite{Krai17}.  In this paper, our treatment includes cases in which the {\it test} body has and does not have {\it thin shell}. For instance, for $n=1$, the {\it thin shell} condition is satisfied for $\beta > 0.1$ . On the other hand, it is well known that the change between the screened and the unscreened case is smooth and  it is not well described by the standard approach~\cite{Hui2009,Brax2012}. The behaviour of the  values of  $\eta$ estimated by other authors  can be divided in two  categories depending on whether the test body is screened (larger $\beta$) or not (lower $\beta$): $\eta$ decreases sharply for lower $\beta$ (unscreened {\it test} body) and slowly with increasing $\beta$ (screened {\it test} body). 

\begin{table}
\renewcommand{\arraystretch}{1.3}
\caption{Thin shell condition for different test bodies surrounded $\rho_{\rm out}=\rho_{\rm vacuum}$ and $M=2.4$ meV }
\begin{center}
\begin{tabular}{K{2.7cm}K{2.7cm}K{2.7cm}}
\hline
\hline
 $n$  &$\beta_{\rm Be}$  &$\beta_{\rm Al}$\\
 $(1)$ & $(2)$ &$(3)$\\ \hline
 $1$ &$3.16$ &$1.78$ \\
 $2$ &$0.46$ &$0.32$ \\
 $3$ &$0.13$ &$0.10$ \\
 $4$ &$0.056$ &$0.032$ \\
\hline
\multicolumn{3}{l}{Values of the couplings $\beta$ (columns (2) and (3) ) associated with} \\ 
\multicolumn{3}{l}{a given $n$ (column (1) ) below which the  {\it thin shell} condition }\\
 \multicolumn{3}{l}{is no longer satisfied within ${\rm Al}$ and ${\rm Be}$ test bodies.} \\
\end{tabular}
\label{TS1}
\end{center}
\end{table}

\begin{table}
\renewcommand{\arraystretch}{1.3}
\caption{Thin shell condition for different test bodies surrounded by $\rho_{\rm out}=\rho_{\rm atmosphere}$ and $M=2.4$ meV }
\begin{center}
\begin{tabular}{K{2.7cm}K{2.7cm}K{2.7cm}}
\hline
\hline
 $n$  &$\beta_{\rm Be}$  &$\beta_{\rm Al}$\\
 $(1)$ & $(2)$ &$(3)$\\ \hline
 $1$ &$0.15$ &$0.10$ \\
 $2$ &$0.043$ &$0.032$ \\
 $3$ &$0.040$ &$0.030$ \\
 $4$ &$0.010$ &$0.0070$ \\
\hline
\multicolumn{3}{l}{Values of the couplings $\beta$ (columns (2) and (3) ) associated with} \\ 
\multicolumn{3}{l}{a given $n$ (column (1) ) below which the  {\it thin shell} condition}\\
 \multicolumn{3}{l}{is no longer satisfied within ${\rm Al}$ and ${\rm Be}$ test bodies.} \\
\end{tabular}
\label{TS2}
\end{center}
\end{table}

\begin{table}
\renewcommand{\arraystretch}{1.3}
\caption{Thin shell condition for the mountain surrounded by  $\rho_{\rm out}=\rho_{\rm atmosphere}$ and two test bodies surrounded by  $\rho_{\rm out}=\rho_{\rm vacuum}$ and $M=10$ eV}
\begin{center}
\begin{tabular}{K{2.7cm}K{2.7cm}K{2.7cm}K{2.7cm}}
\hline
\hline
 $n$  &$\beta_{\rm mountain}$ &$\beta_{\rm Be}$  &$\beta_{\rm Al}$\\
 $(1)$ & $(2)$ &$(3)$&$(4)$\\ \hline
 $1$ & $10$&$10^{7}$ &$10^{7}$ \\
 $2$ & $1$ &$10^{6}$ &$10^{6}$ \\
 $3$ & $0.01$&$10^{5}$ &$10^{5}$ \\
 $4$ & $0.01$&$10^{4}$ &$10^{4}$ \\
\hline
\multicolumn{4}{l}{Values of the couplings $\beta$ associated with a given $n$ (column (1)) below which}\\
\multicolumn{4}{l}{the  {\it thin shell} condition is no longer satisfied within the mountain (column $(2)$), and} \\
\multicolumn{4}{l}{within the ${\rm Al}$ and ${\rm Be}$ test bodies (columns $(3)$ and $(4)$) }\\
\end{tabular}
\label{TS3}
\end{center}
\end{table}

Let us summarize our results. In all the cases analyzed by us, $\eta$ decreases as $n$ increases. Furthermore, it is interesting to compare the effect of the density of the  environment on the estimated  values of $\eta$: 
for a fixed value of $n$ and $\beta$ when the density of the environment {\it increases} by several orders of magnitude,
   the value of $\eta$ {\it decreases} by  several orders of magnitude. On the other hand, the values of $\eta$ are drastically different from the ones obtained using the standard approach for $\beta < 10^{-2}$ and $1 \leq n \leq 4$, while there are differences in two or more orders of magnitude for $\beta > 10^{-2}$.

\subsubsection{The model including the metal encasing}
\label{mencasing}

\begin{figure}
{\includegraphics[scale=0.3]{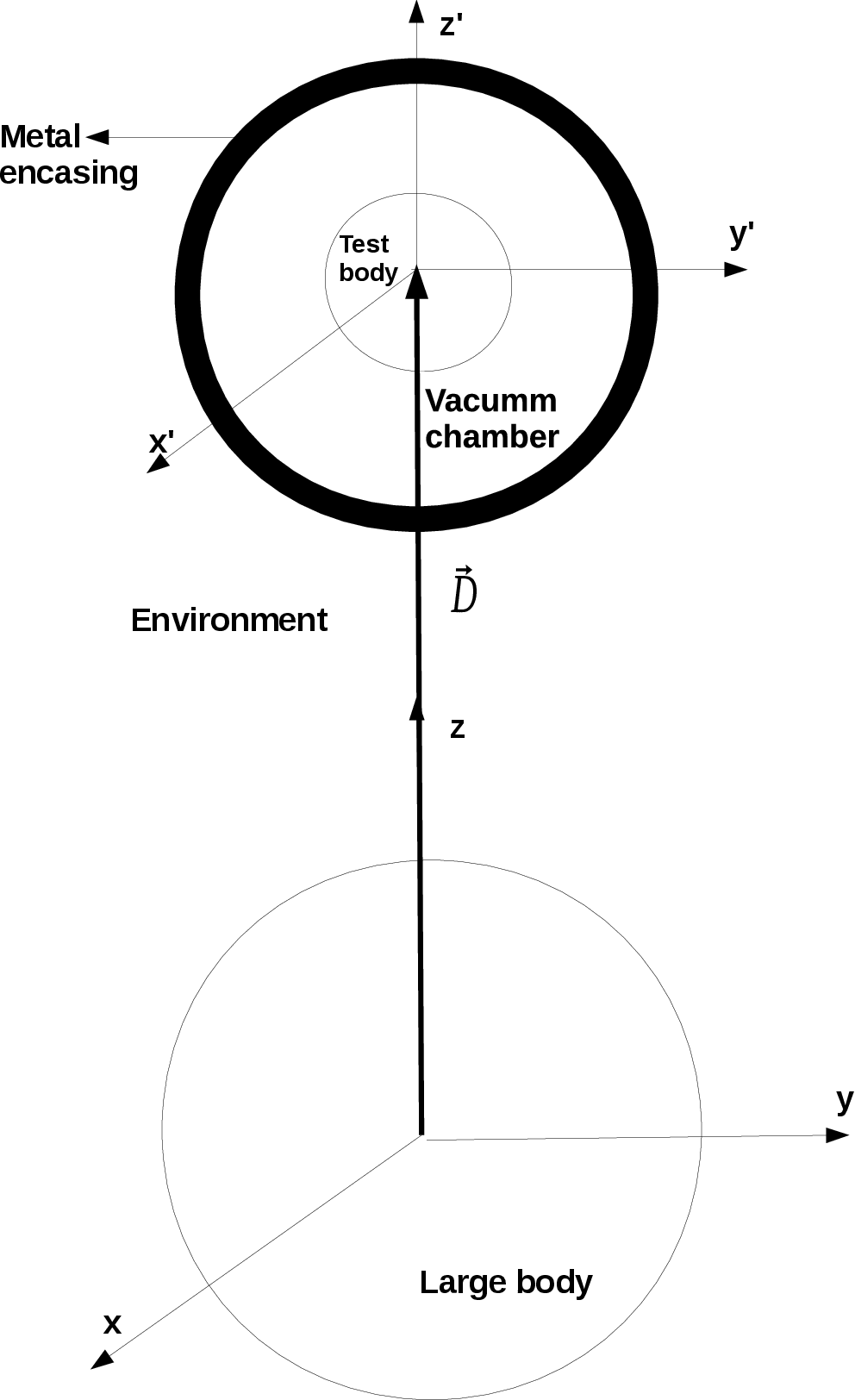}}
\caption{ The two body problem including the metal encasing of the vacuum chamber.}
\label{esquemaconshell}
\end{figure}

\begin{figure}[H]
\begin{center}
\includegraphics[width=8.0cm,height=8.0cm,angle=-90]{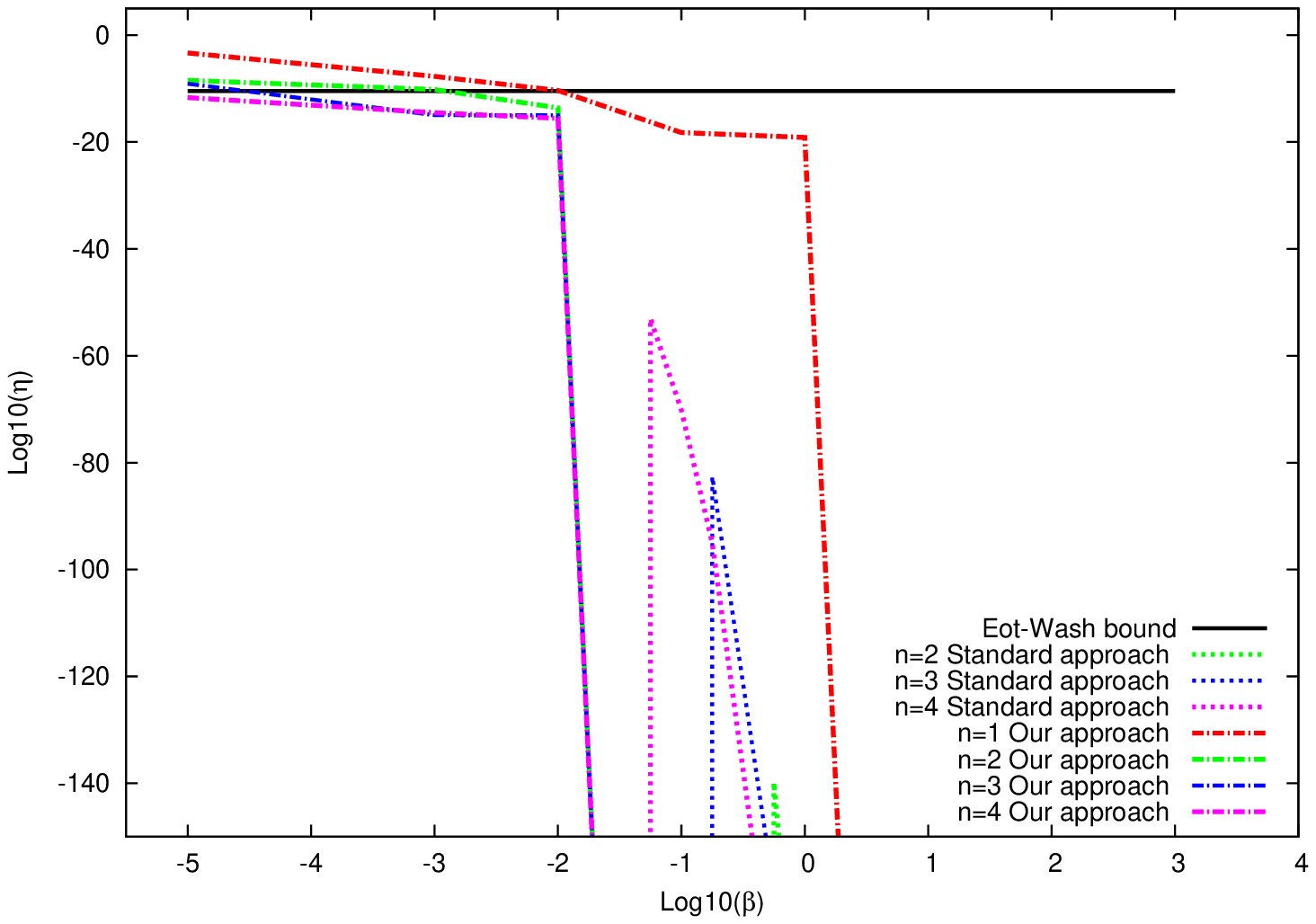}
\caption{The E\"otv\"os parameter $\eta$ (in ${\rm log_{10}}$ scale) as a function of the parameter $\beta$ (in ${\rm log_{10}}$ scale) including the metal encasing of the vacuum chamber as depicted in Fig.~\ref{esquemaconshell} for different positive values of $n$. Here we assume that both the density of the environment surrounding the Earth and inside the vacuum chamber are  $\rho_{\rm out }= 10^{-7} {\rm g \hspace{0.1cm} cm}^{-3}$. 
 Also shown are the predictions computed by Khoury \& Weltman~\cite{KW04} 
with the effect of the metal encasing modeled by multiplying their estimates of $\eta$ by the factor ${\rm sech(2 m_{\rm shell} d)}$.}
\label{competaconshell2}
\end{center}
\end{figure}

\begin{figure}[H]
\begin{center}
\subfloat[$\eta$ results with metal encasing for ${M }= 2.4$ ${\rm meV}$.]
{\includegraphics[width=8.0cm,height=8.0cm,angle=-90]{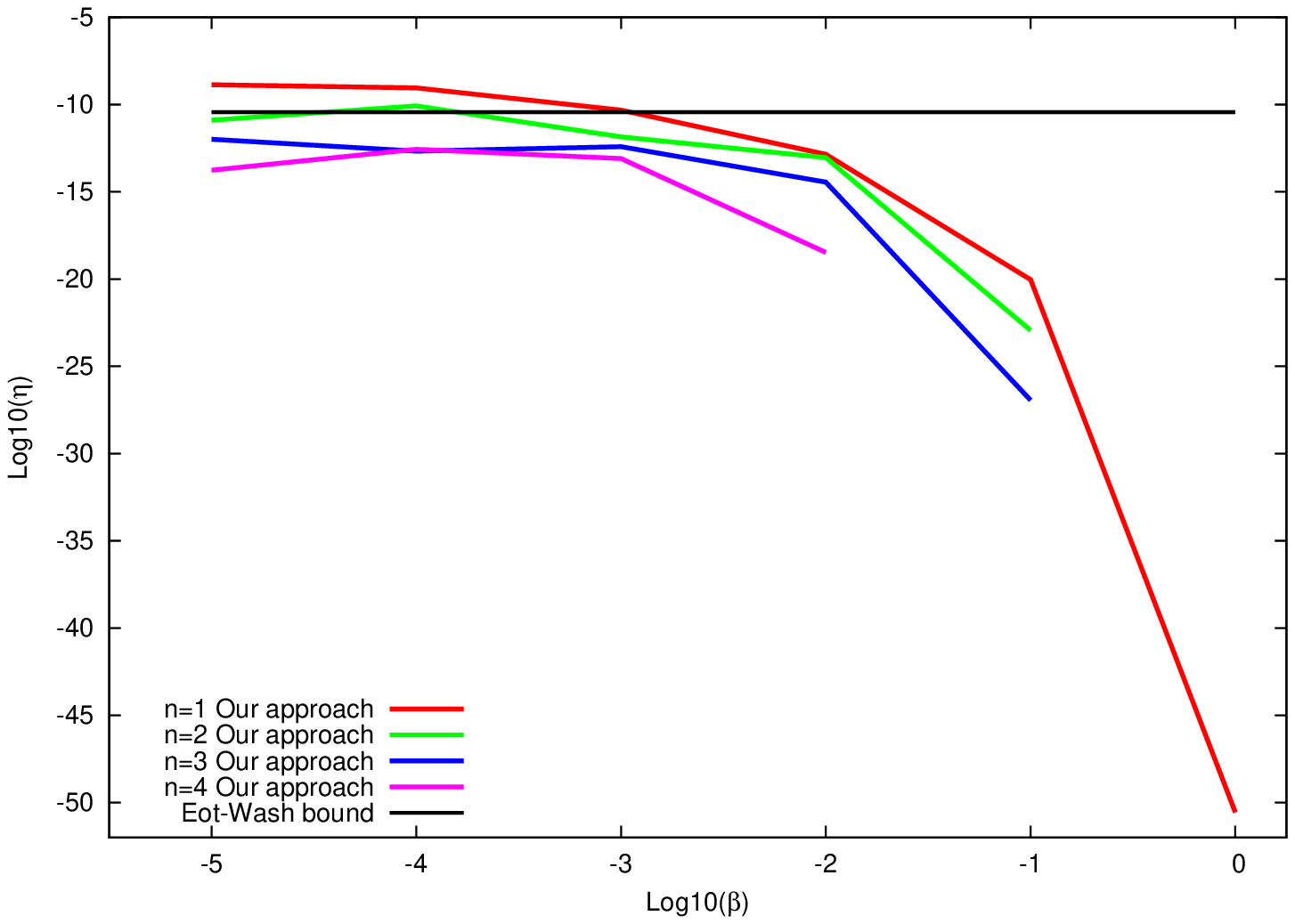}}
\subfloat[$\eta$ results with metal encasing for   ${M = 10 \hspace{0.1cm} {\rm eV}}$. ]
{\includegraphics[width=8.0cm,height=8.0cm,angle=-90]{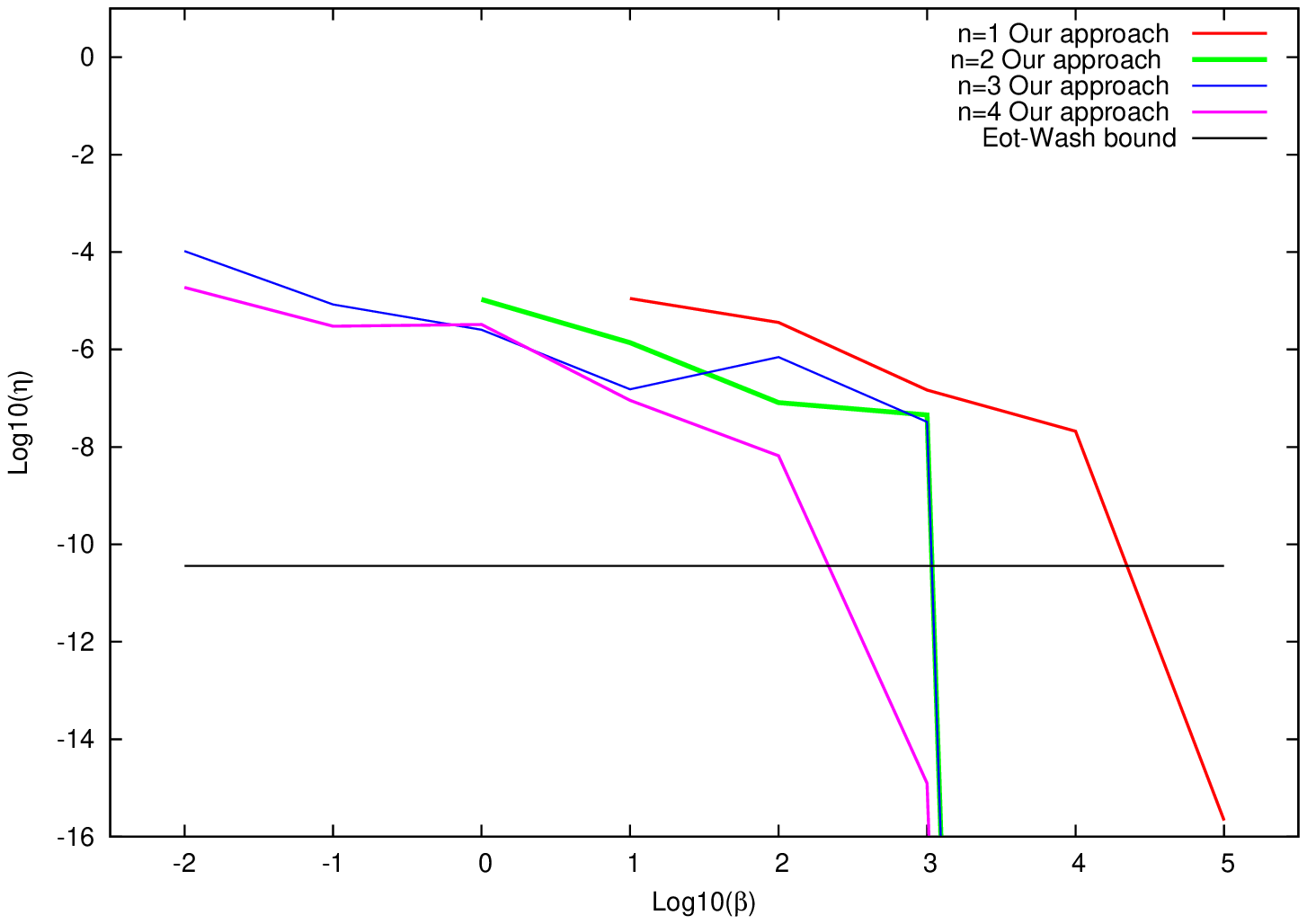}}
\caption{The E\"otv\"os parameter $\eta$ (in ${\rm log_{10}}$ scale) as a function of the parameter $\beta$ (in ${\rm log_{10}}$ scale) including the metal encasing of the vacuum chamber as shown in Fig. \ref{esquemaconshell} for different positive values of $n$. The density of the environment surrounding the Earth $\rho_{\rm out}$ is assumed to be equal the Earth's atmosphere density $\rho_{\rm out }= 10^{-3} {\rm g \hspace{0.1cm} cm}^{-3}$; the density of the environment inside the vacuum chamber is assumed to be $\rho_{\rm vac }= 10^{-7} {\rm g \hspace{0.1cm} cm}^{-3}$;  Left: ${M }= 2.4$ ${\rm meV}$ (cosmological chameleon); Right:  ${M = 10 \hspace{0.1cm} {\rm eV}}$.}
\label{resetaconshell}
\end{center}
\end{figure}

It has been pointed out \cite{Upadhye12,Hamilton15,Schlogel16,Elder16}, that the effect of the metal encasing of the vacuum chamber is determinant 
 for the outcome of experiments that includes such kind of devices. Therefore, we have  made  explicit calculations for $\eta$ by taking into 
account the metal shell surrounding the {\it test} body (see Fig. \ref{esquemaconshell}). In order to compare with other predictions found in the literature 
 we consider first the scenario where the density outside the {\it larger} body is equal to the density inside the vacuum chamber 
(see Fig.~\ref{competaconshell2}) and then analyze the  more realistic situation where the density inside the vacuum chamber is $\rho_{vac}= 10^{-7} {\rm g \hspace{0.1cm} cm}^{-3}$ and the density outside the {\it large} body is equal to the Earth's atmosphere  at  sea level $\rho_{out}= 10^{-3} {\rm g \hspace{0.1cm} cm}^{-3}$ (see Fig.\ref{resetaconshell}). The effect of the metal encasing 
in the  standard  approach has been  estimated  by multiplying the value of $\eta$  (obtained without the encasing)  by the suppressing factor found by Upadhye~\cite{Upadhye12}. That is,  $\eta_{Y} = {\rm sech}(2 m_{\rm shell} d) \eta$, where 
$\eta_Y$ corresponds to the value of $\eta$ when taking into account a Yukawa-like suppression that accounts roughly for the effect of the 
vacuum chamber walls which have the following parameters: $m_{\rm shell}^2 = \frac{n (n+1) M^{n+4} }{\varphi_{\rm shell}^{n+2}}$ and $\varphi_{\rm shell} = \frac{n M^{4+n} M_{pl}}{\beta \rho_{\rm shell}^{1/(n+1)}}$.
 In particular we   used  the values  $d= 1$ {\rm cm} and $\rho_{\rm shell}= 10 \hspace{0.1cm}\hspace{0.1cm}{\rm g}\hspace{0.1cm}{\rm cm^{-3}}$, for the thickness and density of the metal encasing, respectively. Let us remind that in the work 
of Upadhye~\cite{Upadhye12},  the {\it larger} body,  the {\it test} body, the metal encasing  around the {\it test} body, and the  environments are all modeled by {\it planar slabs}\footnote{We remind the reader again the analysis of  Ref.\cite{Upadhye12}  was performed to model the torsion pendulum rather than  the torsion balance.}. 
The  Yukawa-like suppression factor used by other authors was found by fitting the  numerical results to an analytical function. It should be noted that this is  just  an approximation to the calculation and  that a more accurate estimation would be obtained by considering the joint effects of the {\it larger} body, the {\it test} body and the metal shell. We  improved  
 this approximation by including the shell in our previous setup in a self-consistent fashion. 
In Appendix \ref{sec:TBPME} we check the results of our method when considering  the complete experimental setting (with the   full  description of  metal encasing). 
We compare these latter results with our estimates of $\eta$ without considering the metal shell but introducing  a  multiplicative  factor   corresponding to a  Yukawa suppression in the expression of the acceleration. It follows that the difference is very small (see Fig. \ref{sechvsshell}).
 
As noted before,  we   were unable to apply  the energy criterion proposed in Section \ref{sec:energy} to determine if our solutions are better than the ones proposed by the standard approaches  because in those,  the scalar field  configurations of the present experimental setup  have not been calculated explicitly.  Even though there are numerical results of the one body problem surrounded by a metal shell, these can not be  directly  applied to the experimental setup considered  here, since for the present experimental device the large body is outside the vacuum chamber and the test body is surrounded by a metal encasing. Among other  problematic issues, the experimental device analyzed here does not  have  the  spherical or planar symmetries of the situations  that have been  explicitly analyzed in the literature \cite{Upadhye12,Elder16,Schlogel16}.  On the other hand, we have shown in Section \ref{sec:energy} that for  $M=2.4$ {\rm meV} the energy criterion  for the full  fledged  problem   indicates that  the solution for the field   configuration  obtained   without  explicitly   modeling  the encasing that was obtained  in this paper must be preferred     over the  one obtained  by  standard approach, while for the case $M=10$ {\rm eV} the same happens only for a  rather  restricted range of  values of  $\beta$. Table \ref{TS3} shows, for each fixed value of $n$, the values of $\beta$ for which the energy criterion favors the  solution obtained  by our method. Hence we used the energy criterion for the two body problem  without encasing to restrict the  range of   the values  of  model's  parameters for which  our method  yields  trustable estimates  of   the  values  $\eta$   to  be  used  in comparisons with the observational bounds.  Fig. \ref{resetaconshell} left shows,  for each trustable  value of $n$ and $\beta$,  the results     for  the corresponding  values of $\eta$.  For the rest of the cases, we expect to  obtain the appropriate estimates, by considering   some type of  {\it thick shell} approximation to the effective potential, in  future works \cite{Krai17}.

\begin{figure}[H]
\begin{center}
\subfloat[$\eta$ results for $\rho_{\rm out }= 10^{-7} {\rm g \hspace{0.1cm} cm}^{-3}$.]
{\includegraphics[width=7.5cm,height=8.cm,angle=-90]{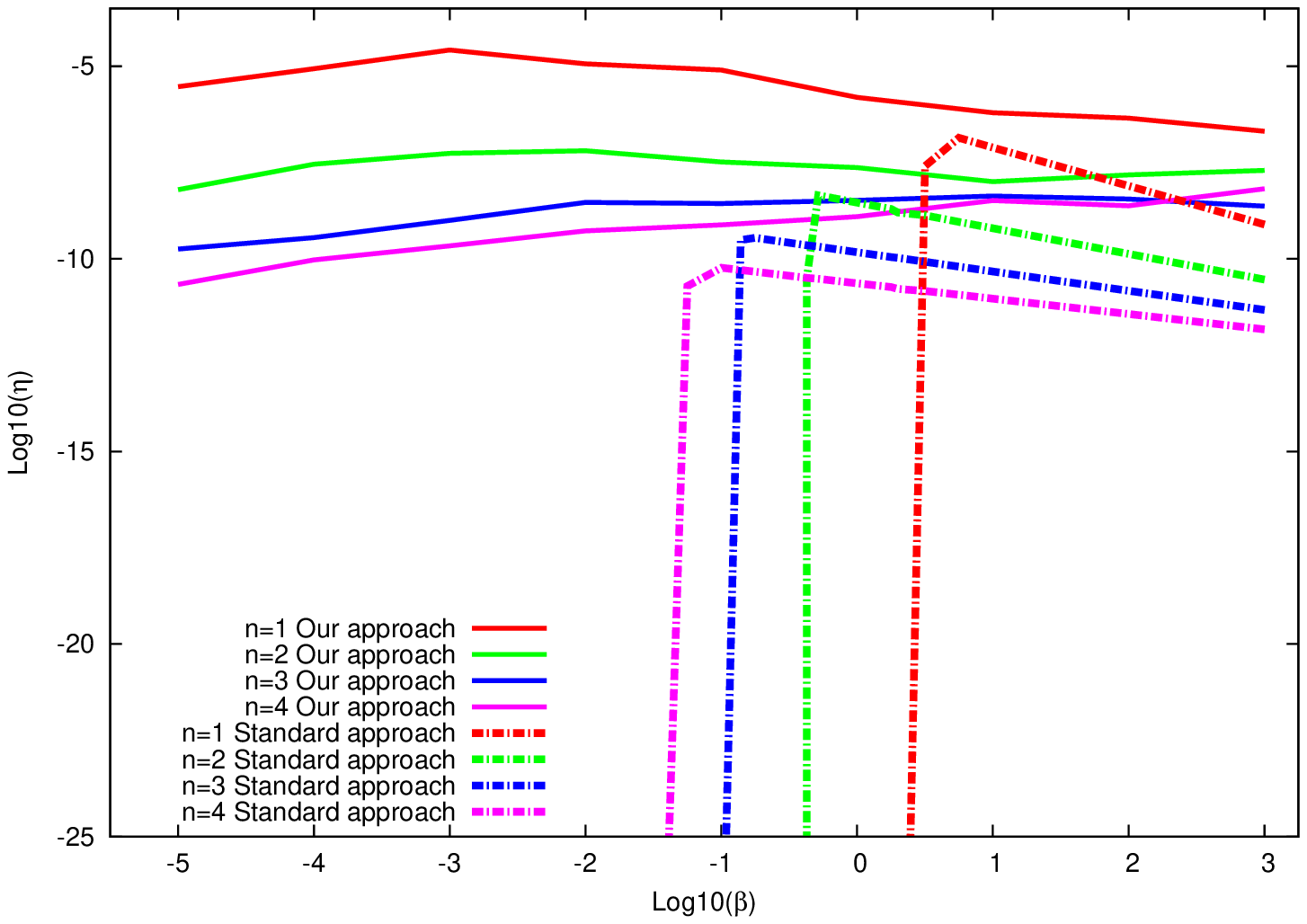}\label{resultadosM1}}
\subfloat[$\eta$ results for $\rho_{\rm out }= 10^{-3} {\rm g \hspace{0.1cm} cm}^{-3}$.]
{\includegraphics[width=7.5cm,height=8.cm,angle=-90]{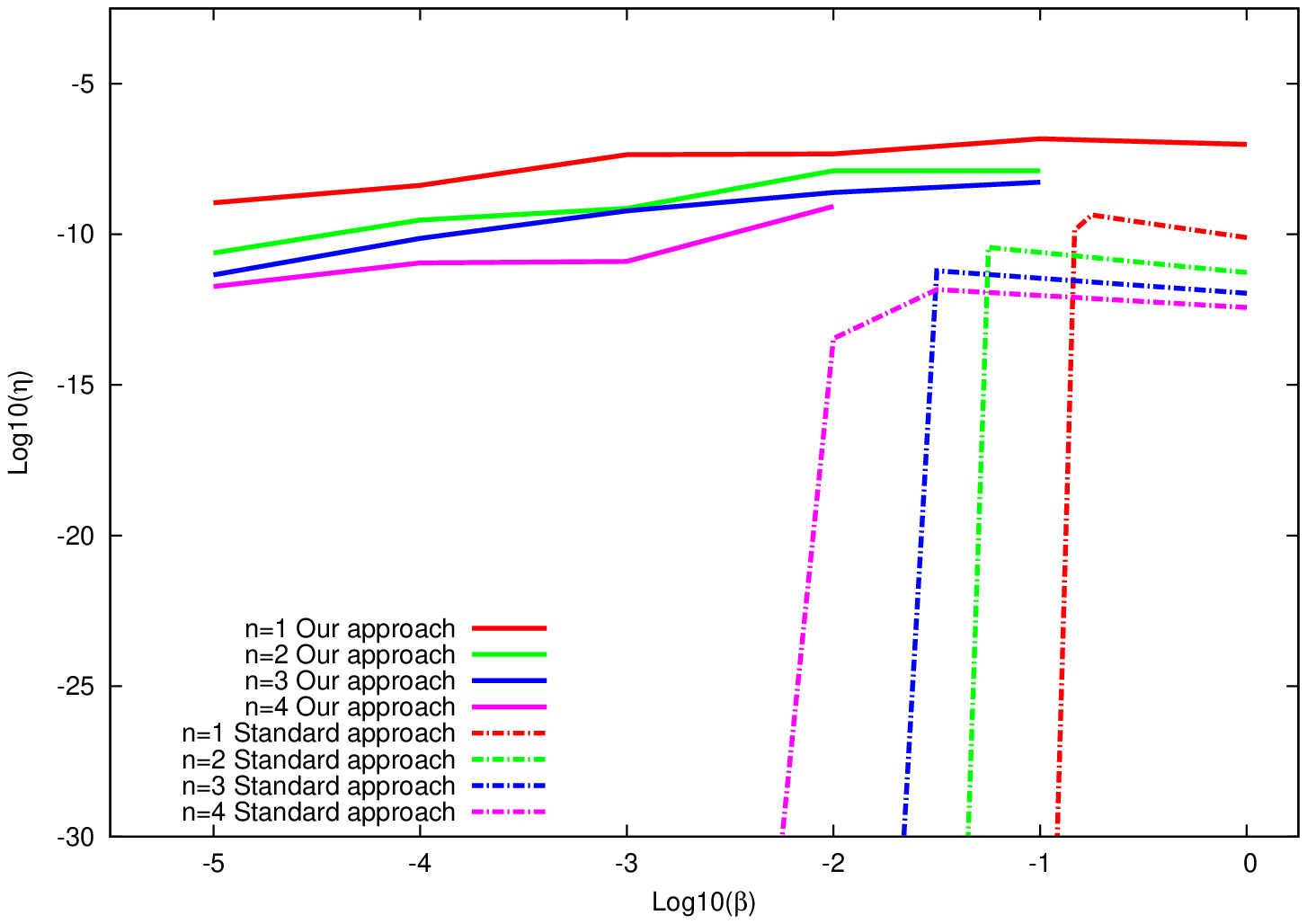}\label{resultadosM01}}
\caption{{The E\"otv\"os parameter $\eta$ (in ${\rm log_{10}}$ scale) as a function of the parameter $\beta$ (in ${\rm log_{10}}$ scale) for different positive values of $n$ and ${M }= 2.4 \times 10^{-3} {\rm eV}$. The density of the environment $\rho_{\rm out}$ is assumed to be equal to: Panel (a) the vacuum's density inside the vacuum chamber  $\rho_{\rm out }=  10^{-7} {\rm g  \hspace{0.1cm} cm}^{-3}$; Panel (b) the Earth's atmosphere density $\rho_{\rm out }= 10^{-3} {\rm g \hspace{0.1cm} cm}^{-3}$. The predictions computed by Khoury \& Weltman \cite{KW04} are also included and labeled as the {\it standard approach}}}
\label{resM}
\end{center}
\end{figure}

 We found that for some values of $n$ and $\beta$, our estimates for the  values for $\eta$ are above the experimental bounds in contrast with the results by other authors. The  main  reason for this   discrepancy  is because other approaches estimated $\eta$ without considering the   suppression   due to the metal encasing 
 {\it ab initio}, which is  already  several order of magnitude below our predictions (see Fig. \ref{resM}). In other words, 
the  results  by other authors are based on  the  suppression of an  initially underestimated ``bare"  value  of  $\eta$.

On the other hand, it follows from the comparison of Figs. \ref{resM} and  \ref{resetaconshell}  that for $M= 2.4$ ${\rm meV}$ and  $\beta > 10^{-3}$ the intensity of the suppression is so important that  it predicts  no violation of the experimental bounds for $\eta$, while for $M = 10 \hspace{0.1cm} {\rm eV}$ the  suppression effect becomes  relevant   only for  $\beta>10$. Nevertheless, the method developed in this paper allows us  to establish the following constraints : i) for $M=2.4$ $ {\rm meV}$ 
and $n=1$ the range $10^{-5} < \beta < 10^{-2}$ is excluded; for the same $M$ and $n=2$ the range  $10^{-5}<\beta < 10^{-3}$ is excluded too
;ii) for $M = 10 \hspace{0.1cm} {\rm eV}$ and $n=1$ the range $10 < \beta < 10^4$ is excluded; for the same $M$ and $n=2$ the ranges 
 $1 < \beta < 10^3$ are excluded; for the same $M$ again and $n=3$ the range $10^{-2} < \beta < 10^3$ is excluded and finally for $n=4$ the range $10^{-2} < \beta < 10^2$ is   excluded  as well. Some of these values were already excluded by other authors\cite{Burrage16,Upadhye12}, but other values were not.  Furthermore, Fig. \ref{resetaconshell} shows that as $M$ increases,  the experimental bounds on  $\eta$ can rule out larger values for $n$ and $\beta$. Finally,  it is worth stressing that the above estimations are  based on   modeling the  vacuum  chamber as a {\it spherical} metal encasing and  the   detailed   consideration of  any asymmetry of the experimental device might  change the results in a important  manner, in one direction or the other. That is, the details of  the geometrical arrangement  might    led  to  an increase or a decrease in the expected   value of $\eta$.

\subsection{Lunar Laser Ranging Experiments}
\label{sec:LLR}

Another  setting  that  serves  to  test the WEP is the LLR which, since 1969, has provided high-precision measurements of the Earth-Moon distance \cite{LLR}.  Reflectors placed on the lunar surface allow the measurement of  the round-trip travel time of short pulses of laser light,  and thus set limits on the differential acceleration of the Earth-Moon system in free fall towards the Sun.

In previous studies~\cite{KW04,TT08,GMM10}, the chameleon effect has been analyzed using the free-fall acceleration of the Moon and the Earth towards the Sun by
assuming  a  constant  solar density. We will base our  analysis on  the same approximation for simplicity in order to apply the methods developed in
the preceding sections  and    also  because as  we  will  next argue  the  approximation should be  a very good  one. Although the   solar  density is  far  from  being constant within the Sun's interior (varying over  several orders of magnitude from the center to the
Sun's surface), one  does not expect that   the results  of the analysis are altered significatively for the situation  at hand.
The  fact is that,  as  shown in those works,  even  when one  describes the Sun in   terms of a  collection of   infinitesimal volumes, one  finds  that   contributions  from the    Sun's  deep interior   become   exponentially suppressed due to the large mass of the chameleon (given that as the density increases, so does the effective mass of the chameleon,
and  thus the effective range of the field decreases) leading to a situation where   the field   in the  Sun's exterior  is in fact  determined  to    a  very large extent   by  a  {\it thin shell}  of matter near the  Sun's  surface. As  the chameleon field in regions  very  close to the  Sun's surface  do not suffer from such a large  suppression,  one  finds that  when focusing  on  the external field,    the approximation where  the field is taken to be as generated almost entirely by this {\it thin shell}
(while the effect of the rest of the Sun is  sub-dominant~\cite{KW04}) becomes a very good one.   In  fact  the  actual numerical   calculations   show that  for the vast majority of values  for $n$ and $\beta$ (including the ones that we  will be interested on), the solar chameleon field  at   very far  locations   is  well estimated  by that  approximation. On the other hand,  when  considering the  force on the small object while  looking for violations of the UFF, the details of the scalar-field profile in the object's interior are important, and  in particular,
the  charged-conductors analogy indicates that  it is essential to  take into  account the  anisotropy of the chameleon field within  the small body's
interior, that results from the presence of the {\it large} body outside.
This should be  clear when  noting that  it is just only  due  to  that anisotropy that a force  on the small body,  directed towards  the {\it large} body,  results from the scalar field  interaction with the small body.
 Finally,  we stress  that   given the   very  small  ratio  between the   Sun's size  and  the  Sun-Earth or Sun-Moon   distances, the non-linearities  that  are important   in the  Sun's  interior  can   be expected to play no significant  role  in determining, together with the small body (Earth/ Moon),  the  behaviour of the scalar field in the neighbourhood  of  such small body.  In view of all the above arguments, it seems reasonable to take the Sun's density as  its average density (as is usually done  in  such analysis
( \cite{KW04,TT08,GMM10}). In a near future we hope to drop this assumption and study the chameleon
two body problem,  in  even more detail,  by developing appropriate analytical and numerical methods. Only such  detailed study  can determine   in a completely  unambiguous manner
the extent to which the assumption about the homogeneity of the Sun's density is in fact  a good approximation or a bad one.

\begin{figure}
{\includegraphics[width=8.5cm,height=8.5cm,angle=-90]{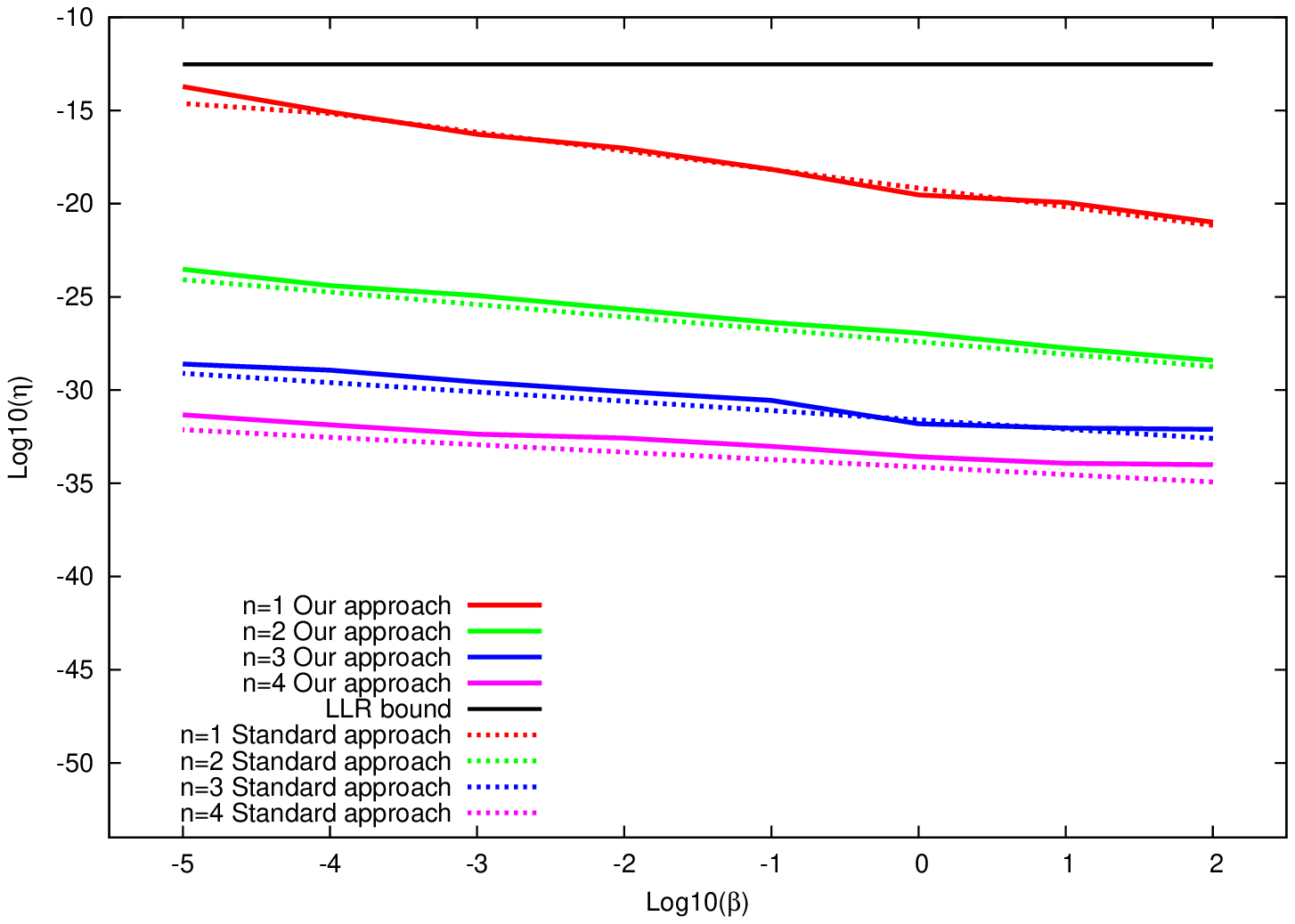}}
\caption{The E\"otv\"os parameter $\eta$ (in ${\rm log_{10}}$ scale) for the LLR experiment as a function of the parameter $\beta$ (in ${\rm log_{10}}$ scale) for different positive values of $n$. All bodies are surrounded by the interstellar medium. ${M }= 2.4 \hspace{0.1cm} {\rm meV}$ (cosmological chameleon). In all cases we compare the results of the calculations performed in this paper with the predictions computed by Khoury \& Weltman \cite{KW04}.}
\label{resLLR}
\end{figure}

\begin{figure}
\subfloat[$n=1$] 
{\includegraphics[width=8.5cm,height=8.5cm,angle=-90]{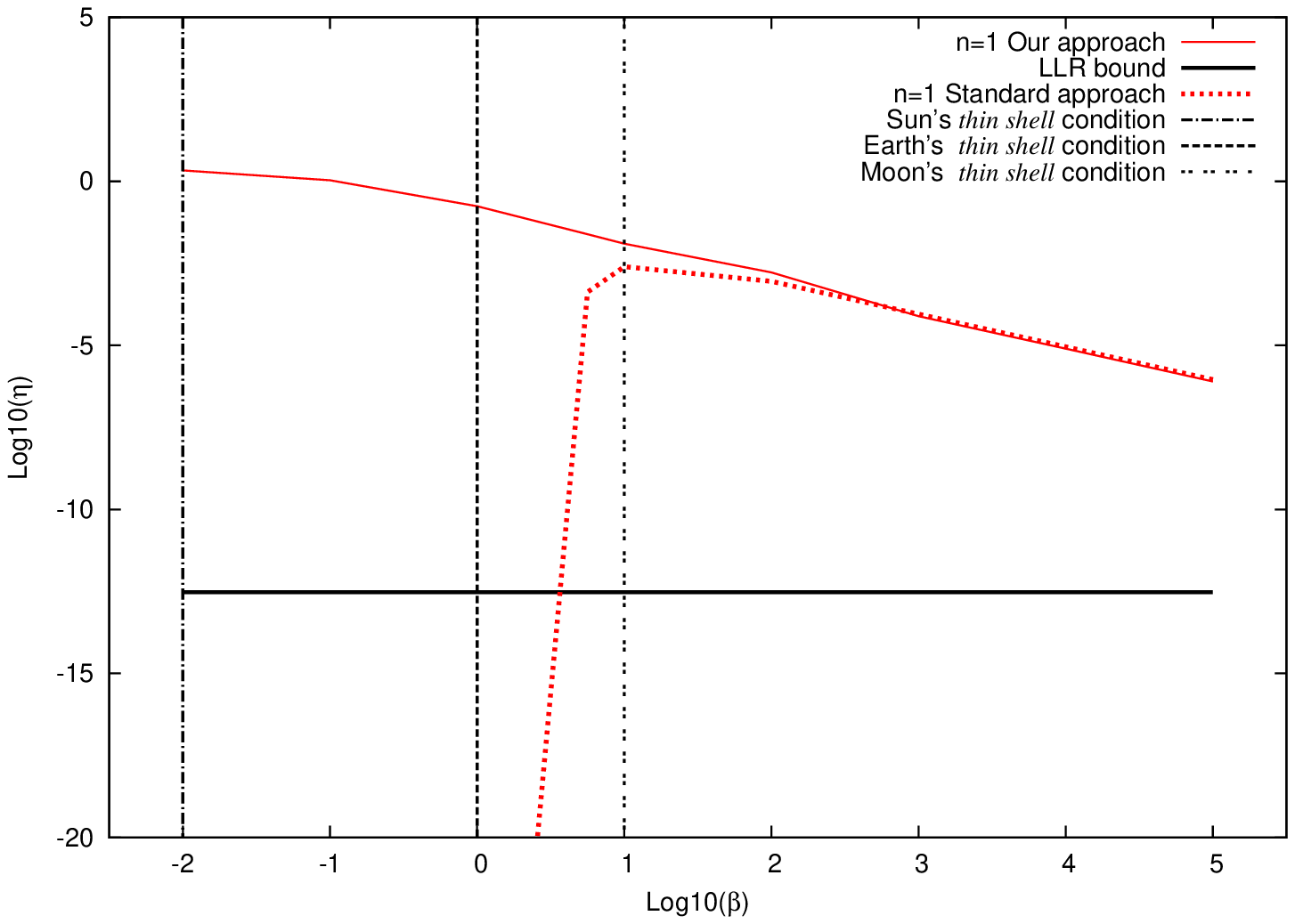}}
\subfloat[$n=2$]
{\includegraphics[width=8.5cm,height=8.5cm,angle=-90]{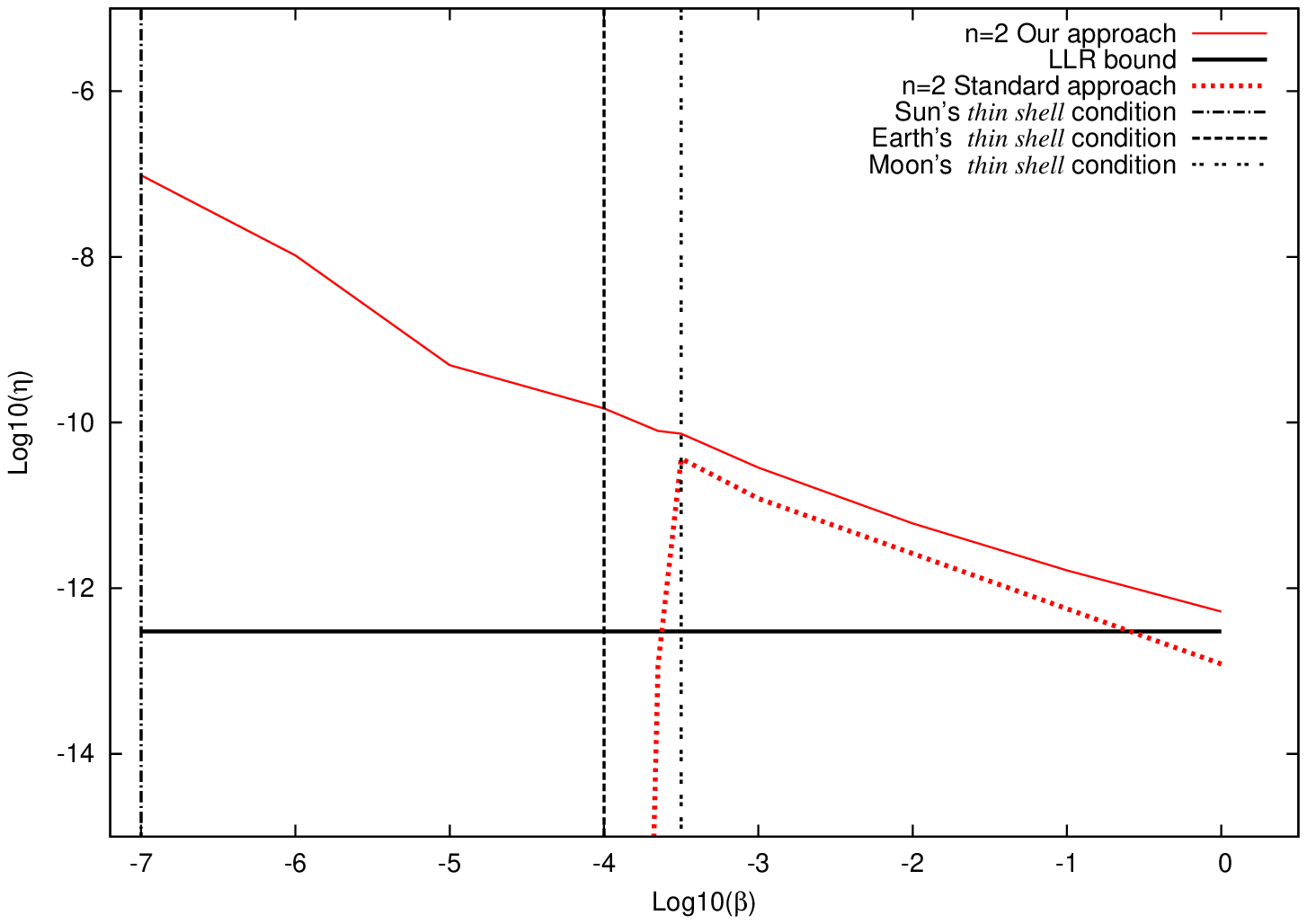}}
\caption{The E\"otv\"os parameter $\eta$ (in ${\rm log_{10}}$ scale) for the LLR experiment as a function of the parameter $\beta$ (in ${\rm log_{10}}$ scale) for different positive values of $n$. All bodies are surrounded by the interstellar medium. ${M = 10 \hspace{0.1cm} {\rm eV}}$. The vertical lines show the values of $\beta$ below which the {\it thin shell condition} is no longer satisfied for the Earth, Moon and Sun. For all values computed in this plot, the energy criterion developed in Section \ref{sec:energy} indicates that our approximation to the effective potential is better than the one used by the standard approach}. 
\label{resLLR3}
\end{figure}

 Figures \ref{resLLR}  and \ref{resLLR3} show the predictions for $\eta$ based on the LLR experiment. 
 For this scenario we  are considering  the Sun as the {\it large} body ($\rho_{SUN}=1.43 \hspace{0.1cm}  {\rm g  \hspace{0.1cm} cm}^{-3}$, $R_{SUN}= 7 \times 10^8$ {\rm m}) and the Earth or the Moon as {\it test} bodies with
the following  properties, respectively, $\rho_{E}=5.5 \hspace{0.1cm} {\rm g  \hspace{0.1cm} cm}^{-3}$, $R_E=6.371 \times 10^6$ {\rm m},  $\rho_{M}=3.34  \hspace{0.1cm}{\rm g  \hspace{0.1cm} cm}^{-3}$, $R_M=1.737 \times 10^6$ {\rm m}.   Furthermore, we assume that  all bodies are surrounded by an environment of constant density equal to the density of the interstellar medium. It is interesting to note that for $M=2.4$  \hspace{0.1cm} {\rm meV} (see Figure~\ref{resLLR}) 
the difference between the  resulting  predictions  from our model and those  obtained  by other authors is  at most one or two orders of magnitude, while for the torsion balance experiments, the  differences were more important, particularly  for  the cases  in which  the {\it test} body is unscreened.   On the other hand,   when considering 
$M = 2.4 \hspace{0.1cm} {\rm meV}$, the predictions indicate   no  conflict   with the  existing  LLR tests  for  violation of the WEP for any of the values of $n$ and $\beta$ considered here.  Again, as in the torsion balance experiment, there is an enhancement of $\eta$ for increasing value of $M$. In fact,  Fig. \ref{resLLR3} shows that for $M = 10 \hspace{0.1cm} {\rm eV}$,  the bounds resulting from LLR can  be  used  to discard the  values of $\beta$ in the range 
$10^{-2}<\beta<10^{5}$ when $n=1$ and $10^{-7}<\beta<1$ when $n=2$. Some of these values were already excluded by other authors\cite{Burrage16,Upadhye12}, while others  are ruled out,   for the first time,   with the present analysis.
We   remind the reader, that we   have used the energy criterion discussed in Section \ref{sec:energy}  in order  to establish when we can trust  the results   obtained  by the method of this paper (which relies on the {\it quadratic} approximation to the effective potential)  over those of the traditional  approach  for the estimates   of $\eta$ in various specific  situations.  For the case $M=2.4$ {\rm meV} we have  found  that the energy functional is always smaller for the field configuration  obtained  with our approach  than for  the one obtained in the standard approach. For the case $M=10$ {\rm eV} the energy criterion favors our results only for the values of $\beta$ where the {\it thin shell} condition holds for the Sun. Accordingly, the results shown in  Fig. \ref{resLLR3} are restricted to the  corresponding  values of $n$ and $\beta$.

\begin{figure}
{\includegraphics[width=8.5cm,height=8.5cm,angle=-90]{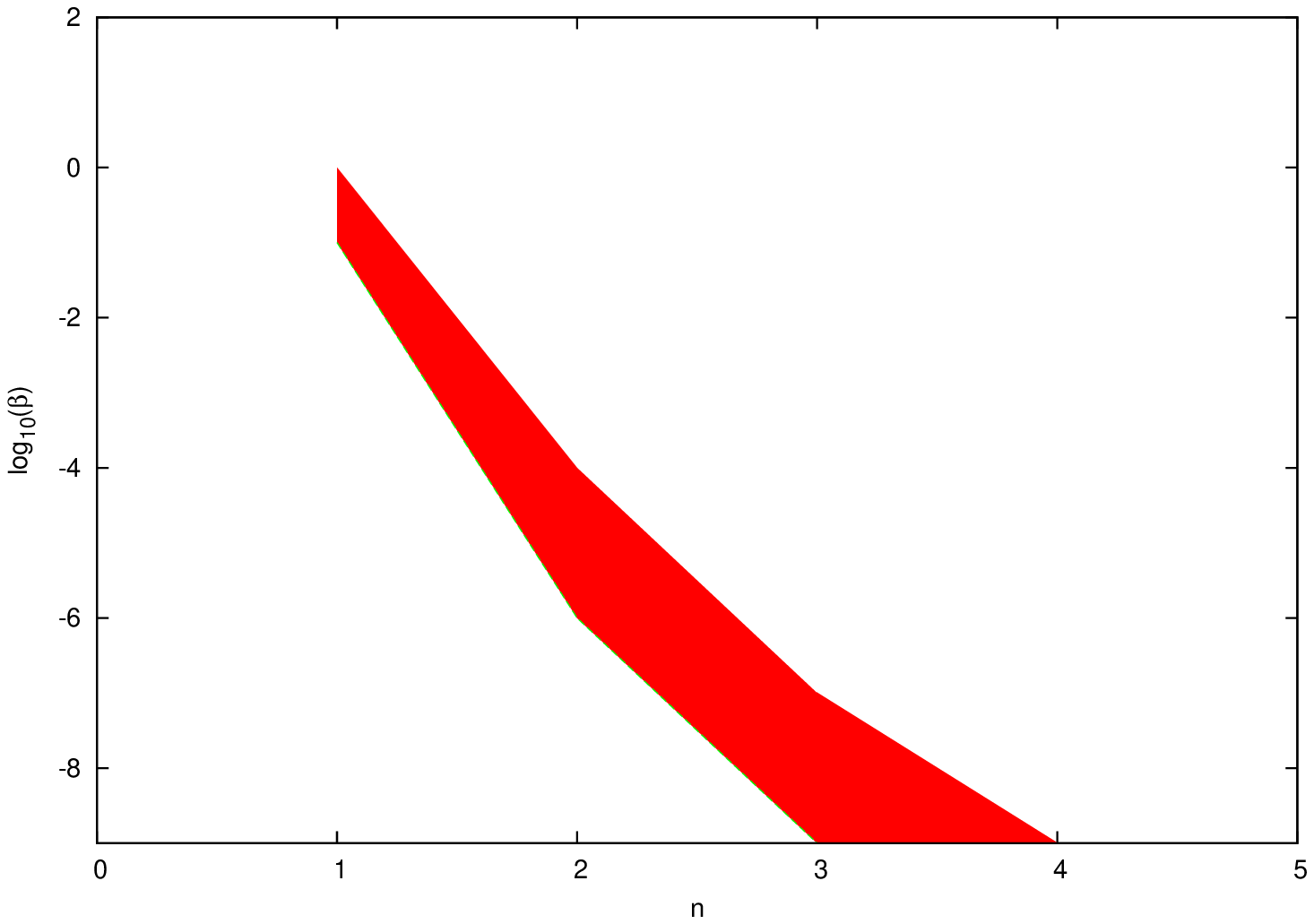}}
\caption{The shaded region in the figure shows the values in the $n - \beta$ plane where our approach has significant differences with the standard approach in the value of $\eta$ for the LLR experiment with ${M = 10 \hspace{0.1cm} {\rm eV}}$.}
\label{nbetaex}
\end{figure}

On the other hand,  the discrepancy in the value of $\eta$ between our approach and the standard one is negligibly small  for $M=2.4$ ${\rm meV}$, while there are important differences for the case $M=10$ {\rm eV} depending on the value of $n$ and $\beta$.  Fig. \ref{nbetaex} shows  the region in the plane $n - \beta$ where the differences are significant. 

\bigskip
 In summary, in the last two  subsections we have analyzed the predictions of the effective violation of the WEP that  would result from the chameleon mediated forces  in the model by considering different possibilities for the outside medium and different values for $M$,  by using Eqs.~(\ref{fuerzaCf2}) and (\ref{fuerzaCf31f}), and compared our results with the experimental bounds  obtained from the  E\"ot-Wash experiment. 
We realized that  when adding the effect of the metal encasing,  a Yukawa-like suppression leads to  a  significant  change in the    estimates. For some  values of the parameters ($\beta > 10^{-3}$) the    corrected   predictions  are within  the  bounds imposed   by the  actual E\"ot-Wash experiment.  
These results suggest that it would be very interesting to carry out   experiments of this   type   with no  suppression   due to  encasing,   as they  would   probably lead  to significant   improvement of   the  bounds on the chameleon  model's  parameters.  
In addition, we considered the LLR experiment, where there is no such suppression  and found  that in this setting  there would  have been  an observed   violation of the experimental bounds in the range  $10^{-2}<\beta<10^5$ when $n=1$, and $10^{-7}<\beta < 1$ when $n=2$, for a fixed $M= 10 \,\ {\rm eV}$.

\subsection{Comparing bounds from other experiments}

\begin{figure}
{\includegraphics[scale=0.2]{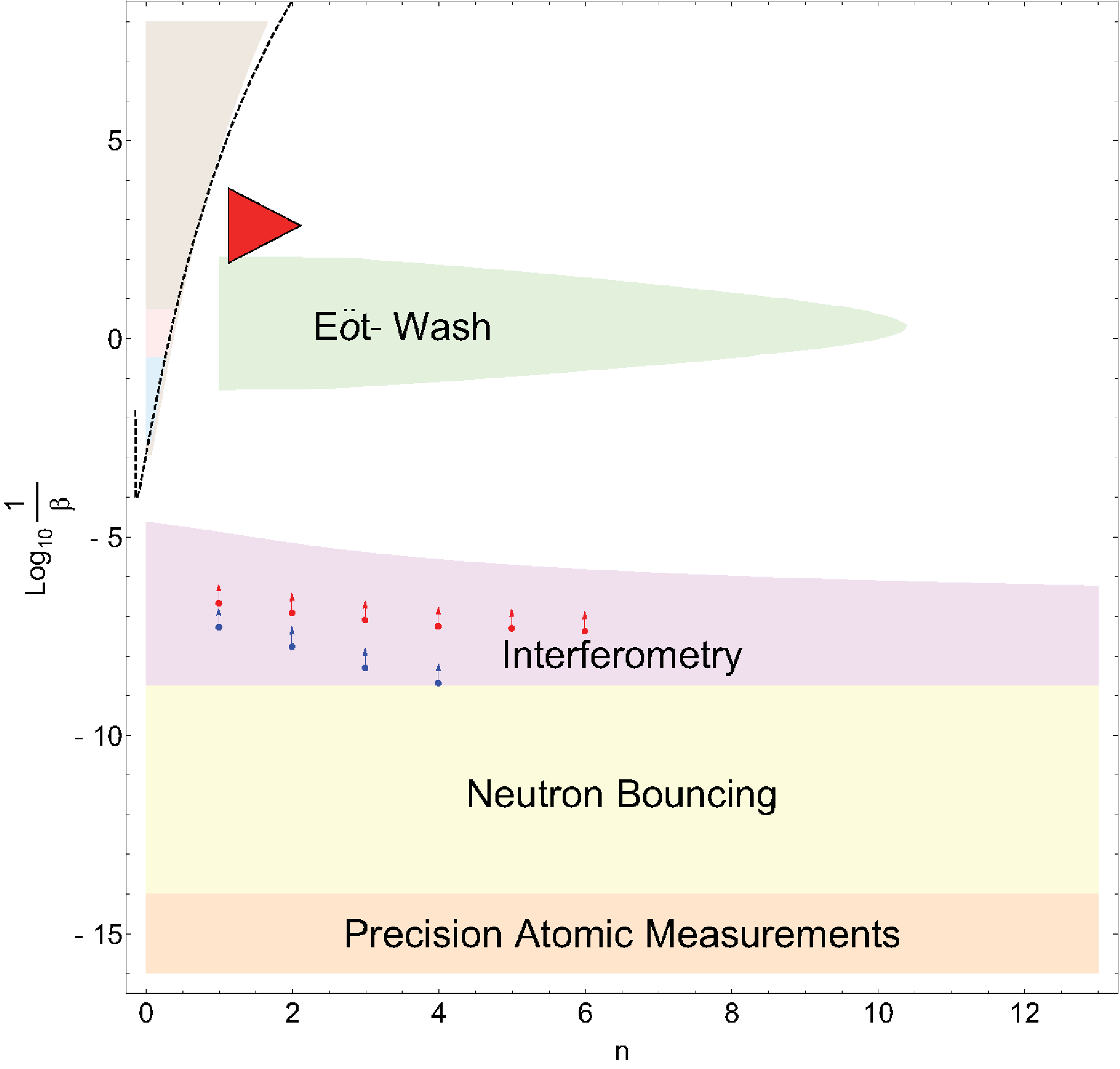}}
\caption{Bounds in the $n-\beta$ plane obtained in this paper (in red) using the torsion balance experiment with combined constraints compiled by Sakstein \& Burrage. 
Here $M=2.4 \hspace{0.1cm} {\rm meV}$.}
\label{compnbeta}
\end{figure}

Recently, Burrage \& Sakstein \cite{Burrage16} reanalyzed the constraints imposed on the chameleon models using different astrophysical
and laboratory data, and found the bounds on the parameter space associated with the plane $n-\beta$ (fixing $M=2.4 \hspace{0.1cm} {\rm meV}$) and $M-\beta$ (fixing $n=1$). Using our analysis we also obtained complementary bounds on those planes but by considering data related 
with the E\"otv\"os type of experiments using a torsion balance and the LLR experiments. We find that the region excluded by our calculations in the $M-\beta$ plane was in fact already excluded by other experiments \cite{Vikram14,Terukina14,Wilcox15,Jain13,Upadhye12,Hamilton15,Brax07}. On the other hand, we can exclude a  new region 
in the $n-\beta$ plane by fixing $M$ to the cosmological value of the chameleon. Fig. \ref{compnbeta} shows the bounds in the $n-\beta$ plane obtained in this paper using the torsion balance experiment (new constraints using LLR were also obtained but only for $n=2$) together with the combined constraints presented in Ref.~\cite{Burrage16} \footnote{We thank J.Sakstein and C. Burrage for providing a modified plot to include our bounds.}. 
 We would like to stress that in the present analysis  we   have  relied on   the 
{\it quadratic} approximation to the effective potential, even for cases when the {\it thin shell} condition  is not satisfied by  the {\it test} body. The reason for this is that in   many such cases  the energy criterion indicates that  our solution is a  better approximation to the  scalar  field  solution   for the  specific  problem than the one proposed by the  standard approach in the case where no encasing of the vacuum chamber is considered. The lack  of  explicit calculations  for the  field  configuration  resulting  from  the standard approach is the reason why we could not apply the energy criterion discussed in Section \ref{sec:energy} to the present  situation. In a forthcoming paper\cite{Krai17}, 
 we expect to explore the situation  using some  adaptive approximation  approximation  to the effective potential  within the {\it test} body   for the relevant values of the   model's parameters. We expect  relatively  small  variation for the resulting estimates  of $\eta$ which   will  not  change the   overall  picture in  a significant way.
 We recall  at this point  that the most stringent bounds in the $n-\beta$ plane are imposed by the following experiments: i) The torsion pendulum of the E\"ot-Wash experiment which has been used to probe fifth forces on sub-millimeter scales~\cite{Upadhye12}, ii) Atomic interferometry~\cite{Hamilton15}, iii) Ultra cold neutron bouncing \cite{Jenke15} and iv) Atomic Precision Tests \cite{BB11}.

\section{Summary  and Conclusions}
\label{sec:conclusions}

In this paper  we  examined,   in  a   critical   manner,  the degree   to which the  chameleon  models evade   existing  bounds on  the  violation  of UFF   as a result of the  famous  {\it thin  shell} effects.   We  were  motivated by the fact that  any  {\it extended} body that  can modify  the  chameleon field  to  such a large  extent  so as to  generate  that kind of nontrivial behaviour of the field   near  their surface, cannot be  considered  as a  ``bona-fide"  test particle.  We have  addressed the issue  with   a new method  that allows the  precise   calculation  of  the full fledged two body system and its  interaction with the chameleon field.
 This  has  allowed  us to  evaluate  the chameleon mediated force between these bodies  and to  determine  the differential acceleration on two {\it test} bodies  towards the  larger  one resulting  from  the chameleon   mediated interaction. As a result, we compared our estimates with the outcome of the WEP experiments on Earth and put constraints  on  the parameter space  of the model.  

The approach that we have followed to compute the force takes into account  the contribution of the two bodies (the larger and smaller ones) 
 to the  static field  configuration  everywhere in space. In order to simplify the calculations we have considered   a scenario where the coupling between the chameleon field and the matter fields is universal.

We have restricted the conclusions   extracted  from  present analysis to those cases where the effective potential can be   well approximated by a  quadratic  expansion around its minimum. However, we  have found that the  condition that   is  taken to indicate  that this approximation is reliable  (namely the {\it thin shell} condition) is not always satisfied for the  values of the model's parameters   for  the  experimental situations considered here. Therefore, we have developed a  criterion,  based on an energy minimization test   in order  to determine in which cases the approximations  we have obtained  for the  profile of the   chameleon  field  in  the two body problem 
is better than the  approximation   for the  scalar field  obtained by the standard approach. Consequently, we have   decided to  only trust   the results of our  method when the energy criterion indicates  it  is   appropriate to do so. 
 Results from Section III show that the method proposed in Sect.II is not appropriate when the large body is {\it not} in 
the thin shell regime. However, when both bodies    are in such  regime or when just  the test body fails  to be in  that regime,  our approach  continues  to  provide  a  better   approximation  than  the one obtained   with  the standard approach.
 In a forthcoming paper\cite{Krai17}, we expect to take one step forward and  incorporate an 
 adaptive approximation to the effective potential when appropriate,  and thus to explore  a  wider range of  values  of the free parameters of the chameleon model. In such way we  expect to set  constrains  for  values of $M$, $n$ and $\beta$  that were  excluded  from   the conclusions  obtained  in this work as a  result  of the  indications coming  from  the energy criterion.  

Furthermore,  apart from the  standard  approximation  made  in  characterizing the effective potential  for  the chameleon  field as  a  quadratic one  around its  minimum, the   relevant approximation that has been performed  is  to cutoff  the series  expansions of the development in spherical harmonics of the chameleon field.  Other approximations were considered  in  order to model the hill,  and the encasing of the vacuum chamber for E\"ot-Wash experiment and the   description of the outside media for each experiment (see Sect. \ref{sec:results}).

In the selection  of   the  adequate  cutoff  in the series, we have followed the approach proposed in Refs.~\cite{Gume1} and~\cite{Gume2}.  The  approach then offers  in principle  a path  for    calculations  to any  desired order. We have found that the force on the {\it test} body  {\it does}  depend on the {\it test} body's  size and composition, making the  model   susceptible to  receive  experimental constraints arising from tests of the WEP,  and thus, our conclusions  differ drastically from  studies which  claim that the chameleon force  is strongly suppressed  whenever the {\it thin shell} regime ensues.  The   specific  origin of  the difference between our calculations and the ones that predict no  effective violation of WEP in chameleon models, is that  in  those works~\cite{KW04,Brax04,MS07}, it is  assumed  that the force mediated by  the chameleon field would be automatically suppressed when  the bodies involved are in the {\it thin shell} regime, to  such an extent as to  make  the chameleon force experimentally irrelevant. By performing a  careful  and  detailed calculation we have shown that this is not always the case.

Regarding the  situation where the coupling between the chameleon and the matter fields is universal,  we must remind the reader that,  as discussed in the Introduction,  it  is  only when  test objects can be    considered  as  true  {\it point} particles which  follow geodesics (as opposed to {\it test} extended bodies)   that one  might  conclude  that  the model  predicts absolutely  no violation of the WEP.  The point is, however, that    such   point {\it test} particles  cannot  be  the ones that are associated with   the {\it thin  shell}  effect, as such effect arises from  a   substantial influence of the  body  on the  scalar field.   The  analysis  of the level  of   validity  of a direct  extension of that notion to  a realistic (and necessarily   extended) body is highly non trivial. The calculations performed in this paper show that this issue is  a rather delicate  one and  in particular that the conclusions  that emerge from it  can be  particularly  relevant for the case of  chameleon models. 

Furthermore, we have  quantitatively  estimated the differential acceleration  for {\it test} bodies  made of Be and Al for  one of the most precise experiments yielding relevant  bounds considering two possibilities for the outside medium, and  found that  the   deviations from  UFF  might  in principle   be    rather substantial. However,  as it  was  pointed to us  by   a  few colleagues  and  by a referee, the actual  E\"{o}t-Wash  experiment  makes use of a metal encasing that   can    substantially  reduce the    prediction for the observable   effect. Consequently, we have  carried out an additional analysis  in which   a  model for such encasing is  included  leading to a solid  estimate  for  the  predicted  values  of  the UFF violations  and  used  those  bounds to place further constraints  on the model.  It was important  to compare the  constraints  emerging for  our analysis  with those obtained by  other experiments that  have set strong bounds on the parameters of the chameleon model. One of the most important results of this paper is that, using experimental bounds from the torsion balance experiment, we were  able to put additional constraints on the  parameters  in the $n-\beta$ and $M-\beta$ planes beyond those placed by other experiments such as i) torsion pendulum, ii) atomic interferometry, iii) cold neutron bouncing and iv) precision atomic measurements.  Thus, our analysis concludes that  i) 
for $M=2.4 \hspace{.1cm} {\rm meV}$ and $n=1$ the new excluded values are  $10^{-5} < \beta < 10^{-2}$, and when  $n=2$  the range $10^{-5}<\beta < 10^{-3}$ is excluded 
as well; ii) for $M = 10 \hspace{0.1cm} {\rm eV}$ and $n=2$ the new excluded values are  $1 < \beta < 10^3$; when  $n=3$ the range $10^{-2} < \beta < 10^3$ is now excluded and finally for $n=4$ the range $10^{-2} < \beta < 10^2$ is now ruled out. 

 Nonetheless, in contrast to  previous works \cite{KW04,Brax04,MS07,Upadhye12,Hui2009}  which   led to the  conclusion  that the  {\it thin shell}  effect in high density environments is  precisely what  prevents the chameleon model to be in conflict with the UFF, our  work  indicates   that the absence of   conflict  between  predictions of  violations of the WEP and the E\"{o}tv\"{o}s-like experiments arises   essentially from the Yukawa-like suppression due to the vacuum's chamber metal shell.  This   observation should  serve as  strong motivation for   our experimental  colleagues  to devise   suitable   laboratory   arrangements. For instance, the construction of experimental setups without the metal encasing or other arrangements that avoid strong Yukawa like suppressions can open the path for obtaining constraints on this class of models that are more stringed than the existing ones.  In summary, experiments where neither  the  {\it large } source  nor  the   {\it test} bodies are   placed  within any   kind of  encasing seem, in principle, to  be   better alternatives  to be used  in  tests of violations of the WEP than current versions of the E\"{o}t-Wash type experiments.  An example of such a set-up is the Microscope Satellite \cite{Microscope} which was recently lunched. The estimated precision  of $\eta$ is improved by two orders of magnitude with respect to  Earth-based experiments.  

We have also considered the LLR experiment as a scenario where the encasing around the {\it test} bodies is absent, and found a new exclusion zone for  
the chameleon coupling, namely,   $10^{-7}<\beta < 1$, when assuming $M=10 \hspace{0.1cm} {\rm eV}$ and $n=2$.

This  work   should  alert   colleagues  who  are  under the   impression that  the  chameleon or {\it thin shell} effect   completely and effectively  protects the model  from  having to confront the strict  experimental bounds
associated with the  tests of  violations of the WEP. In light of our results, the various   models that rely on the chameleon  effect should  only  be  considered  as viable for those specific values of $M$, $n$ and $\beta$ that are not ruled out by  experiments. In this regard, it is important to emphasize that,  just as in this   case, extreme  care  is  needed to asses the viability of other modified  theories of gravity proposed  to  account for  the accelerated expansion of the universe, and  which  often   rely  on the
chameleon or {\it thin shell} effects in order to  avoid the stringent bounds imposed by the {\it classical tests of general relativity}. In fact many of such theories have not been yet studied  in   detail in connection  with the E\"{o}tv\"{o}s type of bounds  considered here,  often because one  takes  for granted that the {\it thin shell} effects will suppress possible  effective violations of the WEP. Moreover, a  good  segment of the community  working in the field  seem   to be under the impression that such violations are not even present given the fact that metric alternative gravity theories incorporate universal  couplings just as
general relativity. That is, such alternative theories couple to matter universally, as the chameleon models considered in the present work, 
where we assumed that  $\beta$  is the same for all the matter species. In view of our results, the viability  of those theories  needs to  be  examined  in detail. Various  types of  popular  $f(R)$ gravity theories are affected  by the previous  discussion. The point is that even if the {\it thin  shell}  effect allows some   those theories  to    avoid  conflict with standard  solar-system tests (as concluded by several authors~\cite{fR}),  the same mechanism can result in  predictions of violations of the WEP   which far larger  than the    experimental  existing bounds as  we have shown in this  work. Therefore, in view of the present work,    it would be prudent   to reconsider  the empirical viability  of  all such theories  in the  context of the  WEP  tests  that  we have analyzed  with a novel  approach.

\section*{Acknowledgements}
We  would  like to thank   P.  Brax, J. Khoury,  A. Upadhye and B. Elder for very valuable    private communications  that helped  us  improve this  manuscript. We are very grateful to J. Sakstein and C. Burrage for providing us a modification of their plot in~\cite{Burrage16} which allowed us to compare 
our bounds with others found in the literature. The authors acknowledge the use of the supercluster Mitzli at UNAM for doing numerical calculations and  thank the people of DGSCA-UNAM for computational and technical support. The authors thank Carolina Negrelli for help with the numerical calculations. L.K. and S.L. are supported  by CONICET grant PIP 11220120100504 and by the National Agency for the Promotion of Science and Technology (ANPCYT) of Argentina grant PICT-201-0081; and with H.V. by grant G140 from UNLP.  M.S. is partially supported by UNAM-PAPIIT grant IN107113 and CONACYT grants CB-166656 and CB-239639. D.S. is supported in part by CONACYT No. 101712,  and PAPIIT- UNAM No. IG100316 México, as well as  sabbatical  fellowships  from  PASPA-DGAPA-UNAM-México, and   from  Fulbright-Garcia Robles-COMEXUS.

\appendix
\section{The expanded solution}
\label{NL}
The results presented in Section~\ref{sec:results} have been calculated taking only the first 
($N=0$, $l=0$) terms in the truncated series
for the expression of the force. Here we show that taking two more terms (up to $N=4$, $l=4$), 
the results are basically the same. We only take two more terms because the larger the number of terms the more unstable
the numerical calculation becomes. On the other hand, the terms in the summations decrease several orders of magnitude when $l\neq0$ and $w\neq0$, so they do not really contribute to the result. 
Eqs.~(\ref{eq1}), (\ref{eq2}),  (\ref{Cinfull1}) and  (\ref{Cinfull}) yield

\begin{subequations}
 \begin{gather}
  b_1\delta_{l0}=C^{\rm out1}_{l}a_l+\sum_{w=0}^{4}C^{\rm out2}_{w}z_{wl},\\
  b_2\delta_{l0}=C^{\rm out2}_{l}x_l+\sum_{w=0}^{4}C^{\rm out1}_{w}y_{wl},\\
  C^{\rm in1}_{l}i_l^\prime(\mu_1R_1)= C^{\rm out1}_{l}k_l^\prime(\mu_{\rm out}R_1)+\sum_{w=0}^{4}C^{\rm out2}_{w}\alpha^{w0}_{l0}
  i_l^\prime(\mu_{\rm out}R_1),\\
  C^{\rm in2}_{l}i_l^\prime(\mu_2R_2)= C^{\rm out2}_{l}k_l^\prime(\mu_{\rm out}R_2)+\sum_{w=0}^{4}C^{\rm out1}_{w}\alpha^{*w0}_{l0}
  i_l^\prime(\mu_{\rm out}R_2);
 \end{gather}
\end{subequations}

Then, using these truncated series ($N=4$) in the expressions for $F_{1\rm z \varphi}$ and $F_{2\rm z \varphi}$, we obtain the same results as those shown in Section~\ref{sec:results}, but we stress that the numerical calculation for the acceleration of the {\it test} body becomes very unstable for small values of $\beta$. 
\bigskip

\section{Exploring the  test particle limit in our specific (idealized)  situation}
\label{sec:Plimit}

In this appendix we show what happens in our proposal when the radius, $R_2$, of the {\it test} body tends to zero while the density $\rho_2$ 
(and $\mu_2$) remain constant. First, we did this numerically within the approach presented in 
Section~\ref{FModel} and confirmed that there are no violations of the WEP 
when the {\it thin shell} condition in the {\it large} body ensues.

 We implement two simplified models that provide more insight about this limit. In the first one,
we take the chameleon field outside the {\it test} body to be that due to the {\it large} body  alone (which is easily found)
and thus we require only to compute the field inside the {\it test} body. 
In the second model, we assume {\it ab initio} that the {\it test} body 
is literally a {\it test point particle}, and thus, that it does not produce any back-reaction on the {\it large} body and the 
environment. Hence, we take the chameleon as generated solely by these two sources. That is, in this case the 
problem reduces to the usual one body problem (OBP) that has been studied 
in the past, and which corresponds to the static and spherically symmetric solution of the chameleon equation with one body surrounded by an 
environment.

 So let us start analyzing the 
first of these two models that we call the simplified two body problem (STBP). Under the STBP we take the OBP solution in the  outside  region and inside the {\it large} body (see Appendix~\ref{sec:OBP} for a review on the OBP solution). However, inside the {\it test} body we consider  the solution that  matches    continuously  on its boundary with the exterior field.  In order to do this  we take the most general regular solution of the differential equation in that  region, which is given by 
\begin{equation}
\quad \varphi_{\rm in 2}(r',\phi',\theta')= \varphi(r',\phi',\theta')=
  \sum_{lm} C_{lm}^{in2} i_l(\mu_2 r') Y_{lm}(\theta',\phi')+\varphi_{c}^{\rm test \,\ body} \hspace{0.25cm} \qquad (r' \le R_{2}) \,\,\,,
\label{phi}
\end{equation}
where $R_2$ is the radius of the {\it test} body, and the primed spherical coordinates are taken with respect to a frame of reference centered at the origin of the {\it test} body. Like in the analysis of Section~\ref{FModel}, inside the {\it test} body we consider only the MSBF's of the first type $i_l(.)$ 
which are regular at $r'=0$ as opposed to 
the MSBF's of second type $k_l(.)$, which are regular at infinity. We have introduced the notation 
$\varphi_{c}^{\rm test \,\ body}=\varphi^{\rm in}_{\rm 2 min}$, for the minimum of $V_{\rm eff}(\varphi)$ inside the {\it test} body, 
and similarly, $\mu_2=m_{\rm eff}^{\rm test\,\ body}$. In fact, the STBP that we analyze 
is {\it axially symmetric} (i.e. symmetric around the $z-axis$ that connects the centers of the two bodies as shown in 
Figure~\ref{figura1}), thus, we need   only the coefficients with $m=0$ (the others vanish identically).

As we indicated,  here we take the chameleon outside the {\it test} body  to  be  well approximated  by the exterior field of the 
 OBP (see Appendix~\ref{sec:OBP}):
\begin{equation}
\label{Outsol}
\varphi^{\rm out}(r)=C \frac{\exp{[-\mu_{\rm out} r]}}{r}+ \varphi_\infty \,\,\,,
\end{equation}
where $\mu_{\rm out}=m_{\rm eff}^{\rm out}$ is the effective mass of the chameleon associated with the environment, 
$\vec{r}^{\,\prime} = \vec r - D \hat z$, $D$ is the distance between the centers of the two bodies (see Figure~\ref{figura1}) and 
 $C$ is a constant that is found when matching the exterior with the interior solution 
of the OBP (see Appendix~\ref{sec:OBP}). Let us remind that in the STBP we  are ignoring the effect of  the {\it test} body on  the exterior solution, thus, 
for the STBP we only require the exterior solution of the {\it large} body as if the {\it test} body were absent. 
The coefficients of the expansion in Eq.~(\ref{phi}) are found by matching this solution with 
the exterior solution Eq.~(\ref{Outsol}) at $R_2$. In order to do so 
we shall employ the following identity to rewrite $\varphi^{\rm out}(r)$ in terms of the coordinate system centered at the {\it test} body:
\begin{equation}
\frac{\exp{[- \mu_{\rm out} |\vec r_2 - \vec r_1|]}}{4 \pi |\vec r_2 - \vec r_1|} = 
\mu_{\rm out} \sum_{l=0}^\infty i_l(\mu_{\rm out} r_2) k_l(\mu_{\rm out} r_1 ) \times \sum_{m=-l}^l 
Y_{lm}(\theta_2,\phi_2) Y_{lm}^*(\theta_1,\phi_1), \nonumber
\end{equation}
as long as $r_2 < r_1$ and being $k_l(.)$ the MSBF of second type mentioned previously. In the current case 
$\vec r_1 = -D \hat z$, $\theta_1=0$, and $\vec r_2 = \vec{r}^{\,\prime}$. As mentioned before 
only the terms with $m=0$ contribute to the solution. The matching condition $\varphi_{\rm in2}(R_{2}) = \varphi_{\rm out}(R_{2})$, leads to
\begin{equation}
C^{\rm in2}_{l0} = \left[\frac{(\varphi_\infty - \varphi_{c}^{\rm tbody})\delta_{l0}}{2l+1} + 
C \mu_{\rm out} i_l(\mu_{\rm out} R_2) k_l(\mu_{\rm out} D)\right] \times \frac{\sqrt{4 \pi (2 l + 1)}} {i_l(\mu_2 R_2)}.  
\label{cl}
\end{equation}
Remarkably,  even at this level of approximation, and despite  considering  the case of   a universal coupling $\beta$, Eq.~(\ref{cl}), 
shows that the field  inside the {\it test} body depends on the {\it test} body properties, like its size $R_2$ and 
composition [both $\mu_2$ and $\varphi_{c}^{\rm tbody}$ depend on $\rho_{\rm tbody}$ according to Eqs.~(\ref{meff}) and (\ref{Phiinout})].

As soon as ones considers {\it test} bodies as extended bodies, they can be an important source of the chameleon, and as such, their composition appears naturally in the detailed  behaviour  of the scalar  field. It is important to emphasize that in this STBP, the approximation involves taking the  scalar field  outside the small body, and  then  using the varying value of that field on the small 
body's boundary to help determine the  value of the  field inside of it. That is done by enforcing continuity of the chameleon field on the  small body's surface, but in the present approximation we cannot impose the continuity of the first derivative of the chameleon there. The reason is that as  the  field  outside  is  fixed  (i.e., it is not allowed to react to the small body)  there are not  enough   undetermined coefficients in the resulting expansion (only the  coefficients  corresponding to the  field  in   small body's interior) to impose  the  two  conditions (continuity of the field  and of  its  normal derivative).   This discontinuity can produce serious consequences in the resulting force when the {\it test} body is of finite size. 
This is why we do not use this simplified model for that purpose. Nonetheless, in the limit  where the   small  body  goes to  zero ($R_2\to 0$) this drawback disappears, simply because  the   exterior  field  does  not  change   over the infinitesimal region occupied by  such point like body.

Now, we want precisely to consider this limit. First, the coefficient of the chameleon in spherical 
harmonics with $l=0$ associated with the solution due to the {\it large} body behaves as follows
 \begin{subequations}
 \begin{gather}
  C^{\rm in1}_{00}\sim\frac{\sqrt{4\pi}\mu_1(1+\mu_{\rm out}R_1)(\varphi_\infty - \varphi_{1\rm min}^{\rm in})\csch(\mu_1R_1)}{\mu_{\rm out}+\mu_1\coth(\mu_1R_1)},\\
  C^{\rm out1}_{00}\sim\frac{\sqrt{4\pi}\exp[\mu_{\rm out}R_1]\mu_{\rm out}( \varphi_{1\rm min}^{\rm in}-\varphi_\infty )\lbrace\mu_1R_1\cosh(\mu_1R_1)-\sinh(\mu_1R_1)\rbrace}{\mu_1\cosh(\mu_1R_1)+\mu_{\rm out}\sinh(\mu_1R_1)},\\
  C^{\rm out2}_{00}\sim0.
 \end{gather}
\end{subequations}
By construction, the chameleon field inside the {\it large} body and outside both bodies does not depend on anything coming from 
the {\it test} body. On the other hand, the coefficient $C^{\rm in2}_{00}$ of Eq.~(\ref{cl}) in the limit $R_2\to 0$ becomes,

\begin{equation}
C^{\rm in2}_{00}\sim\sqrt{4\pi}\Big\lbrace (\varphi_\infty - \varphi_{2\rm min}^{\rm in})+\frac{\exp[\mu_{\rm out}(R_1-D)](\varphi_\infty - \varphi_{1\rm min}^{\rm in})[\sinh(\mu_1R_1)-\mu_1R_1\cosh(\mu_1R_1)]}{D[\mu_1\cosh(\mu_1R_1)+\mu_{\rm out}\sinh(\mu_1R_1)]}\Big\rbrace.
\end{equation}
So, the field $\varphi_{\rm in2}(\vec{r'})$ when $R_2\to 0$ becomes

\begin{equation}
\varphi_{\rm in2} \sim \varphi_\infty+\frac{\exp[\mu_{\rm out}(R_1-D)](\varphi_\infty - \varphi_{1\rm min}^{\rm in})[\sinh(\mu_1R_1)-\mu_1R_1\cosh(\mu_1R_1)]}{D[\mu_1\cosh(\mu_1R_1)+\mu_{\rm out}\sinh(\mu_1R_1)]};
\end{equation}

Thus, it follows from the above equations that in this limit the chameleon field does not depend on the test particle composition,  
and thus, there is no  effective violation of the WEP.
\bigskip

Let us now consider a different  simplification,  where the {\it test} body is taken as a test point particle 
from the very beginning. This is what one usually finds in most of the literature. Since one takes the {\it test} body as test particle and the 
{\it large} body and the environment as spherical sources of different uniform densities, one  takes the following distribution for the 
matter part, $T^m\approx  -\rho= -{\cal M}_2\delta(z-D)\delta(x)\delta(y) - \rho_{\rm in}^1 \Theta(r-R) - \rho_{\rm out} {\tilde \Theta}(r,R) $, 
where ${\cal M}_2$ is the mass of the {\it test} body, and the step function ${\tilde \Theta}(r,R)$ is defined to be unity for 
$R<r<\infty$ and zero otherwise. The  distribution $\rho_{\rm out} {\tilde \Theta}(r,R)$ allows one to describe the environment of density $\rho_{\rm out}$. 
\bigskip

\bigskip

Therefore from Eq.~(\ref{Ueffmin3}) we obtain
\begin{eqnarray}
U_{\varphi} \approx \varphi_{\rm out} ({\vec r}) \frac{ \beta {\cal M}_2}{M_{pl}} + {\rm const.}
\end{eqnarray}
where ${\vec r}$ corresponds to the location of the point-test particle, and 
the constant contains information about the bulk properties of the {\it large} body, the environment 
and the point-test particle, but it is independent of the location of the latter. 
Thus the only contribution for the force arises only from the term 
in the integral of Eq.~(\ref{Ueffmin3}) that is proportional to $\hat\varphi T^m$, which contains the information of the chameleon and the location of the {\it test} body.

In this way, the chameleon force on the point-test particle reads 
\begin{equation}
\label{forcepointp}
{\vec F}= -{\vec \nabla} U_{\varphi}= - \frac{\beta {\cal M}_2}{M_{pl}}  {\vec \nabla} \varphi_{\rm out} \,\,\,,
\end{equation}
again, this is to be evaluated at the location of the point-test particle. Henceforth,  
we recover the same expression considered in~\cite{KW04} for the chameleon force acting on a point test-particle.

Now, in this limit the solution for the chameleon is given by Eq.~(\ref{Outsol}), which is the exterior solution of the one body problem in spherical symmetry. In order to give 
an insight of the E\"otv\"os parameter and the way it can be constrained when a {\it thin shell} exists, it is 
better to take the constant $C$ of Eq.~(\ref{Outsol}) as written in the form Eq.~(\ref{C2}). As explained in 
Appendix~\ref{sec:OBP} below, when there is a {\it thin shell} the exterior solution reads [cf. Eq.~(\ref{C2}) with $f(x)\approx 1$]
\begin{equation}
\varphi^{\rm out}(r) \approx  \varphi_{\infty} - \frac{3\beta {\cal M}}{4\pi M_{pl}}\frac{\Delta R}{R}
\frac{\exp{[-m_{\rm eff}^{\rm out}(r-R)]}}{r} \,\,\,.
\end{equation}
where $\cal M$ is the mass of the {\it large} body. For simplicity let us assume $r\sim R$ and $m_{\rm eff}^{\rm out}(r-R)\ll 1$. Thus
\begin{equation}
\varphi^{\rm out}(r) \approx  \varphi_{\infty} - \frac{3\beta }{4\pi M_{pl}}\frac{\Delta R}{R}
\frac{{\cal M}}{r} \,\,\,.
\end{equation}
Using this expression in (\ref{forcepointp}) the chameleon force on the point-test particle at distance $r=D$ turns out to be
\begin{equation}
{\vec F}_\varphi = - {\hat r} \frac{3  \beta^2}{4\pi M_{pl}^2} \frac{\Delta R}{R} \frac{{\cal M} {\cal M}_2 }{D^2}\,\,\,,
\end{equation}
where ${\hat r}$ is the unit radial vector. Notice that the chameleon force is proportional to the gravitational force acting on the particle 
due to the {\it large} body
\begin{equation}
{\vec F}_g = - {\hat r} \frac{G {\cal M}_{G} {\cal M}_{G,2}}{D^2}\,\,\,,
\end{equation}
where ${\cal M}_{G}$ and ${\cal M}_{G,2}$ are the {\it gravitational} masses of the 
{\it large} body and the point-test particle, respectively, as opposed to their {\it chameleon}-like masses (or ``chameleon charges'')  ${\cal M}$ and ${\cal M}_2$. 
That is, the masses that couple to the chameleon force.

In this way, the magnitude of the acceleration of the point-test particle due to the {\it large} body is
\begin{equation}
a_2=  \frac{|{\vec F}_\varphi + {\vec F}_g|}{{\cal M}_{I,2}}= 
\frac{G {\cal M}_{G}}{D^2}\left(\frac{ {\cal M}_{G,2}}{ {\cal M}_{I,2}} + \frac{6  \beta^2 {\cal M}_2 {\cal M}  }{{\cal M}_{I,2} {\cal M}_{G}} \frac{\Delta R}{R}\right) \,\,\,,
\end{equation}
where ${\cal M}_{I,2}$ is the {\it inertial} mass of the point-test particle and we used $G= 1/(8\pi M_{pl}^2)$.

If the coupling to the chameleon force is not universal then this acceleration reads
\begin{equation}
a_2=\frac{G {\cal M}_{G}}{D^2}\left(\frac{ {\cal M}_{G,2}}{ {\cal M}_{I,2}} + \frac{6 \beta_1\beta_2 {\cal M}_2 {\cal M} }{{\cal M}_{I,2} {\cal M}_{G}} \frac{\Delta R}{R}\right)   \,\,\,.
\end{equation}
where $\beta_1$ is the chameleon coupling to the large body.

A similar expression is obtained for the acceleration of a second point-test particle except that the masses ${\cal M}_2,{\cal M}_{G,2}$ and ${\cal M}_2^I$ 
of the first point-test particle are replaced by the masses ${\cal M}_3,{\cal M}_{G,3}$ and ${\cal M}_3^I$ of the second point-test particle, and similarly $\beta_2$ is 
replaced by $\beta_3$. 

Therefore the relative acceleration of two ``free-falling'' point-test particles under the influence of both gravity and the chameleon due to the 
{\it large} body is
\begin{equation}
|a_2 - a_3|= \frac{G {\cal M}_{G}}{D^2} 
\left|\frac{ {\cal M}_{G,2}}{ {\cal M}_{I,2}}- \frac{ {\cal M}_{G,3}}{ {\cal M}_{I,3}}
 + \left(\frac{\beta_2 {\cal M}_2}{{\cal M}_{I,2}}
   - \frac{\beta_3 {\cal M}_3}{{\cal M}_{I,3}}\right) \frac{6\beta_1 {\cal M} }{{\cal M}_{G}}\frac{\Delta R}{R}\right|\,\,\,.
 \end{equation}
The E\"otv\"os parameter is thus given by
\begin{equation}
\eta= \frac{2|a_2 - a_3|}{a_2 + a_3}
= 2\frac{
\left|\frac{ {\cal M}_{G,2}}{ {\cal M}_{I,2}}- \frac{ {\cal M}_{G,3}}{ {\cal M}_{I,3}}
 + \left(\frac{\beta_2 {\cal M}_2}{{\cal M}_{I,2}}
   - \frac{\beta_3 {\cal M}_3}{{\cal M}_{I,3}}\right) \frac{6 \beta_1 {\cal M} }{{\cal M}_{G}}\frac{\Delta R}{R}  \right| }
{ \frac{ {\cal M}_{G,2}}{ {\cal M}_{I,2}}+ \frac{ {\cal M}_{G,3}}{ {\cal M}_{I,3}}
 + \left(\frac{\beta_2 {\cal M}_2}{{\cal M}_{I,2}}
   + \frac{\beta_3 {\cal M}_3}{{\cal M}_{I,3}}\right) \frac{6 \beta_1 {\cal M} }{{\cal M}_{G}}\frac{\Delta R}{R}  }
 \,\,\,.
 \end{equation}
Taking $\beta_i=0$ ($i=1-3$) we recover the usual expression of this parameter only due to gravity. On the other hand, for $\beta_i\neq 0$ 
but assuming that the inertial and gravitational masses are the same, we obtain the E\"otv\"os parameter due to the chameleon solely:
\begin{equation}
\label{etacham}
\eta= \frac{2|a_2 - a_3|}{a_2 + a_3}
= \frac{\Delta R}{R} \frac{\frac{6 \beta_1 {\cal M} }{{\cal M}_{G}}
\left| \frac{\beta_2 {\cal M}_2}{{\cal M}_{I,2}}
   - \frac{\beta_3 {\cal M}_3}{{\cal M}_{I,3}}\right| }
{ 1 + \left(\frac{\beta_2 {\cal M}_2}{{\cal M}_{I,2}}
   + \frac{\beta_3 {\cal M}_3}{{\cal M}_{I,3}}\right) \frac{3 \beta_1 {\cal M} }{{\cal M}_{G}}\frac{\Delta R}{R}   } \,\,\,.
 \end{equation}
This expression  is  used to stress the importance of the {\it thin shell}  condition $\Delta R \ll R$. If the equality 
between the chameleon-like mass and the inertial mass holds exactly as well, then the relative acceleration between the two point-test 
particles is only   due to  the difference $|\beta_2-\beta_3|$  may not strictly vanish. Thus, if such a difference is of order unity, then the observations $\eta \sim 10^{-13}$ (or $\eta \sim 10^{-11}$ if one takes into account only the effects produced by the local inhomogeneities of the matter near the torsion balance) can be satisfied provided $\Delta R/R$ is 
small enough. This is the chameleon mechanism at work for the point-particle limit. Moreover,  if the couplings $\beta_i$ are universal, 
then $\eta\equiv 0$ regardless of  whether  the {\it large} body is associated with  a {\it thin shell} or not. As we stressed before, in the universal-coupling scenario 
we checked numerically that $\eta\rightarrow 0$ in our full model 
 of Section~\ref{FModel} as we reduced the size of the (extended) {\it test} body.

Summarizing,  the standard assumption is  that the {\it screening} effects 
 produced by the chameleon behaviour, effectively suppress the dependency of $\eta$ on the 
the couplings $\beta_i$ in  models  where  this  parameter is  not universal. Nonetheless,  as we have shown  here  such  supposition  seems to  work only for the   true point-particle limit of {\it test} bodies, but not  at the  more precise  level of the current analysis which is the relevant one for consideration of actual experimental test,  where the {\it test} bodies are not just point particles, but small bodies of finite size.
\bigskip

\section{Composition dependence of the effective force}
\label{sec:detailsforce}

Let us consider the expressions for the terms $F_{1\rm z\varphi}$ and $F_{2\rm z\varphi}$ of Section~\ref{sec:force}, and replace the explicit series expansions 
for the chameleon field found in Section~\ref{FModel}. The result is

\begin{eqnarray}
\label{fuerzaCf}
F_{1\rm z\varphi} &=& \int_{0}^{R_2}\int_{0}^{\pi}\int_{0}^{2\pi}\Big[\frac{(n+1)}{2}\frac{\rho_2\beta}{M_{pl}}\Big] \Big[\sum_l \frac{\partial C^{\rm in2}_{l0}}{\partial D}i_l(\mu_2r')Y_{l0}(\theta',\phi')\Big]r'^2\sin(\theta') d\phi' d\theta' dr' \nonumber \\
&+&\int_{0}^{R_1}\int_{0}^{\pi}\int_{0}^{2\pi}\Big[\frac{(n+1)}{2}\frac{\rho_1\beta}{M_{pl}}\Big] \Big[\sum_l \frac{\partial C^{\rm in1}_{l0}}{\partial D}i_l(\mu_1r)Y_{l0}(\theta,\phi)\Big]r^2\sin(\theta) d\phi d\theta dr \nonumber\\
&+&\int_{R_1}^{\infty}\int_{0}^{\pi}\int_{0}^{2\pi}\Big[\frac{(n+1)}{2}\frac{\rho_{\rm out}\beta}{M_{pl}}\Big]\Big[\sum_l \frac{\partial C^{\rm out1}_{l0}}{\partial D}k_l(\mu_{\rm out}r)Y_{l0}(\theta,\phi)\Big]r^2\sin(\theta) d\phi d\theta dr \nonumber\\
&+&\int_{R_1}^{D}\int_{0}^{\pi}\int_{0}^{2\pi}\Big[\frac{(n+1)}{2}\frac{\rho_{\rm out}\beta}{M_{pl}}\Big] \nonumber\\
&\times&\Big[\sum_l \sum_{w=0}^N\Big\lbrace\frac{\partial C^{\rm out2}_{l0}}{\partial D}\alpha^{l0}_{w0}+C^{\rm out2}_{l0}\frac{\partial\alpha^{l0}_{w0}}{\partial D}\Big\rbrace i_w(\mu_{\rm out}r)Y_{w0}(\theta,\phi)\Big]r^2\sin(\theta) d\phi d\theta dr \nonumber\\
&+&\int_{D}^{\infty}\int_{0}^{\pi}\int_{0}^{2\pi}\Big[\frac{(n+1)}{2}\frac{\rho_{\rm out}\beta}{M_{pl}}\Big] \nonumber\\
&\times&\Big[\sum_l \sum_{w=0}^N\Big\lbrace\frac{\partial C^{\rm out2}_{l0}}{\partial D}\hat{\alpha}^{l0}_{w0}+C^{\rm out2}_{l0}\frac{\partial \hat{\alpha}^{l0}_{w0}}{\partial D}\Big\rbrace k_w(\mu_{\rm out}r)Y_{w0}(\theta,\phi)\Big]r^2\sin(\theta) d\phi d\theta dr \nonumber\\
&-&  \int_{0}^{R_2}\int_{0}^{\pi}\int_{0}^{2\pi}\Big[\frac{(n+1)}{2}\frac{\rho_{\rm out}\beta}{M_{pl}}\Big] \nonumber\\
&\times& \Big[\sum_l \sum_{w=0}^N\Big\lbrace\frac{\partial C^{\rm out1}_{l0}}{\partial D}\alpha^{*l0}_{w0}+C^{\rm out1}_{l0}\frac{\partial \alpha^{*l0}_{w0}}{\partial D}\Big\rbrace i_w(\mu_{\rm out}r')Y_{w0}(\theta',\phi')\Big]r'^2\sin(\theta') d\phi' d\theta' dr', \nonumber \\ 
\end{eqnarray}
 where
\beq
   \begin{split}
    \hat{\alpha}^{lm}_{vw}=& (-1)^{m+v} (2v+1)\sum_{p=|l-v|}^{|l+v|}(2p+1)\Big[\frac{(l+m)!(v+w)!(p-m-w)!}{(l-m)!(v-w)!(p+m+w)!}\Big]^{1/2} \\
    &\times \begin{bmatrix} l & v & p\\ 0 & 0 & 0 \end{bmatrix}\begin{bmatrix} l & v & p\\ m & w & -m-w \end{bmatrix} i_p(\mu_{\rm out}|D|)Y_{p(m-w)}(\theta_D,\phi_D).
   \end{split}
\label{sum2a}
\eeq
The integral $\int_0^{2\pi}\int_{0}^{\pi}Y_{l0}(\theta_j,\phi_j)\sin(\theta_j)d\theta_j d\phi_j$, which appears inside all the terms   in the expression for  $F_{1\rm z \varphi}$, vanishes  when $l\neq0$. Meanwhile, integrals like $\int_0^Ri_l(mr)r^2 dr$ and $\int_R^{\infty}k_l(mr)r^2dr$, and the derivatives $\partial C^{\rm in(out) \,1(2)}_{l0}/\partial D$ converge to finite values for all values of $l$. 
Consequently, the only contribution to the first three integrals of Eq.~(\ref{fuerzaCf}) is the term with $l=0$, while the 
last integrals of Eq.~(\ref{fuerzaCf}) only have no null terms for $w=0$ being $l$ arbitrary in this case. 
In this way, we obtain for the linear term of the force:
\begin{eqnarray}
\label{fuerzaCf2}
F_{1\rm z\varphi} &=& \Big[\frac{(n+1)}{2}\frac{\rho_2\beta}{M_{pl}}\Big] 2\sqrt{\pi}\frac{\partial C^{\rm in2}_{00}}{\partial D}\Big[\frac{\mu_2R_2\cosh(\mu_2R_2)-\sinh(\mu_2R_2)}{\mu_2^3}\Big]\nonumber \\
&+&\Big[\frac{(n+1)}{2}\frac{\rho_1\beta}{M_{pl}}\Big] 2\sqrt{\pi}\Big[\frac{\mu_1R_1\cosh(\mu_1R_1)-\sinh(\mu_1R_1)}{\mu_1^3}\Big]\frac{\partial C^{\rm in1}_{00}}{\partial D} \nonumber\\
&+&\Big[\frac{(n+1)}{2}\frac{\rho_{\rm out}\beta}{M_{pl}}\Big] 2\sqrt{\pi}\frac{\partial C^{\rm out1}_{00}}{\partial D}\Big[\frac{\exp[-\mu_{\rm out}R_1](1+R_1)}{\mu_{\rm out}^3}\Big]\nonumber\\
&+&\Big[\frac{(n+1)}{2}\frac{\rho_{\rm out}\beta}{M_{pl}}\Big] 2\sqrt{\pi}\sum_l\Big\lbrace\frac{\partial C^{\rm out2}_{l0}}{\partial D}\alpha^{l0}_{00}+C^{\rm out2}_{l0}\frac{\partial \alpha^{l0}_{00}}{\partial D}\Big\rbrace \nonumber\\
&\times& \Big[ \frac{-R_1\cosh(\mu_{\rm out}R_1)+D\cosh(\mu_{\rm out}D)}{\mu_{\rm out}^2}+\frac{\sinh(\mu_{\rm out}R_1)-\sinh(\mu_{\rm out}D)}{\mu_{\rm out}^3}\Big]\nonumber\\
&+&\Big[\frac{(n+1)}{2}\frac{\rho_{\rm out}\beta}{M_{pl}}\Big] 2\sqrt{\pi}\sum_l \Big\lbrace \frac{\partial C^{\rm out2}_{l0}}{\partial D}\hat{\alpha}^{l0}_{00}+\frac{\partial \hat{\alpha}^{l0}_{00} }{\partial D}C^{\rm out2}_{l0}\Big\rbrace \frac{\exp[-\mu_{\rm out}R_1](1+\mu_{\rm out}R_1)}{\mu_{\rm out}^3}\nonumber\\
&-&  \Big[\frac{(n+1)}{2}\frac{\rho_{\rm out}\beta}{M_{pl}}\Big] 2\sqrt{\pi} \nonumber\\
&\times& \sum_l \Big\lbrace \frac{\partial C^{\rm out1}_{l0}}{\partial D}\alpha^{*l0}_{00}+\frac{\partial \alpha^{*l0}_{00}}{\partial D} C^{\rm out1}_{l0}\Big\rbrace \Big[\frac{\mu_{\rm out}R_2\cosh(\mu_{\rm out}R_2)-\sinh(\mu_{\rm out}R_2)}{\mu_{\rm out}^3}\Big]. \nonumber \\ 
\end{eqnarray}

The quadratic term can be expressed as:
\begin{eqnarray}
\label{fuerzaCf3}
F_{2\rm z\varphi} &=& -\int_{V_{2}}\Big[\frac{(n+2)}{4}\mu_2^2\Big] \frac{\partial}{\partial D}\Big[\sum_w\sum_l C^{\rm in2}_{l0} i_l(\mu_2r')Y_{l0}(\theta',\varphi')C^{\rm in2}_{w0} i_w(\mu_2r')Y_{w0}(\theta',\phi')\Big]r'^2\sin(\theta') dV_{2} \nonumber \\
&-&\int_{V_{1}}\Big[\frac{(n+2)}{4}\mu_1^2\Big] \frac{\partial}{\partial D}\Big[\sum_w\sum_l C^{\rm in1}_{l0}i_l(\mu_1r)Y_{l0}(\theta,\phi)C^{\rm in1}_{w0}i_l(\mu_1r)Y_{w0}(\theta,\phi)\Big]r^2\sin(\theta) dV_{1} \nonumber\\
&-&\int_{R_1}^{\infty}\int_{0}^{\pi}\int_{0}^{2\pi}\Big[\frac{(n+2)}{4}\mu_{\rm out}^2\Big]\frac{\partial}{\partial D}\Big[\sum_v\sum_l C^{\rm out1}_{l0} k_l(\mu_{\rm out}r)Y_{l0}(\theta,\phi)C^{\rm out1}_{v0} k_v(\mu_{\rm out}r)Y_{v0}(\theta,\phi)\Big]r^2\sin(\theta) d\phi d\theta dr \nonumber\\
&-&\int_{R_1}^{D}\int_{0}^{\pi}\int_{0}^{2\pi}\Big[\frac{(n+2)}{4}\mu_{\rm out}^2\Big]\nonumber\\
&\times& \frac{\partial}{\partial D}\Big[\sum_v\sum_l C^{\rm out1}_{l0} k_l(\mu_{\rm out}r)Y_{l0}(\theta,\phi)C^{\rm out2}_{v0} \sum_{w}^N \alpha^{v0}_{w0}i_w(\mu_{\rm out}r)Y_{w0}(\theta,\phi)\Big]r^2\sin(\theta) d\phi d\theta dr \nonumber\\
&-&\int_{D}^{\infty}\int_{0}^{\pi}\int_{0}^{2\pi}\Big[\frac{(n+2)}{4}\mu_{\rm out}^2\Big]\nonumber\\
&\times& \frac{\partial}{\partial D}\Big[\sum_v\sum_l C^{\rm out1}_{l0} k_l(\mu_{\rm out}r)Y_{l0}(\theta,\phi)C^{\rm out2}_{v0} \sum_{w}^N \hat{\alpha}^{v0}_{w0}k_w(\mu_{\rm out}r)Y_{w0}(\theta,\phi)\Big]r^2\sin(\theta) d\phi d\theta dr \nonumber\\
&-&\int_{R_1}^{D}\int_{0}^{\pi}\int_{0}^{2\pi}\Big[\frac{(n+2)}{4}\mu_{\rm out}^2\Big]\nonumber\\
&\times& \frac{\partial}{\partial D}\Big[\sum_l C^{\rm out2}_{l0} \sum_{u}^N \alpha^{l0}_{u0}i_u(\mu_{\rm out}r)Y_{u0}(\theta,\varphi)\sum_vC^{\rm out2}_{v0} \sum_{w}^N \alpha^{v0}_{w0}i_w(\mu_{\rm out}r)Y_{w0}(\theta,\phi)\Big]r^2\sin(\theta) d\phi d\theta dr \nonumber\\
&-&\int_{D}^{\infty}\int_{0}^{\pi}\int_{0}^{2\pi}\Big[\frac{(n+2)}{4}\mu_{\rm out}^2\Big]\nonumber\\
&\times& \frac{\partial}{\partial D}\Big[\sum_l C^{\rm out2}_{l0} \sum_{u}^N \hat{\alpha}^{l0}_{u0}k_u(\mu_{\rm out}r)Y_{u0}(\theta,\phi)\sum_vC^{\rm out2}_{v0} \sum_{w}^N \hat{\alpha}^{v0}_{w0}k_w(\mu_{\rm out}r)Y_{w0}(\theta,\phi)\Big]r^2\sin(\theta) d\phi d\theta dr \nonumber\\
&+&\int_{0}^{R_2}\int_{0}^{\pi}\int_{0}^{2\pi}\Big[\frac{(n+2)}{4}\mu_{\rm out}^2\Big]\nonumber\\
&\times& \frac{\partial}{\partial D}\Big[\sum_l C^{\rm out1}_{l0} \sum_{u}^N \alpha^{*l0}_{u0}i_u(\mu_{\rm out}r')Y_{u0}(\theta',\phi')\sum_vC^{\rm out1}_{v0} \sum_{w}^N \alpha^{*v0}_{w0}i_w(\mu_{\rm out}r')Y_{w0}(\theta',\phi')\Big]r^2\sin(\theta') d\phi' d\theta' dr' \nonumber\\
\end{eqnarray}

The angular integral that appears inside all the terms in the expression for  $F_{2\rm z \varphi}$ is,
\begin{equation}
\int_0^{2\pi}\int_{0}^{\pi}Y_{l0}(\theta_j,\phi_j)Y_{k0}(\theta_j,\phi_j)\sin(\theta_j)d\theta_j d\phi_j=\frac{4\pi \sqrt{(2l+1)(2k+1)}k!}{4\pi(2k+1)k!}\delta_{kl}=\delta_{kl}.\nonumber
\end{equation}

 Therefore,

\begin{eqnarray}
\label{fuerzaCf31f}
F_{2\rm z\varphi} &=&-\sum_l\frac{(n+2)}{4}\mu_2^2 \frac{\partial}{\partial D}\Big[C^{\rm in2}_{l0}C^{\rm in2}_{l0}\Big]\frac{R_2^3}{2}\Big\lbrace \left[i_l(\mu_2R_2)\right]^2-i_{l-1}(\mu_2R_2)i_{l+1}(\mu_2R_2)\Big\rbrace\nonumber \\
&-&\sum_l\frac{(n+2)}{4}\mu_1^2\frac{\partial}{\partial D}\Big[C^{\rm in1}_{l0}C^{\rm in1}_{l0}\Big]\frac{R_1^3}{2}\Big\lbrace\left[i_l(\mu_1R_1)\right]^2-i_{l-1}(\mu_1R_1)i_{l+1}(\mu_1R_1)\Big\rbrace \nonumber \\
&-&\sum_l\frac{(n+2)}{4}\mu_{\rm out}^2\frac{\partial}{\partial D}\Big[C^{\rm out1}_{l0}C^{\rm out1}_{l0}\Big]\frac{R_1^3}{2}\Big\lbrace\left[k_l(\mu_{\rm out}R_1)\right]^2-k_{l-1}(\mu_{\rm out}R_1)k_{l+1}(\mu_{\rm out}R_1)\Big\rbrace  \nonumber\\
&-&\sum_w\sum_{l}^N \frac{(n+2)}{4}\mu_{\rm out}^2\frac{\partial}{\partial D}\Big[C^{\rm out2}_{w0}C^{\rm out1}_{l0}\alpha^{w0}_{l0}\Big]\Big\lbrace Q_2(l,D)-Q_2(l,R_1)\Big\rbrace \nonumber\\
&-&\sum_w\sum_{l}^N\frac{(n+2)}{4}\mu_{\rm out}^2 \frac{\partial}{\partial D}\Big[C^{\rm out2}_{w0}C^{\rm out1}_{l0}\hat{\alpha}^{w0}_{l0}\Big]\frac{D^3}{2}\Big\lbrace\left[k_l(\mu_{\rm out}D)\right]^2-k_{l-1}(\mu_{\rm out}D)k_{l+1}(\mu_{\rm out}D)\Big\rbrace  \nonumber\\
&-&\sum_l\sum_w\sum_{v}^N\frac{(n+2)}{4}\mu_{\rm out}^2 \frac{\partial}{\partial D}\Big[C^{\rm out2}_{w0}C^{\rm out2}_{l0}\alpha^{w0}_{v0}\alpha^{l0}_{v0}\Big]\Big\lbrace Q_3(v,D)-Q_3(v,R_1)\Big\rbrace  \nonumber\\
&-&\sum_w\sum_l\sum_{v}^N\frac{(n+2)}{4}\mu_{\rm out}^2 \frac{\partial}{\partial D}\Big[C^{\rm out2}_{w0}C^{\rm out2}_{l0}\hat{\alpha}^{w0}_{v0}\hat{\alpha}^{l0}_{v0}\Big]\nonumber\\
&\times& \frac{D^3}{2}\Big\lbrace\left[k_v(\mu_{\rm out}D)\right]^2-k_{v-1}(\mu_{\rm out}D)k_{v+1}(\mu_{\rm out}D)\Big\rbrace  \nonumber\\
&+&\sum_w\sum_l\sum_{v}^N\frac{(n+2)}{4}\mu_{\rm out}^2 \frac{\partial}{\partial D}\Big[C^{\rm out1}_{w0}C^{\rm out1}_{l0}\alpha^{*w0}_{v0}\alpha^{*l0}_{v0}\Big]\nonumber\\
&\times& \frac{R_2^3}{2}\Big\lbrace\left[i_v(\mu_{\rm out}R_2)\right]^2-i_{v-1}(\mu_{\rm out}R_2)i_{v+1}(\mu_{\rm out}R_2)\Big\rbrace \,\,\,,
\end{eqnarray}

where

\begin{eqnarray}
Q_2(l,r) &=& -\frac{2l+1}{4\mu_{\rm out}^3}+\frac{r^3}{2}\Big\lbrace i_l(\mu_{\rm out}r)k_l(\mu_{\rm out}r)+i_{l-1}(\mu_{\rm out}r)k_{l-1}(\mu_{\rm out}r)\Big\rbrace,\nonumber
\end{eqnarray}
\begin{equation}
Q_3(v,r)=\frac{r^3}{2}\Big\lbrace\Big[i_v(\mu_{\rm out}r)\Big]^2-i_{v-1}(\mu_{\rm out}r)i_{v+1}(\mu_{\rm out}r)\Big\rbrace.\nonumber
\end{equation}

As we underlined at the end of Section~\ref{sec:force}, the expressions for $F_{1\rm z\varphi}$ and $F_{2\rm z\varphi}$ show the explicit dependence of the  chameleon mediated force with the composition and size of the {\it test} body through the quantities  $\rho_2$, $C^{\rm in2}_{l0}$, $C^{\rm in1}_{l0}$, $C^{\rm out2}_{l0}$, $C^{\rm out1}_{l0}$ and their derivatives.

\section{The two body problem solution for the chameleon with the metal encasing}
\label{sec:TBPME}

\bigskip

Figure \ref{esquemaconshell} represents  a sketch of the two body problem with a metal encasing around the {\it test} body. As one can see, both, the atmosphere inside the encasing and metal encasing itself, can be described under the same coordinate system that describes the {\it test} body, that is, O'. Accordingly, the solution for the chameleon field is given by,

\begin{equation}
\label{fullsolME}
\quad \varphi=
\begin{cases}
 \varphi_{\rm in1}= \sum\limits_{lm} C_{lm}^{\rm in1} i_l(\mu_1 r) Y_{lm}(\theta,\phi)+\varphi_{1\rm min}^{\rm in} \hspace{1.3cm} 
 (r \le R_1) \\
 \varphi_{\rm out}=\sum\limits_{lm} C_{lm}^{\rm out1} k_l(\mu_{\rm out} r) Y_{lm}(\theta,\phi)+ 
 C_{lm}^{\rm out2} k_l(\mu_{\rm out} r') Y_{lm}(\theta',\phi')+\varphi_{\infty} 
 \hspace{0.7cm} (\rm exterior\,\,solution) \\
 \varphi_{\rm in2}=\sum\limits_{lm} C_{lm}^{\rm in2} i_l(\mu_2 r') Y_{lm}(\theta',\phi')+\varphi_{2\rm min}^{\rm in} \hspace{0.7cm} (r' \le R_2)\\
 \varphi_{\rm vacuum }=\sum\limits_{lm} C_{lm}^{\rm vac1} i_l(\mu_{\rm vac} r') Y_{lm}(\theta',\phi')+C_{lm}^{\rm vac2} k_l(\mu_{\rm vac} r') Y_{lm}(\theta',\phi')+\varphi_{\rm min}^{\rm vac} \hspace{0.7cm} (R_2 \le r' \le R_{\rm vac})\\
 \varphi_{\rm encasing}=\sum\limits_{lm} C_{lm}^{\rm enc1} i_l(\mu_{\rm enc} r') Y_{lm}(\theta',\phi')+C_{lm}^{\rm enc2} k_l(\mu_{\rm enc} r') Y_{lm}(\theta',\phi')+\varphi_{\rm min}^{\rm enc} \hspace{0.7cm} (R_{\rm vac} \le r' \le R_{\rm enc} )
\end{cases}
\end{equation}

where the ``new'' magnitudes $\mu_{\rm vac}=m_{\rm eff}^{\rm vacuum}$, $\mu_{\rm enc }=m_{\rm eff}^{\rm  encasing}$, $\varphi_{\rm min}^{\rm vac}$ and $\varphi_{\rm min}^{\rm enc}$ represents the value of the chameleon field that minimizes the effective potential in each region, and 
$R_{\rm vac},R_{\rm enc}$ are the radii of the vacuum chamber and the metal encasing, respectively. Similarly to the proposal in section \ref{FModel}, the coefficients  $C_{lm}$ of Eq.~(\ref{fullsolME}) are calculated using the following continuity conditions for the field 
and its derivative at the boundaries of the different regions: 
\begin{equation}
\varphi_{\rm in1}(r=R_{1}) = \varphi_{\rm out}(r=R_{1}),\qquad
\partial_r\varphi_{\rm in1}(r=R_{1})=\partial_r\varphi_{\rm out}(r=R_{1}),
\nonumber
\end{equation}
\begin{equation}
\varphi_{\rm in2}(r'=R_{2}) = \varphi_{\rm vacuum}(r'=R_{2}),\qquad
\partial_{r'}\varphi_{\rm in2}(r'=R_{2}) = \partial_{r'}\varphi_{\rm vacuum}(r'=R_{2});
\nonumber
\end{equation} 
\begin{equation}
\varphi_{\rm vacuum}(r'=R_{\rm vac}) = \varphi_{\rm encasing}(r'=R_{\rm vac}),\qquad
\partial_{r'}\varphi_{\rm vacuum}(r'=R_{\rm vac}) = \partial_{r'}\varphi_{\rm encasing}(r'=R_{\rm vac});
\nonumber
\end{equation} 
\begin{equation}
\varphi_{\rm encasing}(r'=R_{\rm enc}) = \varphi_{\rm out}(r'=R_{\rm enc}),\qquad
\partial_{r'}\varphi_{\rm encasing}(r'=R_{\rm enc}) = \partial_{r'}\varphi_{\rm out}(r'=R_{\rm enc});
\nonumber
\end{equation} 
Once the $C_{lm}$ coefficients are obtained, both the energy associated with the
total EMT, and then, the chameleon force; could be compute in a very similar way that has been explained in section \ref{sec:force} and appendix \ref{sec:detailsforce} but adding the new regions.

\begin{figure}
{\includegraphics[width=8.5cm,height=8.5cm,angle=-90]{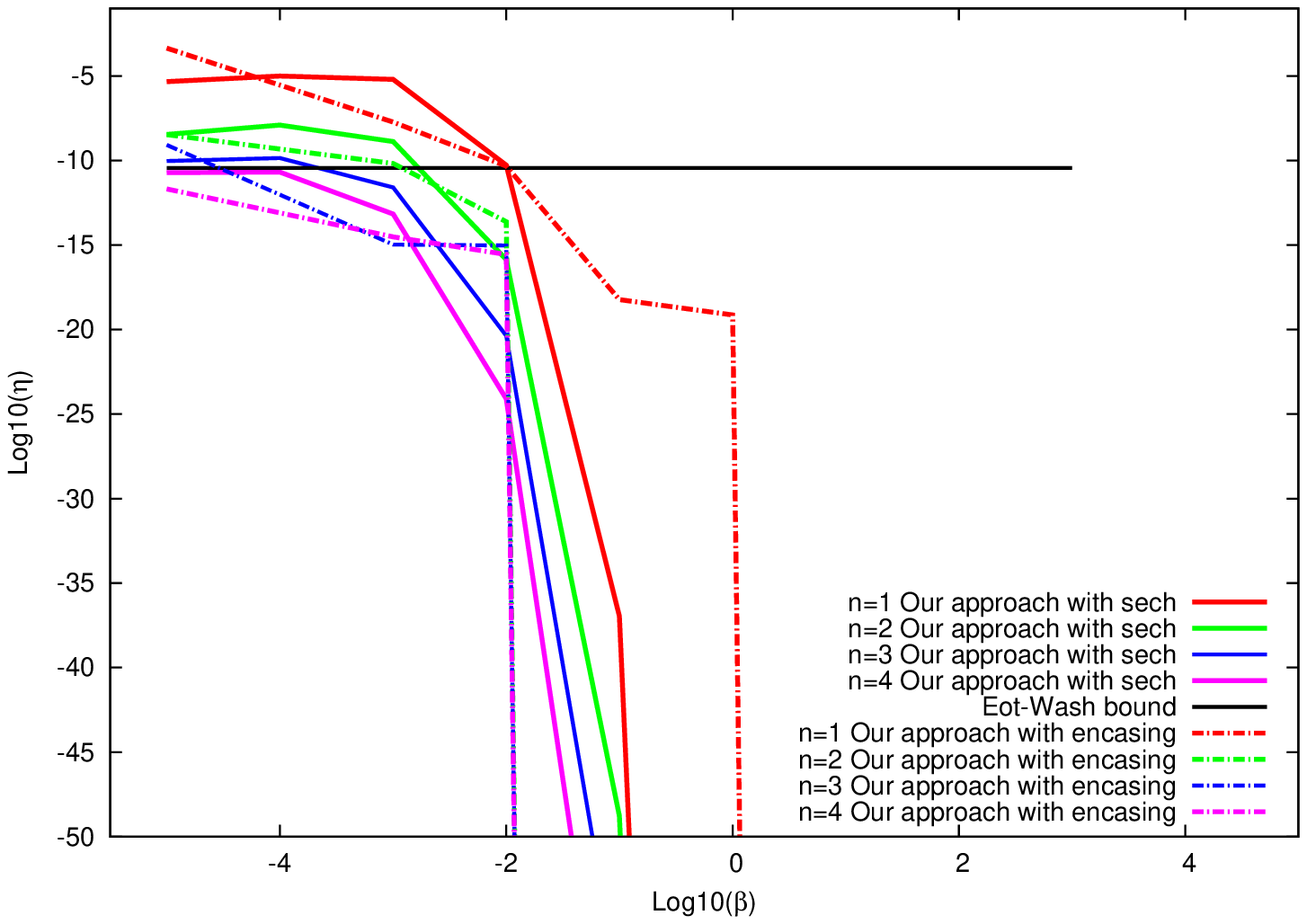}}
\caption{The E\"otv\"os parameter $\eta$ (in ${\rm log_{10}}$ scale) as a function of the parameter $\beta$ (in ${\rm log_{10}}$ scale) including the metal encasing of the vacuum chamber as shown in Fig. \ref{esquemaconshell} for different positive values of $n$ and $M=2.4$ ${\rm meV}$. Here we assume that both the density of the environment surrounding the Earth (and of course, the hill) and inside the vacuum chamber are  $\rho_{\rm out }= 10^{-7} {\rm g \hspace{0.1cm} cm}^{-3}$. We compare our predictions considering the metal encasing in the calculations with those of ours too, but where the effect of the metal encasing is modeled multiplying a factor ${\rm sech(2 m_{\rm shell} d)}$ to the predictions.}
\label{sechvsshell}
\end{figure}

We compare the predictions for the E\"{o}tv\"{o}s parameter obtained under this scheme, with those obtained previously in section \ref{sec:results} (for the E\"{o}t-Wash Torsion Balance Experiments) but applying the Yukawa type suppression ${\rm sech}(2m_{\rm shell}d)$ to these last predictions. We show these results in Fig.\ref{sechvsshell}. For $n=1$ and $\beta>10^{-2}$, the approach including the metal encasing presents an underestimation of the parameters compared to those obtained with the Yukawa suppression; while for $\beta\le 10^{-2}$, it is just the opposite with the exception of $\beta=10^{-5}$. For other the cases, where $n>1$, it is found that in the majority of the analyzed cases there is an underestimation in the scheme that considers the metal encasing against the scheme where the hyperbolic secant is applied. Beyond all these differences, it can be observed that both schemes have a very similar behaviour. For large  values of $\beta$, the suppression due to the metal encasing is very large but becoming small for small values of $\beta$.

\section{The one body problem solution for the chameleon}
\label{sec:OBP}

The quadratic approximation of the potential which consist in assuming
$V_{\rm eff}(\varphi) = \frac{m_{\rm eff}^2}{2}\left(\varphi-\varphi_{\rm min}\right)^2+V_{\rm eff}\left(\varphi_{\rm min}\right)$, is valid when $\varphi$  is not too far from $\varphi_{min}$. Later, in this appendix, we discuss the situations   where  this  approximation is not   a sufficiently  good  one. This is the approximation that we have used throughout this paper. For the one body problem (spherical symmetry) 
we present the analytical solution below.

\begin{itemize}
\item The interior solution ($r<R$) can be expressed:
\begin{equation}
\varphi^{\rm in}(r) = \frac{(\varphi_0 - \varphi_c) \sinh(m_{\rm eff}^{\rm in} r)}{m_{\rm eff}^{\rm in} r} + \varphi_c \qquad (r \le R) \,\,\,,
\end{equation}
where $\varphi_c := \varphi_{\rm min}^{\rm in}$ and the value of the scalar field at the origin is given by 
\begin{equation}
\varphi_0= \varphi_c + 
\frac{(\varphi_\infty - \varphi_c) \left[ 1 + m_{\rm eff}^{\rm out} R \right]}{\frac{m_{\rm eff}^{\rm out} R}{x} \sinh(x) + \cosh(x)} ,\nonumber 
\end{equation}
with $x:= m_{\rm eff}^{\rm in} R$. 
\item The exterior solution ($r>R$) is
\begin{equation}
\label{PhioutTS}
\varphi^{\rm out}(r)= C \frac{\exp{[-m_{\rm eff}^{\rm out} r]}}{r} + \varphi_{\infty} \qquad (r \ge R) \,\,\,,
\end{equation}
where 
\begin{equation}
\label{C2}
C = -\frac{3\beta {\cal M}}{4\pi M_{pl}}\frac{\Delta R}{R} \exp{[m_{\rm eff}^{\rm out} R]} f(x)\,\,\,,
\end{equation}
and 
\begin{eqnarray}
\frac{\Delta R}{R} &=&  -\frac{ 4\pi M_{pl} R (\varphi_c - \varphi_\infty)}{3\beta {\cal M}}= -\frac{(\varphi_c - \varphi_\infty)}{6\beta M_{pl}\Phi_N} \,\,\,,\\
f(x)&:=& \frac{\left[ \cosh(x) - \frac{\sinh(x)}{x}\right] }{\frac{m_{\rm eff}^{\rm out} R}{x} \sinh(x) + \cosh(x)} \,\,\,.
\end{eqnarray}
 $\frac{\Delta R}{R}$ is the so called {\it thin-shell parameter} introduced in~\cite{KW04}, ${\cal M}$ 
and $\Phi_N= \frac{{\cal M}}{8\pi M_{pl}^2 R}$ are the mass and the Newtonian potential at the surface of the body, respectively. These solutions are found by demanding the following requirements: i) regularity condition at $r=0$, i.e., $\varphi^{\rm in}(0)= \varphi_0$, ii) 
continuity and differentiability at $r=R$, iii) the asymptotic condition 
$\varphi^{\rm out}(r \rightarrow \infty) = \varphi_{\rm min}^{\rm out}= \varphi_\infty.$
\end{itemize}

The so called {\it thin-shell} effect appears 
when $1 \ll m_{\rm eff}^{\rm in} R$. So, under this kind of regime the approximate expressions for the field inside and outside the body are: 
\begin{eqnarray}
\varphi^{\rm in}(r) &\approx& \frac{2 (\varphi_\infty - \varphi_c) \exp{[-R m_{\rm eff}^{\rm in}]} \sinh(m_{\rm eff}^{\rm in} r)}{m_{\rm eff}^{\rm in} r} + \varphi_c \qquad (r \le R) \,\,\,, \\
\varphi^{\rm out}(r) &\approx&  R (\varphi_c - \varphi_{\infty}) \frac{\exp{[-m_{\rm eff}^{\rm out}(r-R)]}}{r} + \varphi_{\infty} \,\,\,\qquad \qquad (r \ge R) \,\,\,.
\end{eqnarray}
Note that due to the exponential factor, the $r$ dependence of  $\varphi^{\rm in}(r)$ is strongly  suppressed in regions  well inside the body 
(i.e. $r\ll R$; where $\varphi^{\rm in}(r) \approx \varphi_c$), and it is only within a  {\it thin shell} of size $\Delta R$, which is near the surface of the body, that the field grows exponentially to match the exterior solution $\varphi^{\rm out}(r)$. In addition, when 
$1\ll x$ one verifies that $f(x)\approx 1$ when the density contrast between the body and the environment is high. This is the essence of the ``chameleon effect": the {\it screening} effect that makes the chameleon behave like the electric potential within a conductor.


There is however what is called the {\it thin shell} approximation. This approximation consists in considering the following solutions in three regions
\footnote{In \cite{KW04} (Paper II), $R_{\rm S}$ is denoted $R_{\rm roll}.$} 

\begin{equation}
\label{densOneB2}
\varphi=
\begin{cases}
\varphi_c := \varphi_{\rm min}^{\rm in} \hspace{1cm} 0 \leq r \leq R_{\rm S} \\
\\
 \varphi_{\rm s}(r) \hspace{1cm} R_{\rm S} \leq r \leq  R_{\rm S} + \Delta R= R \\
\\
\varphi^{\rm out}_{\rm thin} (r) \hspace{1cm}   R \leq r  
\end{cases}
\end{equation}
So the interior solution is divided in two parts: one associated to the ``electric conductor'' behaviour of the chameleon $\varphi_c$ where the 
field is almost constant and at the minimum of the potential, and another one corresponding to the {\it thin shell} region $\varphi_s(r) $ where the gradients are 
``high'' and the field start interpolating to the exterior solution $\varphi^{\rm out}_{\rm thin} (r)$ which is supposed to be close to the minimum of the exterior 
effective potential. In this approximation the chameleon solutions 
$\varphi_s$ and $\varphi^{\rm out}_{\rm thin} (r)$ satisfy the corresponding approximate equations:
\begin{eqnarray}
\nabla^2\varphi_s &=& \frac{\partial V_{\rm eff}}{\partial \varphi}\approx \frac{\beta}{M_{pl}} \rho^{\rm in} \,\,\,,\\
\nabla^2\varphi^{\rm out}_{\rm thin}  &=& \frac{\partial V_{\rm eff}}{\partial \varphi}\approx  m_{\rm eff}^{2\,\,{\rm out}} \left(\varphi^{\rm out}_{\rm thin}- \varphi_\infty \right)  \,\,\,,
\end{eqnarray}
where $\varphi_\infty:= \varphi^{\rm out}_{\rm min}$.

The solutions of these two equations are, respectively
\begin{eqnarray}
\varphi_s (r) &=&  \frac{1}{M_{pl}}\left[\frac{\beta\rho^{\rm in} r^2}{6} + \frac{A}{r} + B\right]   \hspace{1cm} (R_{\rm S} \leq r \leq  R) \,\,,\\
\varphi^{\rm out}_{\rm thin} (r)&=& C^{\rm out}_{\rm thin} \frac{e^{-m_{\rm eff}^{\rm out} r}}{r} + \varphi_\infty \hspace{1cm} (R \leq r)
\end{eqnarray}
where $A,B, C^{\rm out}_{\rm thin}$ are integration constants.

When matching the thin shell solution continuously and smoothly with $\varphi_c$ ar $R_{\rm S}$  we find
\begin{equation}
\varphi_s (r)= \frac{\beta\rho^{\rm in}}{6 M_{pl}}\left(r^2 +\frac{2R_{\rm S}^3}{r}\right) -\frac{\beta\rho^{\rm in}}{2 M_{pl}}R_{\rm S}^2 + \varphi_c
 \hspace{1cm} (R_{\rm S} \leq r \leq  R)
\end{equation}

On the other hand, when matching $\varphi_s(r)$ continuously and smoothly with $\varphi^{\rm out}_{\rm thin} (r)$ at the border of the body $r=R$ 
one finds that the constants $C^{\rm out}_{\rm thin}$ and $R_{\rm S}$ are obtained from the following algebraic equations:
\begin{eqnarray*}
C^{\rm out}_{\rm thin} &=& e^{m_{\rm eff}^{\rm out} R}\left\{
\left(\varphi_c-\varphi_\infty\right)R + \frac{\beta\rho^{\rm in} R^3}{6 M_{pl}}\left[ 1- 3 \frac{R_{\rm S}^2}{R^2}+ 2 \frac{R_{\rm S}^3}{R^3}\right]\right\}\,\,\,,\\
&&- \frac{C^{\rm out}_{\rm thin} }{R^2} e^{-m_{\rm eff}^{\rm out} R}\left(R m_{\rm eff}^{\rm out} + 1\right) =  \frac{\beta\rho^{\rm in} R}{3 M_{pl}}\left(1- \frac{R_{\rm S}^3}{R^3}\right)
\end{eqnarray*}
The most interesting approximation for these constants and thus, for the solution itself is when assuming that the thin shell parameter 
$\Delta R/R= 1- R_{\rm S}/R$ is small $\Delta R/R\ll 1$. Moreover, if one assumes $m_{\rm eff}^{\rm out} R\ll 1$ (i.e. the Compton wave length of the chameleon in 
the exterior region is large compared to the body's size) one finds
\begin{equation}
\frac{\Delta R}{R} \approx -\frac{\left(\varphi_c-\varphi_\infty\right) }{6 \beta M_{pl} {\cal M}G/R} = -\frac{\left(\varphi_c-\varphi_\infty\right) }{6 \beta M_{pl} \Phi_N}\,\,\,,
\end{equation}
 where $\Phi_N= G{\cal M} /R$ stands for the Newtonian potential at the surface of the body and ${\cal M}= 4\pi \rho^{\rm in} R^3/3$ is the mass of the body. 

We also find the approximate expression,

\begin{equation}
C^{\rm out}_{\rm thin} \approx R \left(\varphi_c-\varphi_\infty\right) = -\frac{3}{4\pi} \frac{\beta {\cal M}}{M_{pl}} \frac{\Delta R}{R} \,\,\,.
\end{equation}

In this way, we find the approximate solution
\begin{equation}
\varphi^{\rm out}_{\rm thin} (r) \approx  - \frac{\beta}{4\pi M_{pl}} \frac{3 \Delta R}{R} {\cal M} \frac{e^{-m_{\rm eff}^{\rm out}r}}{r} + \varphi_\infty \hspace{1cm} (R \leq r)
\end{equation}
We appreciate the way the $r$ dependence of the exterior solution is suppressed by the thin shell parameter.
\bigskip

One could be also interested in the {\it thick shell} solution corresponding to $R_{\rm S}\rightarrow 0$ or equivalently, $\Delta R \approx R$. In this case 
the most internal region $0 \leq r \leq R_{\rm S}$ ``disappears'' and the solution $\varphi_c\rightarrow\varphi_0$, i.e., it is the value of the field at $r=0$. 
So there is only one interior solution $\varphi_s(r) \rightarrow \varphi_{\rm in}(r) $ and one exterior one $\varphi^{\rm out}_{\rm thick} (r) $, provided respectively 
by
\begin{eqnarray}
\varphi_{\rm in}(r) &=& \frac{\beta \rho^{\rm in}}{6 M_{pl}} r^2 + \varphi_0  \hspace{1cm} (0 \leq r \leq R) \,\,\,,\\
\varphi^{\rm out}_{\rm thick} (r) &=& C^{\rm out}_{\rm thick}  \frac{e^{-m_{\rm eff}^{\rm out} r}}{r} + \varphi_\infty \hspace{1cm} (R \leq r) \,\,\,.
 \end{eqnarray}
When  matching continuously and smoothly both solutions at $r=R$ we determine the values of the constants $\varphi_0$ and $C^{\rm out}_{\rm thick}$. These are obtained from
\begin{eqnarray}
C^{\rm out}_{\rm thick}  &=& R e^{m_{\rm eff}^{\rm out} R}\left\{
\left(\varphi_0-\varphi_\infty\right) + \frac{\beta\rho^{\rm in} R^2}{6 M_{pl}}\right\}\,\,\,,\\
&&- \frac{C^{\rm out}_{\rm thick}}{R^2} e^{-m_{\rm eff}^{\rm out} R}\left(R m_{\rm eff}^{\rm out} + 1\right) =  \frac{\beta\rho^{\rm in} R}{3 M_{pl}} \,\,\,.
\end{eqnarray}
Again, when assuming $m_{\rm eff}^{\rm out} R\ll 1$ one finds
\begin{eqnarray}
C^{\rm out}_{\rm thick}  & \approx & R \left\{ \left(\varphi_0-\varphi_\infty\right) + \frac{\beta\rho^{\rm in} R^2}{6 M_{pl}}\right\}\,\,\,,\\
&& C^{\rm out}_{\rm thick} \approx  -\frac{\beta\rho^{\rm in} R^3}{3 M_{pl}} \,\,\,.
\end{eqnarray}

So,
\begin{equation}
\varphi_0 \approx \varphi_\infty -\frac{\beta\rho^{\rm in} R^2}{2 M_{pl}} = \varphi_\infty - 3\beta M_{pl} \Phi_N \,\,\,.
\end{equation}

Finally,
\begin{eqnarray}
\varphi^{\rm out}_{\rm thick} (r) &\approx&  -\frac{\beta\rho^{\rm in} R^3}{3 M_{pl}}  \frac{e^{-m_{\rm eff}^{\rm out} r}}{r} + \varphi_\infty \\
&=& -\frac{\beta {\cal M} }{4\pi M_{pl}}  \frac{e^{-m_{\rm eff}^{\rm out} r}}{r} + \varphi_\infty \hspace{1cm} (R \leq r) \,\,\,.
\end{eqnarray}
Thus, when comparing this solution with the approximate solution for $\varphi^{\rm out}_{\rm thin} (r)$ we see that the $r$ dependence of the {\it thick shell} exterior solution 
is {\it not} suppressed by the thin shell parameter.\\

 In our approach, we always approximate the resulting effective potentials (in each of the  various  bodies and surrounding media present in the problem) up to quadratic order in $\varphi$ around their corresponding 
minima. Thus,  our effective potential is  described,  in   each region  by  quadratic expressions  in $\varphi$, for the whole domain ${\mathbb{R}}^3$.

  In particular, inside the body $0<r<R$  we  use   a single  quadratic expression 
while, as mentioned above, in the KW approach, the potential is taken  to  be linear in $\varphi$ in the region $R_{\rm roll}<r<R$. Furthermore, in order to determine  which of the two approximations  is  better  we rely  on the energy functional  criteria    described  in section III   which  takes into account the exact form  of the effective potential. It follows from the results of  Section \ref{sec:energy} that when the large body does not satisfy the {\it thin shell} condition as characterized  within the standard approach, the latter approximation to the effective potential is better than ours. In a future work \cite{Krai17} we will consider 
an adaptive  method interpolating  between the two  approximations  and  use the energy criteria developed in this work to establish if   does it provide a   better  approximation for all   values of the parameters.

\end{document}